\documentclass[fleqn,usenatbib,useAMS]{mnras}

\usepackage{float}
\usepackage{natbib}
\usepackage{hhline}
\usepackage{multirow}
\usepackage{graphicx}
\usepackage{ae,aecompl}
\usepackage{deluxetable}
\usepackage{amssymb, amsmath}
\usepackage[usenames, dvipsnames]{xcolor}
\usepackage{xcolor,colortbl}
\usepackage{CJKutf8}
\usepackage{fontawesome5}
\usepackage{hyperref}
\usepackage{xurl}

\definecolor{orcidlogocol}{HTML}{A6CE39}

\hypersetup{colorlinks=true,
            citecolor=MidnightBlue,
            linkcolor=MidnightBlue,
            filecolor=magenta,
            urlcolor=cyan}
\DeclareGraphicsExtensions{.pdf,.png,.jpg}

\defcitealias{Huang2018b}{Paper~I}	
\defcitealias{Huang2018c}{Paper~II}

\def\aj{{AJ}}

\def\apj{{ApJ}}
\def\apjs{{ApJS}}
\def\mnras{{MNRAS}}

\def\pasp{{PASP}}

\def\h{\hskip -3 mm}

\def\lax{{$\mathrel{\hbox{\rlap{\hbox{\lower4pt\hbox{${\sim}$}}}\hbox{$<$}}}$}}
\def\gax{{$\mathrel{\hbox{\rlap{\hbox{\lower4pt\hbox{${\sim}$}}}\hbox{$>$}}}$}}
\def\simlt{\lower.5ex\hbox{$\; \buildrel < \over {\sim} \;$}}
\def\simgt{\lower.5ex\hbox{$\; \buildrel > \over {\sim} \;$}}

\def\etal{{\ et al.}}

\def\mlratio{{$M_{\star}/L_{\star}$}}
\def\snratio{{$\mathrm{S}/\mathrm{N}$}}

\def\smag{mag~arcsec$^{-2}$}
\def\ser{{S\'{e}rsic}}
\def\cmodel{\texttt{CModel}}

\def\mstar{{$M_{\star}$}}
\def\logms{{$\log_{10} (M_{\star}/M_{\odot})$}}

\def\logmvir{{$\log_{10} M_{\mathrm{vir}}$}}

\def\logmmax{{$\log_{10} (M_{\star,\mathrm{max}}/M_{\odot})$}}

\def\logmcmodel{{$\log_{10} (M_{\star,\mathrm{CMod}}/M_{\odot})$}}
\def\logm10{{$\log (M_{\star,10\ \mathrm{kpc}}/M_{\odot})$}}
\def\logm30{{$\log (M_{\star,30\ \mathrm{kpc}}/M_{\odot})$}}
\def\logm50{{$\log (M_{\star,50\ \mathrm{kpc}}/M_{\odot})$}}
\def\logm100{{$\log (M_{\star,100\ \mathrm{kpc}}/M_{\odot})$}}

\def\mvir{{$M_{\mathrm{vir}}$}}
\def\mhalo{{$M_{\mathrm{vir}}$}}

\def\mh200b{{$M_{\mathrm{200b}}$}}
\def\mh200c{{$M_{\mathrm{200c}}$}}

\newcommand{\maper}[1]{\ensuremath{M_{\star, {#1}\ \rm kpc}}}
\newcommand{\logmaper}[1]{{$\log (M_{\star, {#1}\ \rm kpc}/M_{\odot})$}}

\newcommand{\menve}[2]{{$M_{\star, [#1, #2]}$}}

\def\insitu{{\textit{in situ}}}
\def\exsitu{{\textit{ex situ}}}

\def\sigms{{$\sigma_{M_{\star}}$}}

\def\sigmh{{$\sigma_{M_{\mathrm{vir}}}$}}
\def\sigmvir{{$\sigma_{M_{\mathrm{vir}}}$}}

\def\mmax{{$M_{\star,\mathrm{max}}$}}
\def\mcmodel{\ensuremath{M_{\star,\mathrm{cmod}}}}
\def\masap{{$M_{\rm vir,\ ASAP}$}}

\def\smdpl{\texttt{SMDPL}}
\def\mdpl2{\texttt{MDPL2}}

\def\asap{\texttt{ASAP}}
\def\redm{\texttt{redMaPPer}}
\def\camira{\texttt{CAMIRA}}

\def\ellipse{\texttt{Ellipse}}
\def\iraf{\texttt{IRAF}}
\def\hscpipe{\texttt{hscPipe}}
\def\topn{{Top-$N$}}

\def\dsigma{{$\Delta\Sigma$}}
\def\rdsigma{{$R \times \Delta\Sigma$}}
\def\chisq{{$\chi^2$}}
\def\sqdeg{{deg$^2$}}

\definecolor{LightGray}{gray}{0.85}
\definecolor{Tab1}{RGB}{114, 158, 206}
\definecolor{Tab2}{RGB}{255, 158,  74}
\definecolor{Tab3}{RGB}{103, 191,  92}
\definecolor{Tab4}{RGB}{174, 199, 232}
\definecolor{Tab5}{RGB}{255, 187, 120}
\definecolor{Tab6}{RGB}{152, 223, 138}
\definecolor{Tab7}{RGB}{255, 152, 150}
\definecolor{Tab8}{RGB}{197, 176, 213}
\definecolor{hpurple}{HTML}{7E16DF}

\newcommand{\obsSym}{\ensuremath{{\mathcal{O}}}}
\newcommand{\haloSym}{\ensuremath{{\mathcal{M}}}}
\newcommand{\slope}{\ensuremath{\alpha}}
\newcommand{\intercept}{\ensuremath{\pi}}
\newcommand{\hmf}{\Phi(\mu)}
\newcommand{\scatterObsSymMhalo}{\ensuremath{\sigma_{\mathcal{O} | \mathcal{M}}}}
\newcommand{\scatterMhaloObsSym}{\ensuremath{\sigma_{\mathcal{M} | \mathcal{O}}}}
\newcommand{\eg}{{\it e.g.,\/}}

\newcommand{\xxx}[1]{\textcolor{red}{\textbf{XXX}}}

\title[Outer Galaxy Mass as a Halo Mass Proxy]{
    The Outer Stellar Mass of Massive Galaxies: A Simple Tracer of Halo Mass with Scatter 
    Comparable to Richness and Reduced Projection Effects}

\author[S. Huang et al.]{
        Song Huang (黄崧)\ \href{https://orcid.org/0000-0003-1385-7591}{\textcolor{orcidlogocol}{\faOrcid}}$^{1,2}$\thanks{E-mail: sh19@astro.princeton.edu (SH)},
        Alexie Leauthaud\ \href{https://orcid.org/0000-0002-3677-3617}{\textcolor{orcidlogocol}{\faOrcid}}$^{2}$,
        Christopher Bradshaw\ \href{https://orcid.org/0000-0003-0833-573X}{\textcolor{orcidlogocol}{\faOrcid}}$^{2}$,
        \newauthor
        Andrew Hearin\ \href{https://orcid.org/0000-0003-2219-6852}{\textcolor{orcidlogocol}{\faOrcid}}$^{3}$,
        Peter Behroozi\ \href{https://orcid.org/0000-0002-2517-6446}{\textcolor{orcidlogocol}{\faOrcid}}$^{4}$,
        Johannes Lange\ \href{https://orcid.org/0000-0002-2450-1366}{\textcolor{orcidlogocol}{\faOrcid}}$^{2, 5}$,
        Jenny Greene\ \href{https://orcid.org/0000-0002-5612-3427}{\textcolor{orcidlogocol}{\faOrcid}}$^{1}$,
        \newauthor
        Joseph DeRose\ \href{https://orcid.org/0000-0002-0728-0960}{\textcolor{orcidlogocol}{\faOrcid}}$^{6}$,
        Joshua S. Speagle (沈佳士)\ \href{https://orcid.org/0000-0002-5065-9896}{\textcolor{orcidlogocol}{\faOrcid}}$^{7, 8, 9}$,
        Enia Xhakaj$^{2}$\\
        $^{1}$Department of Astrophysical Sciences, Peyton Hall,
              Princeton University, Princeton, NJ 08540, USA \\
        $^{2}$Department of Astronomy and Astrophysics, University of California
              Santa Cruz, 1156 High St., Santa Cruz, CA 95064, USA\\
        $^{3}$Argonne National Laboratory, Argonne, IL 60439, USA\\
        $^{4}$Department of Astronomy and Steward Observatory, University of Arizona,
              Tucson, AZ 85721, USA\\
        $^{5}$Kavli Institute for Particle Astrophysics and Cosmology and Department of Physics, 
            Stanford University, CA 94305, USA\\
        $^{6}$Lawrence Berkeley National Laboratory, 1 Cyclotron Road, Berkeley, CA 93720, USA\\
        $^{7}$Department of Statistical Sciences, University of Toronto, Toronto, M5S 3G3, Canada\\
        $^{8}$David A. Dunlap Department of Astronomy \& Astrophysics, University of Toronto, 
            Toronto, M5S 3H4, Canada\\
        $^{9}$Dunlap Institute for Astronomy \& Astrophysics, University of Toronto, 
            Toronto, M5S 3H4, Canada
        }

\date{Accepted 2021. Received 2021; in original form 2021 Sep 1}
\pubyear{2021}

\begin{document}

\begin{CJK*}{UTF8}{gbsn}
\label{firstpage}
\pagerange{\pageref{firstpage}--\pageref{lastpage}}

\maketitle

\begin{abstract}
    
    Using the weak gravitational lensing data from the Hyper Suprime-Cam Subaru Strategic Program 
    (HSC survey), we study the potential of different stellar mass estimates in tracing halo mass.
    We consider galaxies with \logms{}$>11.5$ at $0.2 < z < 0.5$ with carefully measured light 
    profiles, and clusters from the \redm{} and \camira{} richness-based algorithms.
    We devise a method (the ``\topn{} test'') to evaluate the scatter in the halo mass-observable
    relation for different tracers, and to inter-compare halo mass proxies in four number density 
    bins using stacked galaxy-galaxy lensing profiles.
    This test reveals three key findings. 
    Stellar masses based on \cmodel{} photometry and aperture luminosity within $R<$30 kpc are 
    poor proxies of halo mass. 
    In contrast, the stellar mass of the outer envelope is an excellent halo mass proxy. 
    The stellar mass within $R=[50,100]$ kpc, \menve{50}{100}, has performance comparable to the
    state-of-the-art richness-based cluster finders at \logmvir{}$\gtrsim 14.0$ and could be a
    better halo mass tracer at lower halo masses. 
    Finally, using N-body simulations, we find that the lensing profiles of massive halos selected
    by \menve{50}{100} are consistent with the expectation for a sample without projection or
    mis-centering effects. 
    Richness-selected clusters, on the other hand, display an excess at $R\sim 1$ Mpc in their 
    lensing profiles, which may suggest a more significant impact from selection biases.
    These results suggest that \mstar{}-based tracers have distinct advantages in
    identifying massive halos, which could open up new avenues for cluster cosmology.
    The codes and data used in this work can be found here: 
    \href{https://github.com/dr-guangtou/jianbing}{\faGithub}

\end{abstract}


\section{Introduction}
    \label{sec:intro}
    
    With the rapid developments of multi-wavelength sky surveys, galaxy clusters have become
    increasingly important for studies of cosmology and the galaxy-halo connection.
    As the rare highest density peaks of the matter density distribution, galaxy clusters have long
    been recognised as powerful probes of the mean cosmic matter density ($\Omega_{\rm m}$), the
    amplitude of the power spectrum ($\sigma_{8}$), and the cosmic expansion (e.g.,
    \citealt{Evrard1989, Peebles1989, White1993, Viana1996, Wang1998, Wagoner2021}; see
    \citealt{Allen2011, Kravtsov2012, Weinberg2013} for recent reviews).
    The abundance, spatial distribution, and total mass distributions of galaxy clusters encode
    valuable cosmological information (see, e.g., \citealt{Haiman2001, Holder2001, Vikhlinin2009b,
    Rozo2010, Benson2013, Mantz2014, Bocquet2019, Abbott2020, To2021a, To2021b, Wu2021}). 
    Galaxy clusters are also promising laboratories for studying the boundaries of dark matter halos
    (e.g., \citealt{Diemer2014, More2015b, More2016, Chang2018, Shin2019, Zurcher2019, Tomooka2020,
    Xhakaj2020})\footnote{Please see
    \href{http://www.benediktdiemer.com/research/splashback/}{Benedikt Diemer's webpage} for a more
    complete list of reference on this topic.}, and for investigating halo assembly bias (e.g.,
    \citealt{Tinker2012, Miyatake2016, Zu2017}).
    To achieve these goals, a reliable ``cluster finder'' that can identify galaxy clusters 
    is fundamental. 
    In addition the identification of clusters, it is also critical to be able to measure the 
    halo masses of clusters, as well as to calibrate halo mass--observable scaling relations.

    Thanks to the advent of large optical surveys such as the Sloan Digital Sky Survey (SDSS,
    \citealt{York2000, SDSSDR7, SDSSDR16})\footnote{\url{https://www.sdss.org/}}, 
    the Dark Energy Survey (DES, \citealt{DES2016, Abbott2018,
    DES2021})\footnote{\url{https://www.darkenergysurvey.org/}}, and the Hyper Suprime-Cam Subaru
    Strategic Program (HSC-SSP, \citealt{Miyazaki2012, HSC-SSP, HSC-DR1,
    HSC-DR2})\footnote{\url{https://hsc.mtk.nao.ac.jp/ssp/}}, optical cluster finders
    are widely used to construct cluster samples (e.g., \citealt{Kepner1999,
    GladdersYee2000, Koester2007, Hao2010, Wen2012, Rykoff2014, Oguri2018, Aguena2021, Wen2021,
    Zou2021}), and weak gravitational lensing is also regarded
    as the most promising approach for calibrating mass-observable relations
    (e.g., \citealt{Leauthaud2010, Becker2011, vonderLinden2014, Applegate2014, 
    Applegate2016, Okabe2016, Grandis2019}; also see \citealt{Umetsu2020b} for a recent review).
    Among optical cluster finders, red--sequence based methods such as \redm{} (e.g.,
    \citealt{Rykoff2014, Rozo2014, Rozo2015a, Rozo2015b, Rykoff2016}) and \camira{} (e.g.,
    \citealt{Oguri2014, Oguri2018}) are among the most widely-used in the literature.

    While red-sequence cluster finders enjoy many successes, these methods are subject to numerous
    potential sources of systematic error, such as anisotropic selection biases (including both
    projection bias and orientation bias; e.g., \citealt{NohCohn2012, Dietrich2014, Osato2018,
    Herbonnet2019}) and mis-centering (e.g., \citealt{Saro2015, Zhang2019b}).
    Projection effects arising from structures surrounding the clusters in the line-of-sight
    direction raise a number of especially challenging difficulties (e.g., \citealt{Cohn2007,
    Erickson2011, Farahi2016, Zu2017, Busch2017, Costanzi2019, Sunayama2019, Sunayama2020}). 
    In particular, projection effects can significantly complicate the calibration of 
    the mass-richness relation, which in turn impacts cosmological inference 
    (e.g., \citealt{Erickson2011, Costanzi2019, Sunayama2020, Wu2021}).
    In \citet{DES2020}, the authors conclude that projection effects alone can lead to a $\sim 20$\%
    over-estimate of halo mass in a given richness bin, and could lead to a  ``tension'' with a
    Planck 2018 cosmology (e.g., \citealt{PLANCK2020}). 
    However, it is difficult to precisely evaluate the impact of projection effects on red-sequence
    cluster finders, because such a quantification requires realistic mock catalogues of cluster
    galaxies with red-sequences that are consistent with observations, which is not an easy task
    (e.g., \citealt{DeRose2019, Korytov2019}).

    In this context, it is of great interest to study potential alternative methods that might 
    suffer less from projection effects. 
    One example is to use the light from massive central galaxies (or the brightest cluster galaxy,
    BCG). The stellar mass of the BCG follows a well-established stellar-halo
    mass relation (SHMR, e.g., \citealt{Leauthaud2012, Tinker2017, Kravtsov2018}; also see
    \citealt{Wechsler2018} for a recent review) with moderate scatter at the high-mass end (e.g.,
    \citealt{More2009, Leauthaud2012, Reddick2013, Zu2015, Lehmann2017, Kravtsov2018}).
    Historically, BCG luminosity or stellar mass has not been considered as a competitive halo mass
    proxy, but optical surveys have also struggled to accurately measure BCG total luminosity (e.g.,
    \citealt{Bernardi2013, Huang2018b}). 
    Recently, deep imaging surveys have showed that total BCG luminosity may correlate well with 
    halo mass (e.g., \citealt{Huang2018c, SampaioSantos2021}).
    In \citet{Huang2020}, for example, the authors showed that a simple phenomenological model based
    on the stellar masses of BCGs measured within two apertures further reduces the scatter in the
    halo mass -- observable relation. 
    In addition, recent work has also highlighted the connection between the diffuse envelope around
    a BCG (often referred to as the Intra-Cluster Light, or ICL) and dark matter halo mass (e.g.,
    \citealt{Montes2018, Montes2019, Zhang2019b, Furnell2021}).

    In this paper, we use data from the HSC survey to quantify the potential of using BCG light to
    identify massive clusters. 
    We design a so-called \topn{} test to evaluate their relative performance with respect to the
    red-sequence methods. 
    The \topn{} test compares the stacked galaxy--galaxy lensing profiles (the excess surface
    density profiles, or the \dsigma{} profiles) of ``clusters'' selected by different halo mass
    proxies in fixed number density bins (e.g., \citealt{Reyes2008}). 
    We model these lensing signals using cosmological simulations and evaluate the scatter of the
    halo mass-observable relations.  
    Section \ref{sec:method} explains the philosophy behind the \topn{} test and the methodology 
    for estimating the scatter in mass-observable relations. 
    Section \ref{sec:data} presents the data and Section \ref{sec:measure} presents key
    measurements, including different \mstar{} measurements based on 1-D mass profiles and
    galaxy-galaxy lensing profiles. 
    Section \ref{sec:proxies} presents the different proxies that we test. 
    Our results are presented in Section \ref{sec:result} and discussed in Section
    \ref{sec:discussion}. 
    Finally, we summarise and conclude in Section \ref{sec:summary}.

    We assume $H_0$ = 70~km~s$^{-1}$ Mpc$^{-1}$, ${\Omega}_{\rm m}=0.3$, and ${\Omega}_{\rm
    \Lambda}=0.7$.
    Stellar mass (\mstar{}) is derived using a Chabrier Initial Mass Function (IMF;
    \citealt{Chabrier2003}).
    We adopt $M_{\rm vir}$ for dark matter halo mass as defined in \citealt{BryanNorman1998}.
    We use $\mathcal{M}\equiv \log_{10} (M_{\rm vir}/M_{\odot})$ and 
    $\mathcal{O}\equiv \log_{10} \rm Observable$ to indicate the ten-base logarithms of
    halo mass and observables.
    We also use \sigmvir{}$\equiv \sigma_{\log_{10} M_{\rm vir}}$ for the scatter of 
    halo mass and \sigms{}$\equiv \sigma_{\log_{10} M_{\star}}$ for the scatter of stellar mass. 
    
\begin{figure*}
\includegraphics[width=\textwidth]{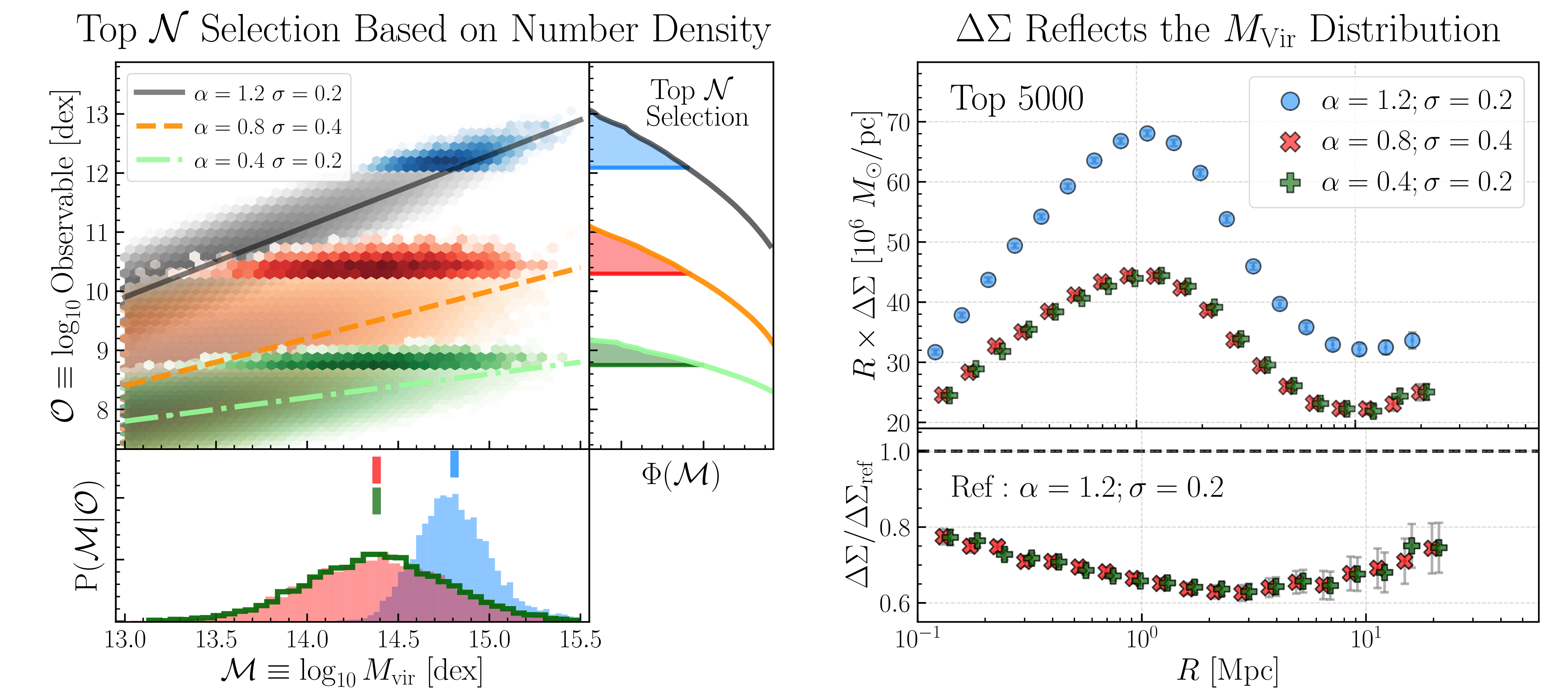}
\caption{
    \textbf{Left:} coloured hexbins and corresponding lines illustrate
        three mock \mvir{}--observable relations with different slope and scatter values
        (grey: $\alpha=1.2$ and \scatterObsSymMhalo{}$=0.2$; 
        orange: $\alpha=0.8$ and \scatterObsSymMhalo{}$=0.4$;
        light green: $\alpha=0.4$ and \scatterObsSymMhalo{}$=0.2$).
        Hexbins with darker colours highlight the top 5000 objects for each observable.
        The right panel shows the number density distributions of these observables
        and highlights the \topn{} selections using shaded regions with corresponding 
        colours.
        The bottom panel shows the \mvir{} distributions of the three \topn{} bins.
        We use short vertical bars to indicate the mean \mvir{} values.
    \textbf{Right:} stacked \rdsigma{} profiles of the three \topn{} samples using the 
        \mdpl2{} simulation. 
        The bottom panel shows the ratios of lensing profiles using the sample with the steepest
        slope (blue) as a reference.
    The \texttt{Jupyter} notebook for reproducing this figure can be found here:
    \href{https://github.com/dr-guangtou/jianbing/blob/master/notebooks/figure/fig1.ipynb}{\faGithub}.
    }
    \label{fig:theory_1}
\end{figure*}

\section{Methodology and Modelling Framework}
    \label{sec:method}

    This section explains the basic idea of the \topn{} test and presents our modelling framework 
    for estimating the scatter in \mvir{}-observable relations.
    
\subsection{Philosophy of the \texorpdfstring{\topn{}}{TopN} Test}
    \label{sec:topn_intro}
    
    Cosmological simulations permit a precise prediction for how \dsigma{} of dark matter halos
    scales with true halo mass, \mvir{}.
    Since simulated halos can easily be rank-ordered by their mass, it is a trivial matter to use a
    cosmological simulation to generate a prediction for the \dsigma{} profiles of samples of dark
    matter halos that have been stacked according to number density.
    The philosophy behind the \topn{} test is to capitalise upon this predictive power.
    When analysing observational data, of course one does not have direct access to true halo mass, 
    and so one must instead rely upon an observational proxy. 
    In the analogous manner as can be done for simulated halos, observed galaxy clusters can be
    arranged into stacked samples according to any particular halo mass proxy, and so it is equally
    straightforward to measure the \dsigma{} profile of clusters as a function of the number density
    defined by the choice of proxy. 
    When the halo mass proxy presents a scaling relation with \mvir{} that has low scatter and a
    steep slope, then the associated stacked samples will exhibit a lensing amplitude that scales
    steeply with number density, and the stacks will furthermore exhibit \dsigma{} profiles whose
    shape closely resembles the profile of \mvir{}-ranked stacks of simulated halos of the
    corresponding number density.
    For example, by comparing the stacked \dsigma{} profile of the top 100 most massive galaxies
    with the \dsigma{} profile of the top 100 richest clusters selected in the same survey volume,
    one can compare which of these proxies is more ``\mvir{}-like''.
    In this manner, the \topn{} test compares $\Delta\Sigma$ for cluster samples defined in bins of
    fixed number density, and uses such comparisons to inform the optimal choice of halo mass proxy.
    
    Figure \ref{fig:theory_1} illustrates the main idea of the \topn{} test using halos from the
    MultiDark Planck 2 (\mdpl2{})\footnote{\url{https://www.cosmosim.org/cms/simulations/mdpl2/}} 
    simulation. 
    In this exercise, each halo is characterised by its true halo mass, \haloSym{}$\equiv
    \log_{10}M_{\rm Vir},$ and additionally by three hypothetical observables, \obsSym{}$\equiv
    \log_{10}{(\rm Observable)}$.
    We assume that each \obsSym{} follows a $\log$-linear scaling relation with \mvir{} that is
    characterised by a value for the slope, $\alpha$, and by a level of Gaussian scatter in
    \obsSym{} at fixed \haloSym{}, \scatterObsSymMhalo{} (e.g., \citealt{Lieu2016},
    \citealt{Ziparo2016}, \citealt{Evrard2014, Farahi2018}).
    We then select the top $N=5000$ objects using \obsSym{} to rank-order the clusters.
    The value of $N$ translates into a fixed volume number density threshold shown on the
    right sub-panel using the number density distributions of these observables.
    When comparing the \mvir{} distributions of the \topn{}-selected samples (bottom panel), the
    halo mass proxies with smaller $\alpha$ and/or larger \scatterObsSymMhalo{} result in \mvir{}
    distributions with larger \scatterMhaloObsSym{} and lower mean \mvir{}.
    This selection at fixed number density (\topn{} selection) yields a \mvir{} distribution that
    reflects the properties of the underlying \mvir{}-observable relation.
    Figure \ref{fig:theory_1} also shows the \emph{stacked} \dsigma{} profiles of these \topn{}
    samples.
    The \topn{} sample with a higher mean \mvir{} and a lower value of \scatterMhaloObsSym{} has a 
    \dsigma{} profile with larger overall amplitude.
    We can therefore use the \emph{stacked} \dsigma{} profiles of different \topn{} samples to probe
    their underlying \mvir{}--observable relations. 
    \citet[][]{Reyes2008} applied a similar method to develop improved halo mass tracers of
    clusters.
    
    The right panel of Figure \ref{fig:theory_1} illustrates that the ratio of \dsigma{} profiles
    exhibits scale-dependent features that reveal subtle differences in other halo properties, and
    also in large-scale environment. 
    Our use of \topn{} tests in this paper will additionally leverage the discriminating power of
    this scale-dependence when assessing various halo mass proxies.

    Finally, Figure \ref{fig:theory_1} also reveals a degeneracy between the slope, $\alpha,$ and
    the scatter, \scatterObsSymMhalo{}, such that different combinations of $\alpha$ and
    \scatterObsSymMhalo{} can produce identical \mvir{} distributions.
    We discuss this degeneracy in the next section (\S\ \ref{sec:comp_scatters}).

\subsection{Modelling Methodology}
    \label{sec:model}

    To quantitatively interpret differences between \dsigma{} profiles, we develop a simple
    forward-modelling method based on data from the \mdpl2{} and the Small MultiDark Planck
    (\smdpl{})\footnote{\url{https://www.cosmosim.org/cms/simulations/smdpl/}} N-body simulations
    (e.g., \citealt{Klypin2016}).
    We assume a $\log$-linear \mvir{}-observable relation with a constant Gaussian scatter. 
    We use this model to estimate the scatter in the mass-proxy relation, and also to
    infer the underlying \mvir{} distribution of different samples.

\begin{figure*}
    \centering
    \includegraphics[width=\textwidth]{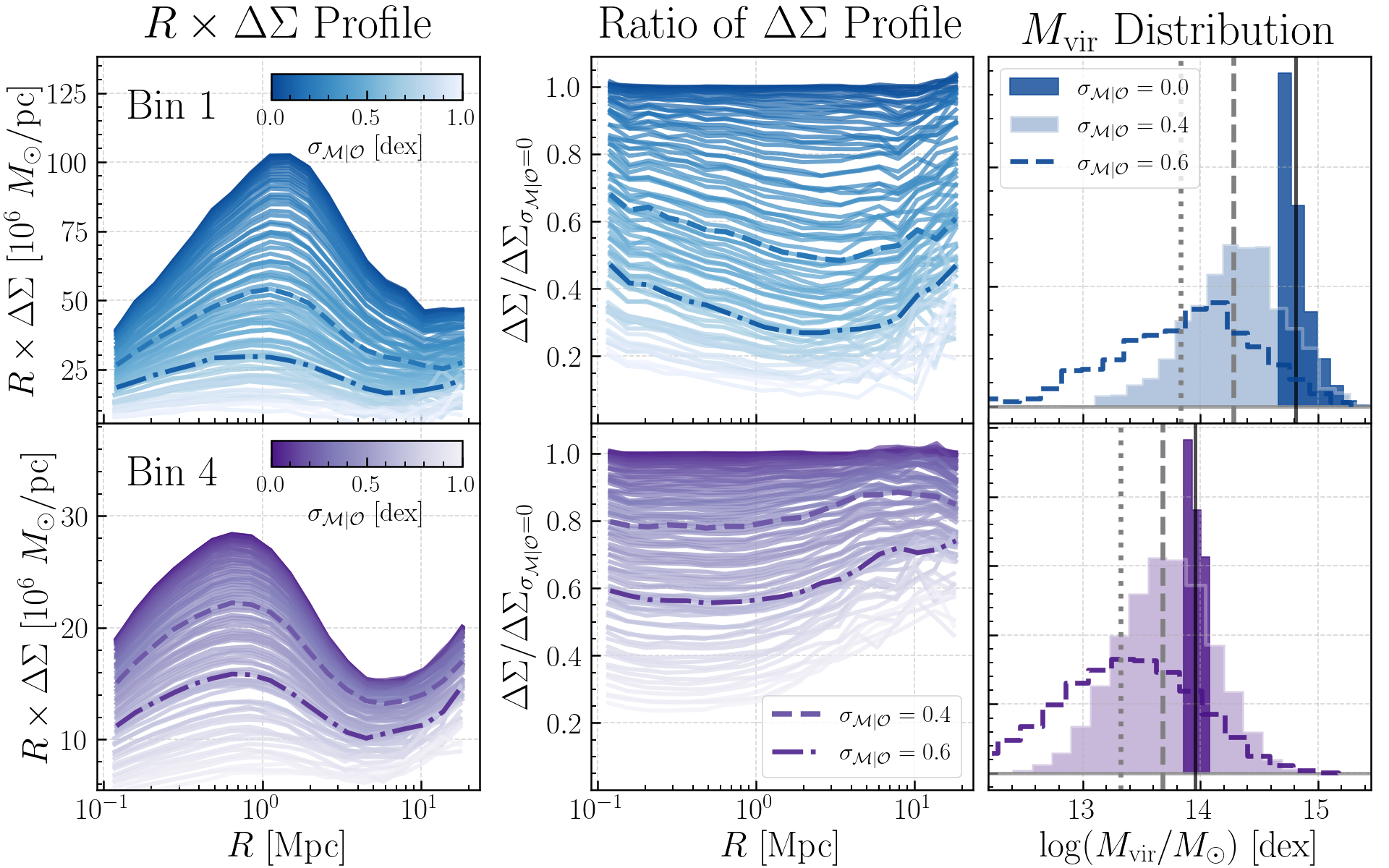}
    \caption{
        The \dsigma{} profiles and \mhalo{} distributions of different \topn{} samples with
        a wide range of scatter values ($0<$\scatterMhaloObsSym{}$<1$). 
        This combines data from both \mdpl2{} and \smdpl{}. 
        The top and bottom rows display the first and last \topn{} bins (the most and least
        massive in average \mhalo{}) used in this work (see \S\ \ref{sec:binning}).
        Their number density thresholds correspond to $0.2 < z < 0.5$ \redm{} clusters in HSC
        \texttt{S16A} area with $35 < \lambda < 150$ (Bin 1) and $6 \leq \lambda < 10$ (Bin 4).
        \textbf{Left}: the \rdsigma{} profiles of the \topn{} samples with different
        $\sigma_{\langle s \mid \mu\rangle}$ values. 
        We highlight the profiles corresponding to $\sigma_{\langle s \mid \mu\rangle}=0.4$
        (dashed line) and $=0.6$ (dot-dashed line).
        \textbf{Middle}: the ratios between the \dsigma{} profiles of the \topn{} samples with
        non-zero scatter and the ``perfect'' sample ($\Delta\Sigma_{\sigma=0}$). 
        \textbf{Right}: $\log M_{\rm vir}$ distributions of the ``perfect'' \topn{} sample and
        the $\sigma_{\langle s \mid \mu\rangle}=0.4$ and $=0.6$ samples. 
        Grey vertical lines indicate the mean $\log M_{\rm vir}$ for each distribution.
        The \texttt{Jupyter} notebook for reproducing this figure can be found here:
        \href{https://github.com/dr-guangtou/jianbing/blob/master/notebooks/figure/fig2.ipynb}{\faGithub}.
    }
    \label{fig:mdpl2}
\end{figure*}

\subsubsection{The relationship between \scatterMhaloObsSym{}, \scatterObsSymMhalo{}, and $\alpha$}
    \label{sec:comp_scatters}

    In this section, we use a simple analytic model to explore the connection between the scatter
    and slope of the $\log$-linear \mvir{}-observable relation and the amplitude of the \dsigma{}
    profiles.
    
    The shape of the halo mass function (HMF) directly influences the characteristics of a
    \mvir{}--observable relation \citep[\eg{}][]{Tinker2008}.
    We will first use an analytic form of the HMF to demonstrate the relation between 
    \scatterMhaloObsSym{} and \scatterObsSymMhalo{}.
    We approximate the HMF using the following exponential functional form suitable for the 
    high--\mvir{} end:

    \begin{equation}
        \hmf{} \equiv \frac{dn(\haloSym{})}{d\mu{}}  = \exp \left(\beta_0 - \beta_1 \haloSym{} - \frac{\beta_2}{2}
        \haloSym{}^2 \right).
        \label{eq:quadratic_hmf}
    \end{equation}

    \noindent For large values of mass, the HMF declines rapidly with \mhalo{} ($\beta_1 > 0$) with
    a steepening slope ($\beta_2 > 0$)\footnote{For the \mdpl2{} HMF at $z\sim 0.4$, we adopt the 
    best--fit parameter values of $\beta_{0}=-0.558$, $\beta_{1}=0.670$, and $\beta_{2}=2.959$.}. 
    We model the halo mass proxy \obsSym to follow a $\log$-linear relation with a constant
    $\log$-normal scatter value:

    \begin{equation}
        \obsSym = \mathcal{N}(\slope \haloSym + \intercept,\ \scatterObsSymMhalo).
        \label{eq:lognormal_obs_given_mhalo}
    \end{equation}

    The \scatterObsSymMhalo{} value here is often quoted as the ``scatter of the SHMR'' (e.g.,
    \citealt{More2011, Leauthaud2012, Reddick2013, Behroozi2013})
    and has been frequently used to infer physical information about galaxy formation (e.g.,
    \citealt{Gu2016, Matthee2017, Tinker2017c, Wechsler2018}).
    Yet it is the scatter of \mvir{} at fixed observable, \scatterMhaloObsSym{}, that we estimate in
    observations using the \topn{} test.
    We now briefly discuss the connection between \scatterObsSymMhalo{} and \scatterMhaloObsSym{}.
    First, the probability density of the observable $\obsSym$ is given by,

    \begin{equation}
        P(\obsSym{}) \equiv \int_{0}^{\infty} \hmf{} P(\obsSym{} | \haloSym{}) d\haloSym{}
    \end{equation}
    At fixed \obsSym{}, the mean value of \haloSym{} is,

    \begin{equation}
    \begin{aligned}
        \langle \haloSym{} | \obsSym \rangle
        &= \frac{1}{P(\obsSym)}
            \int_{0}^{\infty} \hmf{} P(\obsSym{} | \haloSym{}) \haloSym{} d\haloSym{} \\
        &= \frac{\left( \frac{\obsSym- \intercept}{\slope} - \beta_1 \left(\frac{\scatterObsSymMhalo}{\slope}\right)^2 \right)}{ 1 + \beta_2 \left(\frac{\scatterObsSymMhalo}{\slope}\right)^2}
        \label{eq:mean_of_mu}
    \end{aligned}
    \end{equation}

    \noindent The three components of $\langle \haloSym{} | \obsSym \rangle$ are:

    \begin{enumerate}

        \item The mean relation between the observable and halo mass, $(\obsSym- \intercept) / \slope$.

        \item A shift due to the Eddington bias caused by the linear slope of the HMF, $-\beta_1
            (\frac{\scatterObsSymMhalo}{\slope})^2$. In the case of $\beta_1 > 0$, this shift is to
            lower \haloSym{} as there are more low \haloSym{} objects up-scattered into the
            selection.

        \item A second shift is caused by the curvature of the HMF, $(1
            + \beta_2 (\frac{\scatterObsSymMhalo}{\slope})^2)^{-1}$. 
            Again, $\beta_2 > 0$ results in more low \haloSym{} objects and thus a shift to lower 
            \haloSym{}.

    \end{enumerate}

    For the scatter in \haloSym{} at fixed \obsSym{}, we have

    \begin{equation}
    \begin{aligned}
        \scatterMhaloObsSym{}
        &= \frac{1}{P(\obsSym{})}
            \int_{0}^{\infty} \hmf{} P(\obsSym{} | \haloSym{}) ( \haloSym{}  - \langle \haloSym{}
            \rangle )^2 d\haloSym{} \\
    	&= \frac{\scatterObsSymMhalo}{\sqrt{\beta_2 \scatterObsSymMhalo^2 + \slope^2}}
        \label{eq:scatter_of_mu}
    \end{aligned}
    \end{equation}

    \noindent In the case of a power law halo mass function ($\beta_2 = 0$), this expression reduces
    to the commonly seen $\scatterObsSymMhalo / \slope$. 
    The positive $\beta_2$ of the HMF decreases this scatter. 
    Finally, the higher moments of $P(\obsSym{})$ such as the skewness or excess kurtosis confirm
    that $P(\haloSym{} | \obsSym{})$ follows a Gaussian distribution for the approximated HMF in
    Equation \ref{eq:quadratic_hmf}.

    We now rewrite Equation \ref{eq:scatter_of_mu} in a more practical form that makes it clear
    that \scatterMhaloObsSym{} depends on the {\em ratio} of \scatterObsSymMhalo{} and \slope.
    This is obvious in the case of a power law mass function ($\beta_2 = 0$) and is also true
    for the more general quadratic form (\ref{eq:quadratic_hmf}),

    \begin{equation}
        \scatterMhaloObsSym{}
        = \frac{\scatterObsSymMhalo}{\sqrt{\beta_2 \scatterObsSymMhalo^2 + \slope^2}}
        = \left(\beta_2 + (\frac{\slope}{\scatterObsSymMhalo})^2\right)^{-1/2}
        \label{eq:ratio_is_what_matters}
    \end{equation}

    This equation shows that, for a given \topn{} selection, two \mvir{}--observable relations with
    the same $\scatterObsSymMhalo / \slope$ ratio will have the same value \scatterMhaloObsSym{}. 
    We demonstrate this in the right panel of Figure \ref{fig:theory_1} by populating \mdpl2{} halos
    with mock observables that follow different \mvir{}--observable relations.
    The two observables whose \mvir{}--observable relations share the same value of $ \alpha /
    \scatterObsSymMhalo = 2$ (red and green) lead to the same \mvir{} distributions in the top
    $N=5000$ sample and result in almost identical stacked \dsigma{} profiles.

    As shown above, our \topn{} test only probes the observed \mvir{} distribution, and cannot even
    in principle distinguish between {\em i)} mass-observable relations with steep slopes and high
    scatter versus {\em ii)} mass-observable relations with shallow slopes but low scatter. 
    We note that this degeneracy is unimportant for the present work, because here we only wish to
    perform $relative$ comparisons between different proxies for \mvir{}. 
    In this paper, we are not concerned with inferring the specific values of either $\alpha$ or 
    \scatterObsSymMhalo{}; rather, our principal focus is on a comparative assessment of which
    observable proxies \obsSym{} supply the strongest discriminating power regarding \mvir{}. 
    For a few specific cases of \obsSym{} that are of special importance, we will estimate the value
    of \scatterObsSymMhalo{} by using the same value for the slope as is conventionally assumed in
    the literature on \mvir{}--observable relations.
    This allows us to inter-compare different proxies using \scatterObsSymMhalo{}, and also to make
    comparisons with previous results from observations or hydro simulations. 
    However, we reiterate that in such cases, the specific value of our estimate on
    \scatterObsSymMhalo{} will strictly depend upon the assumed values for the slope $\alpha,$ as
    this mathematical degeneracy cannot be evaded.

\subsubsection{Estimating \scatterMhaloObsSym{}}
    \label{sec:estimate_scatter}

    The discussion in \S\ \ref{sec:comp_scatters} demonstrates that we can compare the
    \scatterMhaloObsSym{} values of two \topn{} samples using their stacked \dsigma{} profiles:
    the sample selected by the ``better'' \mhalo{} tracer should yield a \dsigma{} profile with
    higher amplitude.
    More importantly, we can also estimate \scatterMhaloObsSym{} from \dsigma{} profiles by
    comparing with model profiles built from simulations using the same number density selection.

    We build our model by populating halos in simulations with mock observables that follow
    $\log$-linear relations (Equation \ref{eq:lognormal_obs_given_mhalo}) with fixed slope value at
    $\alpha = 1$ (see justification in previous section) but with different \scatterObsSymMhalo{}
    values.
    In each realisation, we derive the best-fit $\haloSym| \obsSym$ relation and estimate the
    \scatterMhaloObsSym{} value for the same pre-defined number density bins (\topn{} bins) used in
    the observations.
    For each \topn{} bin, we calculate the stacked \dsigma{} profiles and store the underlying
    \mhalo{} distributions at different \scatterMhaloObsSym{} values. 
    We adopt a densely sampled grid of \scatterMhaloObsSym{} values between 0.0 and 1.0 dex.

    Observations suggest that at high-\mhalo{} and low redshift, \scatterObsSymMhalo{}$=0.2$ dex
    (e.g., \citealt{More2011, Leauthaud2012, Reddick2013, Behroozi2013, Tinker2017}). 
    Given the slope of the SHMR, this means that \scatterMhaloObsSym{} is expected to be in the
    $\sim 0.4$-0.6 dex range (e.g., Figure 5 \& 7 of \citealt{Wechsler2018}). 
    Therefore it is essential to cover a large range in \scatterMhaloObsSym{} values.
    
    We use the
    \href{https://halotools.readthedocs.io/en/latest/api/halotools.mock_observables.mean_delta_sigma.html}{\texttt{mean\_delta\_sigma}}
    function from \texttt{halotools} \citep{Hearin2017}\footnote{
    \url{https://github.com/astropy/halotools}} to calculation the stacked \dsigma{} profiles based
    on the algorithm described in the Appendix B of \citet{Lange2019} in comoving coordinates.
    We then convert them into physical coordinates before comparing to observations.
    We use 50 (10) millions down-sampled particles from the \mdpl2{} (\smdpl{}) simulations for the 
    calculation and choose the \texttt{Z} direction as the line-of-sight.
    The \mdpl2{} simulation has a large box size of 1 Gpc$/h$ that helps sample the very
    high-\mhalo{} end of HMF.
    However, its particle mass resolution ($1.51 \times 10^{9} M_{\odot}/h$) is not sufficient to
    resolve the $< 10^{12.5} M_{\odot}/h$ halos presented in samples with large
    \scatterMhaloObsSym{} values.
    In contrast, the \smdpl{} simulation has a much better mass resolution ($9.63 \times 10^{7}
    M_{\odot}/h$) for calculating accurate \dsigma{} profiles for less massive halos but does not
    have sufficient volume (box size $=0.4$ Gpc$/h$) to sample the very  high-\mvir{} end.
    Therefore, we combine the predictions from the \mdpl2{} and \smdpl{} simulations.
    We use the \mdpl2{} simulation to cover the $0.00 <$\scatterMhaloObsSym{}$<0.65$ dex range
    with a 0.01 dex grid, and use \smdpl{} to cover the $0.65 <$\scatterMhaloObsSym{}$<1.0$ dex
    range with a 0.05 dex grid size.
    Using the overlapping $\scatterMhaloObsSym{}$ range, we confirm the two simulations provide
    \dsigma{} profiles that are consistent within their statistical uncertainties.
    We use the $z=0.364$ snapshot from \mdpl2{} and the $z=0.404$ snapshot from \smdpl{}, which are
    the closest ones to the mean redshift ($\sim 0.4$) of the HSC sample. 
    
    Figure \ref{fig:mdpl2} shows the predicted \dsigma{} profiles as a function of 
    \scatterMhaloObsSym{} in two number density bins. 
    In addition to the expected decreasing \dsigma{} amplitudes with increasing
    \scatterMhaloObsSym{} values, we also see scale dependent differences in the ratios of the
    predicted \dsigma{} profiles.
    We highlight the \dsigma{} profiles with \scatterMhaloObsSym{}$=0.4$ and 0.6 dex along with
    their \mhalo{} distributions.
    
    We estimate \scatterMhaloObsSym{} by matching the observed \dsigma{} profiles to the model
    predictions. 
    For an observed lensing profile ($\Delta\Sigma_{\rm O}$) and its covariance matrix 
    ($\boldsymbol{C}$), we use the $\chi^2$ statistic to evaluate how well a given model profile 
    ($\Delta\Sigma_{\rm M}$) describes the observations:

    \begin{equation}
        \chi^{2}=(\Delta\Sigma_{\rm M}-\Delta\Sigma_{\rm O})^{\top}
                \boldsymbol{C}^{-1}(\Delta\Sigma_{\rm M}-\Delta\Sigma_{\rm O}).
        \label{eq:chi2}
    \end{equation}

    \noindent
    This ignores the uncertainties in the theoretical \dsigma{} profiles (\eg{} due to the
    stochasticity when populating halos or sample variance) which are negligible compared to the
    observed uncertainties. 
    Appendix \ref{app:fitting} includes further details about the fitting process.
    In this remainder of this paper, we will use \sigmvir{}$\equiv \sigma_{\log M_{\rm vir}} \equiv
    $\scatterMhaloObsSym{} to refer to the scatter of halo mass at given observable.

    This model only focuses on the underlying distribution of \mvir{} which is {\em not} the only 
    factor that determines the \dsigma{} profile. 
    For example, our model does not account for mis-centering or baryonic physics -- these and other
    effects will be discussed in later sections.
    To evaluate the impact of satellite galaxies on \dsigma{} profile, we also provide a special 
    version of the model that can match the observed stellar mass function (SMF) and clustering 
    signals of HSC massive galaxies. 
    We provide more details of this model in Appendix \ref{app:hsc_model}.

    We include massive satellite galaxies in our model when comparing to HSC massive galaxies since
    they should present in the observed sample.
    For richness-selected clusters, we remove the satellite galaxies from the model before
    calculating the \dsigma{} profiles.
    We assume the cluster finders identify the correct central galaxies even though that is not 
    always the case. 
    However, including satellites or not in our models does not have any impact on our results. 

\begin{figure*}
    \centering
    \includegraphics[width=\textwidth]{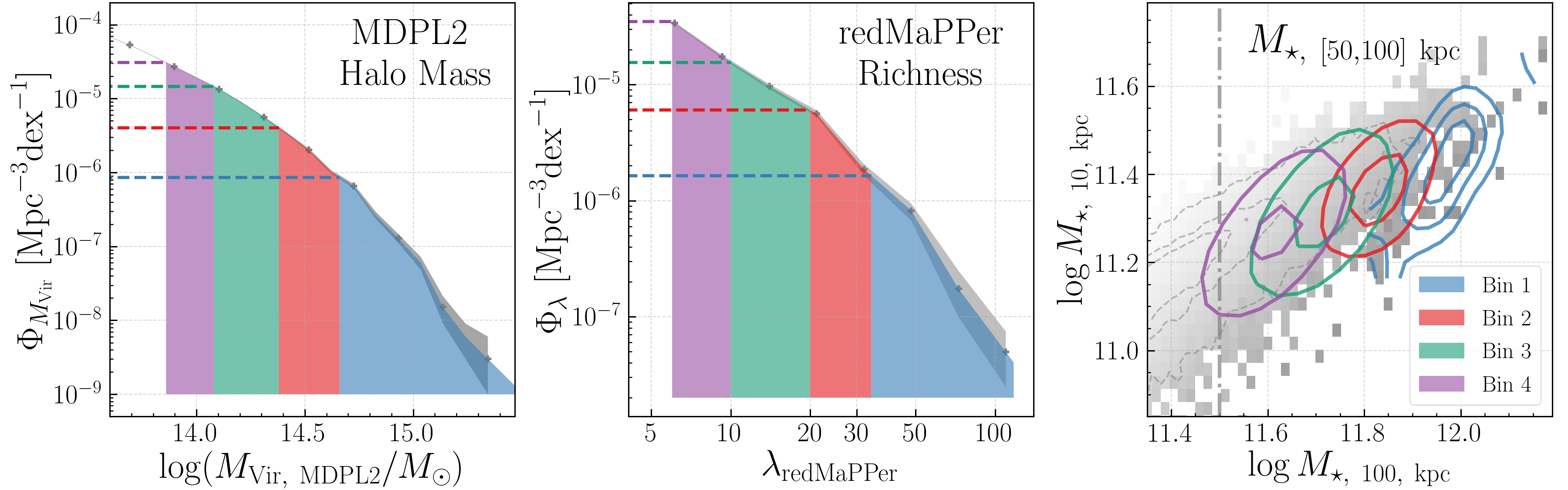}
    \caption{
        \textbf{Left}: the HMF of the \mdpl2{} simulation and the \mvir{} range of our \topn{}
        tests for an ``ideal'' (zero scatter) \topn{} selection. 
        The four coloured regions highlight the \mvir{} ranges of the four \topn{} bins.
        Dashed lines with corresponding colours label the number density boundaries of each bin. 
        \textbf{Middle}: the richness function of the \texttt{S16A} \redm{} clusters used in this
        work. 
        Coloured regions reflect the $\lambda$ ranges of all four bins.
        We describe the choice of these bins in \S\ \ref{sec:binning} and list their properties
        in Table \ref{tab:summary}.
        \textbf{Right}:
        The grey shaded region and the dashed-line contours show the distribution of HSC
        massive galaxies over the \maper{100}-\maper{10} plane.
        The dot-dash line highlights \logmmax{}$=11.5$, a conservative \maper{100} completeness
        limit.
        The four solid-line contours with corresponding colours highlight the distributions
        of \menve{50}{100}-selected galaxies in the four \topn{} bins.
        The \texttt{Jupyter} notebook for reproducing this figure can be found here:
        \href{https://github.com/dr-guangtou/jianbing/blob/master/notebooks/figure/fig3.ipynb}{\faGithub}.
    }
    \label{fig:density_bins}
\end{figure*}

\section{Data}
    \label{sec:data}

    In this section, we introduce the imaging data  (\S\ \ref{sec:hsc}), the HSC massive
    galaxy sample (\S\ \ref{sec:galaxy_sample}), and the richness-selected galaxy cluster
    catalogues (\S\ \ref{sec:cluster_sample})

\subsection{Hyper Suprime-Cam Survey Subaru Strategic Program}
    \label{sec:hsc}

    In this work, we use $\sim 137$ \sqdeg{} of deep optical images from the \texttt{WIDE}
    layer of the \texttt{S16A} release of the Hyper Suprime-Cam Subaru Strategic Program
    (HSC-SSP, or HSC survey; e.g. \citealt{HSC-SSP, HSC-DR1, HSC-DR2})
    \footnote{\url{https://hsc.mtk.nao.ac.jp/ssp/}} - an ambitious cosmology survey using the
    8.2-m Subaru Telescope.
    HSC multi--band ($grizY$) images have impressive depth ($\sim$3--4 mag deeper than
    SDSS), superb seeing conditions (the mean $i$-band seeing has a $\sim 0.58$ arcsec Full-Width
    Half-Maximum, or FWHM), and fine pixel resolution (0.168 arcsec), all making it ideal
    for studying galaxy structure and performing galaxy-galaxy lensing measurements.

    We use the coadd images produced by \texttt{hscPipe 4.0.2}.
    \texttt{hscPipe} is a specifically modified version of the Large Synoptic Survey
    Telescope (LSST) pipeline (e.g.\ \citealt{Juric2015}; 
    \citealt{Axelrod2010}\footnote{{\url{https://pipelines.lsst.io}}} for HSC\footnote{
        The most recent version of \texttt{hscPipe} can be found here:
        \url{https://hsc.mtk.nao.ac.jp/pipedoc_e/}.
    }.
    Please see \citet{HSC-PIPE} for more details about the data reduction process and
    \citet{SynPipe} for its photometric performance.
    We also make use of the photometric redshift (photo-$z$) measurements of HSC galaxies from the
    \href{https://github.com/joshspeagle/frankenz}{\texttt{frankenz}} \citep{Speagle2019} algorithm.
    Please see \citet{HSC-PHOTOZ} for a summary of its performance.
    For galaxy-galaxy lensing measurements, we use the public shape catalogue for 
    \texttt{S16A}\footnote{The \texttt{S16A} weak lensing shape catalogue is released here:
    \url{https://hsc-release.mtk.nao.ac.jp/doc/index.php/s16a-shape-catalog-pdr2/}}
    based on the $i$-band images and the re-Gaussianization algorithm (\citealt{HirataSeljak2003}).
    HSC Y1 cosmology (e.g., \citealt{Hikage2019, Hamana2020}) and other cluster lensing analyses
    (e.g., \citealt{Umetsu2020}) used the same catalogue.
    Please see \citet{HSC-PIPE}, \citet{HSC-WLCAT}, and \citet{HSC-WLCALIB} for details about the
    shape measurements and lensing calibration.
    All galaxies and clusters used in this work are filtered through the bright star masks (see
    \citealt{HSC-STAR} for details) used in the HSC \texttt{S18A} or \texttt{PDR2} data release to
    avoid the contamination from saturated stars.
    Please refer to \citet{Huang2018b, Huang2018c, Huang2020} for further details about the
    HSC data. 
    All imaging data, along with the photometric and the photo-$z$ catalogues, have been released to
    the public\footnote{\url{https://hsc.mtk.nao.ac.jp/ssp/data-release}}.
    
\subsection{HSC Massive Galaxy Sample}
    \label{sec:galaxy_sample}

    Using the \texttt{S16A} data, we select a sample of massive galaxies at $0.19 < z < 0.52$.
    In this redshift range, we can resolve their inner light profile ($r<10$ kpc) but also have the
    depth to explore the faint outskirt ($r \sim 100$ kpc).
    This is the same sample used in \citet{Huang2020}. Please refer to \citet{Huang2020} for a 
    detailed description of the sample, here we only provide a brief summary.

    The sample contains 24926 massive galaxies selected using a cut on the \cmodel{}-based 
    \mstar{}, $\mcmodel{} \geq 10^{11.2} M_{\odot}$.
    The \mcmodel{} is based on the \mlratio{} estimated by five-band SED fitting using
    \texttt{iSEDfit} (see \citealt{Moustakas2013}).
    All galaxies have valid 1-D surface brightness profiles measured in $i$-band with empirical
    background correction that enables non-parametric \mstar{} measurements out to $>100$ kpc.
    During the extraction of the 1-D profiles, $\sim 9$\% of the original sample was excluded 
    due to contamination by nearby objects. 
    We treat this as a small decrease of the effective survey area in this work.
    Among all 24051 galaxies, 15558 have useful spec-$z$.
    However, when making a cut using the 100 kpc aperture \mstar{}, at $\maper{100} \geq 10^{11.5}
    M_{\odot}$, 4429 of the 4848 galaxies ($\sim 91$\%) have spec-$z$.
    And 2190 of the 2299 galaxies ($\sim 95$\%) with $\maper{100} \geq 10^{11.6} M_{\odot}$ have
    spec-$z$.

    Using this sample, we have uncovered remarkable structural diversity in the outer stellar halo
    (\citealt{Huang2018b}) and shown that the stellar mass distribution has a connection with
    \mvir{} (\citealt{Huang2018c, Huang2020}).
    We have also compared these observed 1-D stellar mass density profiles with those from
    state-of-the-art hydro-simulations to gain insights into their assembly history
    (\citealt{Ardila2021}).
    We release the catalogue of this massive galaxy sample here:
    \href{https://zenodo.org/record/4902141}{\faDatabase}.

\subsection{Red Sequence Cluster catalogues}
    \label{sec:cluster_sample}

    Taking advantage of the well-defined ``red-sequence'' of low-$z$ galaxy clusters and the
    potential low-scatter nature of the \mvir{}-richness scaling relation (e.g., \citealt{Rozo2009,
    Rykoff2012}), richness-based cluster finders provide a promising way to identify massive halos
    in imaging data.
    Here we will evaluate two cluster catalogues based on red-sequence algorithms using the \topn{}
    test by estimating the \sigmvir{} values of these two cluster samples and directly comparing
    their \dsigma{} profiles to those from \mstar{}-based \mvir{} proxies.
    The massive galaxy sample and the cluster samples are independently selected from the same 
    HSC footprint but there is no one-to-one correspondence between the two catalogues. 
    For example, there are massive galaxies not contained in either cluster catalogues. 
    We will briefly discuss this in \S\ \ref{sec:discussion}.
    
\subsubsection{\redm{} Clusters}
    \label{sec:cluster_redmapper}

    \redm{} \citep{Rykoff2014, Rozo2014, Rozo2015a, Rozo2015b}
    \footnote{\url{http://risa.stanford.edu/redmapper/}} is a popular cluster finding algorithm
    based on the richness of red-sequence galaxies.
    It has been applied to several large imaging surveys including SDSS (e.g.,
    \citealt{Rykoff2014}), DES (\citealt{Rykoff2016, McClintock2019}), and HSC.
    The \mvir{}--richness relation of \redm{} clusters has been investigated in multiple
    works (e.g., \citealt{Saro2015, Farahi2016, Simet2017, Melchior2017, Baxter2018, Murata2018,
    McClintock2019})

    We use an internal version of the \redm{} cluster catalogue for \texttt{S16A} data
    (Kawinwanichakij \& Rykoff, private communication) based on the updated \texttt{Python} version
    of \redm{}\footnote{\url{https://github.com/erykoff/redmapper}}.
    The algorithm is similar to that used in \citet{Rykoff2016} with minor modifications.
    At $0.19 < z < 0.52$, it contains 2409 clusters with $\lambda \geq 5$ and 227 with $\lambda \geq
    20$.
    Of these clusters, 1623 have spec-$z$ (from a variety of sources) and the rest have a
    high-quality photo-$z$ from their red-sequence.
    The sample has a median photo-$z$ bias of $\delta_{z} \sim 0.0012$ (0.0008), a scatter of
    $\sigma_{z}/(1 + z) \sim 0.011$ (0.007), and a 4-$\sigma$ outlier fraction of $\sim 0.7$\%
    (0.5\%) for $\lambda \geq 5$ ($\geq 20$) clusters, showing performance consistent with that of
    the DES catalogue \citep{McClintock2019}.
    We confirm that only using the photo-$z$ from \redm{} does not affect relevant conclusions.
    Regarding the completeness of the cluster sample, \citet{McClintock2019} estimates that
    the DES limiting magnitude is deep enough for $0.2 L_{\star}$ galaxies at $z \sim 0.7$,
    and that the galaxy sample for \redm{} is $>90$-95\% complete.
    Given the deeper imaging in HSC, it is safe to expect even better completeness at $z<0.52$.

    In addition to the richness, \redm{} provides a list of candidates of the central galaxy along
    with their central probability ($P_{\rm cen}$).
    We choose the galaxy with the highest $P_{\rm cen}$ as the centre of the cluster.
    Using a sub-sample of X-ray detected clusters, \citet{Zhang2019b} analyses \redm{} mis-centring
    in DES for clusters with $\lambda > 20$. 
    They find $\sim 83$\% of the clusters are well-centred.
    In the HSC \redm{} sample, 66\% (77\%) of clusters have central galaxies with $P_{\rm cen} >
    0.8$ (0.7).

    We also use the 364 SDSS \texttt{DR8} \redm{} clusters (\citealt{Rykoff2014}) with $\lambda_{\rm
    SDSS} \geq 20$ in the \texttt{S16A} footprint to show that the results found for the HSC \redm{}
    clusters also hold for SDSS \redm{} catalogue (see Appendix \S\ \ref{app:sdss_redm}).
    The SDSS sample is only complete at $z < 0.33$.
    In Appendix \S\ \ref{app:des_redm}, we compare the stacked \dsigma{} profiles of HSC and DES
    \redm{} (\eg{} \citealt{Chang2018, McClintock2019}) clusters in the same redshift ($0.2 < z <
    0.5$) and richness ($20 \leq \lambda < 100$) bins and show they are consistent with each other.

\subsubsection{\camira{} Clusters}
    \label{sec:cluster_camira}

    \camira{}\footnote{\url{https://www.slac.stanford.edu/~oguri/cluster/}} is a
    red-sequence cluster finding algorithm developed by \citet{Oguri2014}.
    It has been applied to SDSS (\citealt{Oguri2014}) and HSC (\citealt{Oguri2018}) data.
    Unlike \redm{}, \camira{} does not have a richness-dependent radius, and instead counts red
    galaxies with $L \geq 0.2L_{\star}$ within a fixed $R\leq 1 h^{-1}$ Mpc.
    Its \mvir{}-richness relation has been calibrated using a variety of methods \citep{Murata2019,
    Chiu2020a, Chiu2020b}.

    Here we use the public \texttt{S16A} \camira{} catalogue that contains 998 (263) clusters with
    $N_{\rm Mem} \geq 10$ ($\geq 20$) for $0.19 < z < 0.52$.
    Among them, 725 clusters have spec-$z$ measurements for their central galaxies.
    Our \camira{} sample has a median photo-$z$ bias of $\delta_{z} \sim -0.0042$ ($-0.0036$), a
    scatter of $\sigma_{z}/(1 + z) \sim 0.013$ (0.009), and a 4-$\sigma$ outlier fraction of
    $\sim 1.4$\% (0.9\%) for the $N_{\rm Mem} \geq 10$ ($\geq 20$) clusters.
    Similar to \redm{}, only using the photo-$z$ has no impact on any key results.
    The \camira{} catalogues shows excellent completeness when compared to X-ray clusters
    ($\gtrapprox 0.8$; see \citealt{Oguri2018} \S\ 5.3) or mock galaxy catalogues ($> 0.8$ for
    $M_{200c} > 5 \times 10^{13} h^{-1} M_{\odot}$ clusters at $0.3 < z < 0.6$; see
    \citealt{Oguri2018} \S\ 6).

    \camira{} assigns a central galaxy to each cluster without providing a central probability.
    \citet{Oguri2018} investigated the off-centre distance ($R_{\rm off}$) distribution using
    matched X-ray clusters.
    While the distribution centred at $R_{\rm off} \approx 0.0$ Mpc, $\sim 30$\% of the 
    clusters are offset from the X-ray peak with their $R_{\rm off}$ distribution described by a
    $\sigma=0.26 \pm 0.04 h^{-1}$ Mpc Gaussian component.

    We also test the internal \texttt{S18A}, \texttt{S19A}, and \texttt{S20A} \camira{} catalogues
    within the \texttt{S16A} footprint.
    Differences in the data reduction process (e.g., background subtraction, deblending)
    cause subtle differences in the cluster detection and richness measurements.
    However, we verify that these updates do not change any conclusions.

\section{Measurements}
    \label{sec:measure}

    Here we briefly introduce the methodologies behind the key measurements used in the \topn{}
    tests: the 1-D surface stellar mass density profiles ($\mu_{\star}$, \S\ \ref{sec:1d_prof})
    and the galaxy-galaxy lensing \dsigma{} profiles ($\S$ \ref{sec:dsigma}).

\subsection{1-D Surface Mass Density Profiles}
    \label{sec:1d_prof}

    Our method for extracting 1-D $\mu_{\star}$ profiles is presented in previous work
    \citep{Huang2018b, Huang2018c, Ardila2021}.
    We refer readers to these papers for full technical details and only provide a brief summary
    here.

    Using the \ellipse{} isophotal analysis function from \iraf{}, we extract 1-D $i$-band
    surface brightness profiles after aggressively masking out nearby contamination and
    empirically correcting for the local background.
    In addition to the mask, the strategy of taking the median of flux density values along the
    isophote after 3-$\sigma$ clipping makes our 1-D profile robust against the high density of
    faint objects around massive galaxies (\citealt{Ardila2021}).
    With background subtraction, the 1-D profile is stable above $\sim 28$ \smag{}, roughly
    corresponding to $r\sim 100$ kpc for our sample.
    The inner $\sim 5$-6 kpc of the profile is smeared by the seeing.

    We then convert the $i$-band surface brightness profile to the $\mu_{\star}$ profile using the
    average $i$-band \mlratio{} derived from SED fitting after applying corrections for
    galactic extinction and cosmological dimming.
    We ignore the \mlratio{} gradient in this work.
    Low-$z$ massive galaxies have shallow but negative colour gradients (e.g., \citealt{Huang2018b,
    Wang2019, Montes2021}), which suggests the average \mlratio{} will underestimate the \mstar{} in
    the central region and overestimate it in the outskirts.
    However, the lack of clear dependence of colour gradients on \mstar{} (\citealt{Huang2018b})
    suggests this systematic will not influence the conclusions of this work.
    We release the massive galaxy catalogue along with the 1-D $\mu_{\star}$ profiles here: 
    \href{https://zenodo.org/record/5259075}{\faDatabase}.

\subsection{Galaxy-Galaxy Lensing Measurements}
    \label{sec:dsigma}

    The galaxy-galaxy (g-g) lensing measurements done here follow almost exactly those in 
    \citet{Speagle2019} and \citet{Huang2020}, which are themselves based on the methodology 
    presented in \citet{Leauthaud2017}.
    This method subtracts lensing signals around a large number of random positions to achieve
    unbiased measurements (\citealt{Singh2017}). 
    The equations used to derive the \dsigma{} profile are given in Appendix
    \ref{app:dsigma_detail}.
    Compared to these earlier works, we provide a new recipe for the $f_{\rm bias}$ factor that more
    accurately accounts for the photo-$z$ dilution effect (see Equation \ref{eq:fbias})\footnote{The
    typical $f_{\rm bias}$ factor value is at the $\sim 1$-2\% level, and has no impact on the
    results of this work.}.

    We measure \dsigma{} in 11 physical logarithmic radial bins from 200 kpc to 10 Mpc using 
    the \texttt{S16A} weak lensing shape catalogue (\citealt{HSC-WLCAT, HSC-WLCALIB}). 
    We adopt the \texttt{frankenz}\footnote{\url{https://github.com/joshspeagle/frankenz}} photo-$z$
    for source galaxies. 
    For the photo-$z$ quality cut, we use the ``basic'' cut ($\chi^{2}_{5} < 6$) in 
    \citet{Speagle2019} that removes about 5\% of source galaxies with unreliable photo-$z$.
    Most of them are at very low redshift so will not contribute to the lensing signals in this work.
    The lens-source separation criteria are: $z_{\rm s} - z_{\rm L} \ge 0.1$ and
    $z_{\rm s} > z_{\rm L} + \sigma_{s,68}$, where $\sigma_{s,68}$ is the 1$\sigma$ uncertainty
    of the source photo-$z$.
    We confirm that other photo-$z$ quality cuts and slightly different lens-source separation
    criteria do not affect any results.

    We use both jackknife resampling in 40 pre-defined sub-regions and bootstrap resampling with
    2000 iterations to estimate the covariance matrix and the uncertainties of the \dsigma{}
    profiles.
    The two methods lead to fully consistent results.

    We use \texttt{v0.2} of the \texttt{Python} g-g lensing code \texttt{dsigma}
    \footnote{\url{https://github.com/johannesulf/dsigma}} to calculate \dsigma{} profiles, and we
    release the \dsigma{} measurements for our massive galaxies and clusters here:
    \href{https://zenodo.org/record/5259075}{\faDatabase}.

\section{Halo Mass Proxies and Bins}
    \label{sec:proxies}

    This section introduces different \mvir{} proxies in our \topn{} tests.
    We broadly grouped these observables into \mstar{}- and richness-based categories. 
    For \mstar{}-based proxies, we include \mstar{} based on the default HSC photometry for extended
    objects (\S\ \ref{sec:mcmodel}), a series of \mstar{} measures from the 1-D $\mu_{\star}$
    profiles (\S\ \ref{sec:maper} \& \S\ \ref{sec:menvelope}), and a linear combination of different
    aperture \mstar{} measures (\S\ \ref{sec:masap}).
    And for richness-based methods, we include clusters from both \redm{} and \camira{}. 
    We also describe our number density bins and show the estimated scatter by comparing to the model 
    described in \S\ \ref{sec:estimate_scatter}.
    
\subsection{Proxies}

\subsubsection{\cmodel{} stellar mass}
    \label{sec:mcmodel}

    \cmodel{} is the default photometric model for extended objects in both the SDSS and HSC
    surveys, and will continue to be used in future imaging surveys.
    \cmodel{} attempts to describe the 2-D flux distribution of all extended objects using a
    combination of an exponential and a de Vacouleurs component (e.g., \citealt{HSC-PIPE}).
    It is an efficient and flexible model and can provide statistically robust colour measurements
    down to very faint magnitudes (e.g., \citealt{SynPipe}).
    However, \cmodel{} does not always provide accurate total flux measurements, especially for
    massive galaxies whose surface brightness profiles can not be described with the underlying
    assumptions.
    In both the SDSS and HSC surveys, \cmodel{} photometry significantly underestimates the flux in
    the extended outskirts of massive, early-type galaxies (e.g., \citealt{Bernardi2013,
    Huang2018b}).
    In addition to the intrinsic limitations associated with the assumed model, systematics in 
    critical steps in the data reduction process such as background subtraction and object 
    deblending often interfere with \cmodel{} fitting, making it even more challenging to accurately
    recover the total flux.
    These issues becomes especially pronounced with deep imaging surveys such as HSC.

    Because \cmodel{} is the default photometry provided for extended objects in many imaging
    surveys, it is worth testing using the \topn{} methodology. 
    We will quantify the impact of \cmodel{} photometry errors on the use of  \cmodel{} masses 
    as a halo mass proxy. 
    The \cmodel{} stellar mass will be labelled as \mcmodel{}.
    
    We note that the \cmodel{} photometry used here is from HSC \texttt{S16A} and an old
    version of \hscpipe{} (\texttt{v4}).
    Although the updated \hscpipe{} includes multiple improvements and modifications, they do not
    solve the aforementioned issues for bright galaxies.
    We compare the \cmodel{} magnitudes of our sample using \texttt{S16A}, {\tt S18A}, and {\tt
    S20A} data release\footnote{The {\tt S18A} release applies a much improved background
    subtraction around bright object.  But the well preserved low surface brightness envelopes
    around massive galaxies make object deblending more challenging. This global background
    correction algorithm was then turned off in the following release {\tt S20A}.}. 
    We find no systematic difference between these measurements, hence our results about \mcmodel{}
    should apply to all HSC data release.
    
\subsubsection{Aperture \mstar{} From 1-D Profiles}
    \label{sec:maper}

    In \citet{Huang2018b}, we showed that the \mstar{} within a 100 kpc aperture is a better
    estimate of the ``total'' \mstar{} of massive galaxies than \mcmodel{}. 
    We also demonstrated in \citet{Huang2018c, Huang2020} that changing the aperture used to measure
    \mstar{} changes the \mstar{}--\mvir{} relation.
    Here, we measure \mstar{} in apertures of 10, 30, 50, 75, 100, and 150 kpc in our \topn{} tests,
    and we evaluate how each performs as a proxy for \mvir{}. 
    Throughout the paper, we will label these ``aperture \mstar{}" measurements with \maper{10},
    \maper{100}, etc. 

    In practice, we integrate the 1-D $\mu_{\star}$ profile after accounting for the isophotal shape
    of the galaxy to get the ``curve-of-growth'' (CoG) of \mstar{}, which describes the relation
    between the semi-major axis length of an elliptical aperture and the enclosed \mstar{}.
    Interpolation of the CoG provides the measurements of different aperture \mstar{}.
    We note that the $\mu_{\star}$ profile outside 100 kpc becomes less reliable due to background
    subtraction issues which affects the accuracy of aperture \mstar{} using larger radii.
    Given the imaging depth of HSC data, we do not recover substantial amount of \mstar{} beyond 100
    kpc.
    The mean difference between \maper{150} and \maper{100} is only $\sim 0.02$ dex, while the
    maximum difference is $\sim 0.15$ dex. 

    The intrinsic $\mu_{\star}$ of massive galaxies certainly extends beyond the HSC surface
    brightness limit for individual galaxies (e.g., \citealt{Wang2019, Zhang2019, Montes2021,
    Kluge2021}), hence the true total \mstar{} is beyond the reach of our current aperture \mstar{}
    measurements.
    We attempt to account for the ``missing \mstar{}'' by fitting a 1-D \ser{} model to the
    $\mu_{\star}$ profile between 50 and 100 kpc. We use this model to predict the mass beyond
    the regime in which it can be measured with HSC.
    This model (assuming it correctly predicts the true profile) confirms that there is little
    \mstar{} beyond 100 kpc.
    Using this technique, the predicted average difference between \maper{300} and \maper{100} is
    only $\sim\,0.05$ dex.

    Using the CoG, we also measure the radius that contains 50\%, 80\%, and 90\% of the maximum
    \mstar{} measured by the 1-D profile (\mmax{}).
    We denote these radii as $R_{50}, R_{80}, R_{90}$.
    The ``half-mass'', or effective radius ($R_{50}$), provides another way to define apertures.
    For example, we can measure \mstar{} out to $2\times R_{50}$ or $4\times R_{50}$. 
    Aperture masses using $R_{50}$ will be labelled as $M_{\star, 2R_{50}}$, $M_{\star, 4R_{50}},$
    etc. 
    We briefly explore the result of using these radii-based proxies in Appendix \ref{app:size}.

\subsubsection{Stellar Mass of the Outer Envelope}
    \label{sec:menvelope}

    In \citet{Bradshaw2020}, by studying simulated data the authors noticed that the ``\exsitu{}''
    component (the stellar mass that formed outside the halo of the main progenitor) of massive
    galaxies seems to have a tighter relation with \mvir{} than either the ``\insitu{}'' component
    or the total \mstar{}.
    This is also consistent with the modelling results from \citet{Huang2020}.

    While we cannot separate the \exsitu{} component from the \mstar{} distribution directly when
    using observational data alone, recent simulations and observations suggest that the \exsitu{}
    stars dominate the outskirts of massive galaxies (e.g., \citealt{Lackner2012,
    RodriguezGomez2016, Pulsoni2021}).
    It is therefore interesting to test whether the stellar mass in the outer envelope is a useful
    \mvir{} proxy using the \topn{} tests.

    Here we simply define this ``outer envelope'' (or outskirt) \mstar{} as the difference between
    two aperture \mstar{}.
    For example, we will use \menve{50}{100} to denote the stellar mass between 50 and 100 kpc, and
    $M_{\star,\ [2,4]R_{50}}$ to denote the stellar mass between $2 \times R_{50}$ and $4 \times
    R_{50}$. 
    It is not obvious a prior which combination of radial boundaries will provide the best
    \exsitu{} \mstar{} proxy, and so we will explore a range of different definitions of the outer
    envelope.

    Many of the massive galaxies in \citet{Huang2020} are the central galaxies (or the brightest
    cluster galaxy, BCG) of a galaxy cluster. 
    Their ``outer envelope'' is also sometimes called the ICL.  
    We avoid this terminology because:
    1) the photometric definition of ICL is often ambiguous and arbitrary (e.g.,
    \citealt{Kluge2021}), and
    2) not all massive galaxies in our sample live in clusters (i.e., \mvir{}$\geq 10^{14}
    M_{\odot}$).  Therefore we prefer to use the more general term -- outer envelope -- to describe
    the outer structure of all massive galaxies.
    
    To estimate the outer envelope \mstar{} from the 1-D surface brightness profile, we
    assume a fixed \mlratio{} value and isophotal shape that represents the inner region.
    These low-$z$ massive galaxies on average show shallow negative optical colour (hence \mlratio{})
    and axis ratio gradients. 
    In our case, the colour gradient means we could slightly over-estimate the outer envelope stellar
    mass while the axis ratio gradient could lead to an under-estimation.
    We ignore these minor systematics here and will look into more accurate outer envelope \mstar{}
    measurement in future work.
    We have performed \topn{} tests using the luminosity of the outskirts and this does not impact
    any of our main conclusions.

\subsubsection{\asap{} model}
    \label{sec:masap}

    In \citet{Huang2020}, we presented a phenomenological model (\asap{}) that connects a linear
    combination of \maper{10} and \maper{100} to the \mvir{} of the host halo of massive galaxies
    (\maper{100}{}$\geq 10^{11.5} M_{\odot}$).
    We constrained this model using the SMFs for \maper{10} and \maper{100} along with the \dsigma{}
    profiles of galaxies in twelve 2-D bins over the \maper{100}-\maper{10} plane\footnote{Note that
    \citet{Huang2020} used the notation \mmax{} to refer to \maper{100}.}.
    In \citet{Ardila2021}, we provided an updated \asap{} recipe to predict the \mvir{} of a
    massive galaxy based on its \maper{100} and \maper{10}:

    \begin{equation}
        \begin{aligned}
        \log M_{\mathrm{vir}} &=3.26 \times\left(\log M_{\star}^{100}-11.72\right) \\
        &-2.46 \times\left(\log M_{\star}^{10}-11.34\right) \\
        &+13.69.
        \end{aligned}
        \label{eq:asap}
    \end{equation}

    In \citealt{Huang2020}, we showed that the \asap{} model scaling relation summarized by
    Eq.~\ref{eq:asap} can predict the {\em average} \mvir{} of massive halos better than
    \maper{100}. 
    Throughout this paper, we will use the label \masap{} to refer to the left-hand side of
    Eq.~\ref{eq:asap}, and we will investigate the scatter exhibited by individual galaxies for this
    \asap{} prediction.

\begin{table*}
\resizebox{0.7\textwidth}{!}{%
\small
\begin{tabular}{|c|cccc|}
\hline
\rowcolor[HTML]{d8dcd6} Property   & Bin 1   & Bin 2   & Bin 3  & Bin 4 \\ \hhline{|=====|}

$N_{\rm Sample}$     &   50  &  197  &  662  &  1165  \\  

$\log_{10} M_{\rm vir,\ \rm MDPL2}$  & [14.66, 15.55] & [14.38, 14.66) & [14.08, 14.38) & [13.86, 14.08) \\

$n(>M_{\rm vir})$  & $5.11\times 10^{-7}$ & $2.52\times 10^{-6}$ & $9.29\times 10^{-6}$ & $2.12\times 10^{-5}$ \\ \hhline{|=====|}

\multirow{2}{*}{$\lambda_{\rm redMaPPer}$}  &  [35, 120] &  [20, 35)  & [10, 20) &  [6, 10)  \\
& \sigmvir{}$=0.27\pm0.02$ & $0.38\pm0.02$ & $0.39\pm0.02$ & $0.58\pm0.02$ \\ \hline
                                            
\multirow{2}{*}{$N_{\rm CAMIRA}$}  &  [35, 75) & [21, 35) & [12, 21) &  \\
& \sigmvir{}$=0.30\pm0.03$ & $0.36\pm0.01$ & $0.50\pm0.02$ & {} \\ \hhline{|=====|}

\multirow{2}{*}{$\log_{10} M_{\star, \rm CModel}$}   & [11.88, 12.19] & [11.77, 11.88) & [11.67, 11.77) & [11.60, 11.67) \\ 
& \sigmvir{}$=0.60\pm0.04$ & $0.64\pm0.04$ & $0.87\pm0.06$ & $0.82\pm0.03$ \\ \hline

\multirow{2}{*}{$\log_{10} M_{\star, 30\ \rm kpc}$} & [11.77, 12.00] & [11.69, 11.77) & [11.61, 11.70) & [11.53, 11.61) \\ 
& \sigmvir{}$=0.52\pm0.04$ & $0.57\pm0.03$ & $0.61\pm0.03$ & $0.61\pm0.02$ \\ \hline

\multirow{2}{*}{$\log_{10} M_{\star, 50\ \rm kpc}$} & [11.86, 12.10] & [11.76, 11.86) & [11.66, 11.76) & [11.58, 11.66) \\ 
& \sigmvir{}$=0.46\pm0.04$ & $0.53\pm0.03$ & $0.59\pm0.02$ & $0.61\pm0.02$ \\ \hline

\multirow{2}{*}{$\log_{10} M_{\star, 100\ \rm kpc}$} & [11.93, 12.18] & [11.83, 11.93) & [11.71, 11.83) & [11.63, 11.71) \\ 
& \sigmvir{}$=0.38\pm0.02$ & $0.51\pm0.02$ & $0.56\pm0.02$ & $0.60\pm0.02$ \\ \hline

\multirow{2}{*}{$\log_{10} M_{\star, 150\ \rm kpc}$} & [11.96, 12.21] & [11.85, 11.96) & [11.73, 11.85) & [11.64, 11.73)  \\ 
& \sigmvir{}$=0.37\pm0.03$ & $0.47\pm0.03$ & $0.56\pm0.02$ & $0.57\pm0.03$ \\ \hhline{|=====|}

\multirow{2}{*}{$\log_{10} M_{\star, [50, 100]}$} & [11.20, 11.60] & [11.00, 11.20) & [10.80, 11.00) & [10.63, 11.00)  \\ 
& \sigmvir{}$=0.36\pm0.02$ & $0.43\pm0.02$ & $0.44\pm0.02$ & $0.48\pm0.02$ \\ \hline

\multirow{2}{*}{$\log_{10} M_{\rm Vir,\ ASAP}$} & [14.45, 15.28] & [14.09, 14.45) & [13.80, 14.11) & [13.60, 13.80)  \\ 
& \sigmvir{}$=0.38\pm0.03$ & $0.44\pm0.02$ & $0.48\pm0.02$ & $0.56\pm0.02$ \\ \hline

\end{tabular}%
}
\caption{
	Summary of results from the \topn{} test results in four number density bins. 
	The first three rows summarize the basic properties of each bin.
	$N_{\rm sample}$ is the number of HSC galaxies in each bin. 
	$\log_{10}M_{\rm vir, MDPL2}$ shows the corresponding halo mass range in this number density bin
	based on the \mdpl2{} simulation. 
	This is the \mvir{} range for an ideal (zero scatter) \topn{} selection.
	$n(>M_{\rm Vir})$ is the volume number density of halos above the lower-\mvir{} threshold.
	Subsequent rows contain the key results for different halo mass proxies. 
	The first row shows the range of the observed properties in the four bins.  The second row shows
	the best--fit scatter of \mvir{} at fixed observable ($\sigma_{\mathcal{M}|\mathcal{O}}$) of
	\mvir{} in each bin along with its uncertainty.
	For a complete summary table of all the properties we tested, please see this \texttt{Jupyter} notebook \href{https://github.com/dr-guangtou/jianbing/blob/master/notebooks/topn_result_summary.ipynb}{\faGithub} 
	}
\label{tab:summary}
\end{table*}

\subsubsection{Richness}
    \label{sec:proxy_richness}

    We compare these \mstar{}-based proxies to the cluster richness by two popular red-sequence
    cluster finders: \redm{} and \camira{} (introduced in \S\ \ref{sec:cluster_redmapper} and \S\
    \ref{sec:cluster_camira}).
    Calibrations of the \mvir{}--richness relations suggest that the richness of red-sequence
    galaxies is a very promising \mvir{} proxy (e.g., \citealt{Melchior2017, Murata2018,
    McClintock2019}).
    
    Theoretically speaking, richness measurements should outperform \mstar{}-based \mvir{} proxies
    for massive halos if the majority of their satellites have not merged onto the central galaxies.
    The $\log$-linear slopes of \mvir{}--richness relations are typically $> 1.0$ (e.g.,
    \citealt{Saro2015, Mantz2016, Farahi2016, Simet2017, Baxter2018, Melchior2017, McClintock2019}),
    while the slopes of \mvir{}--\mstar{} relations are usually around 0.3-0.5 at $z<0.5$ (e.g.,
    \citealt{RodriguezPuebla2017, Tinker2017, Moster2018, Kravtsov2018, Huang2020}).
    Recent calibrations of \mvir{}--richness relations also suggest modest intrinsic scatter values
    at the high-\mvir{} end (at $\sim 25$\% or 0.1-0.2 dex level; e.g., \citealt{Rykoff2014,
    Saro2015, Simet2017}).
    Both of these observations support the idea that richness should be a superior \mvir{} proxy to
    the \mstar{} of the central galaxy.
    Therefore a \topn{} comparison between \mstar{}- and richness-based proxies can help confirm
    this expectation, or reveal new insights.

    For \redm{}, we denote its richness as $\lambda_{\rm redMaPPer}$. 
    For \camira{}, we use $N_{\rm CAMIRA}$ to represent its richness measurement.

\begin{figure}
    \centering
    \includegraphics[width=0.47\textwidth]{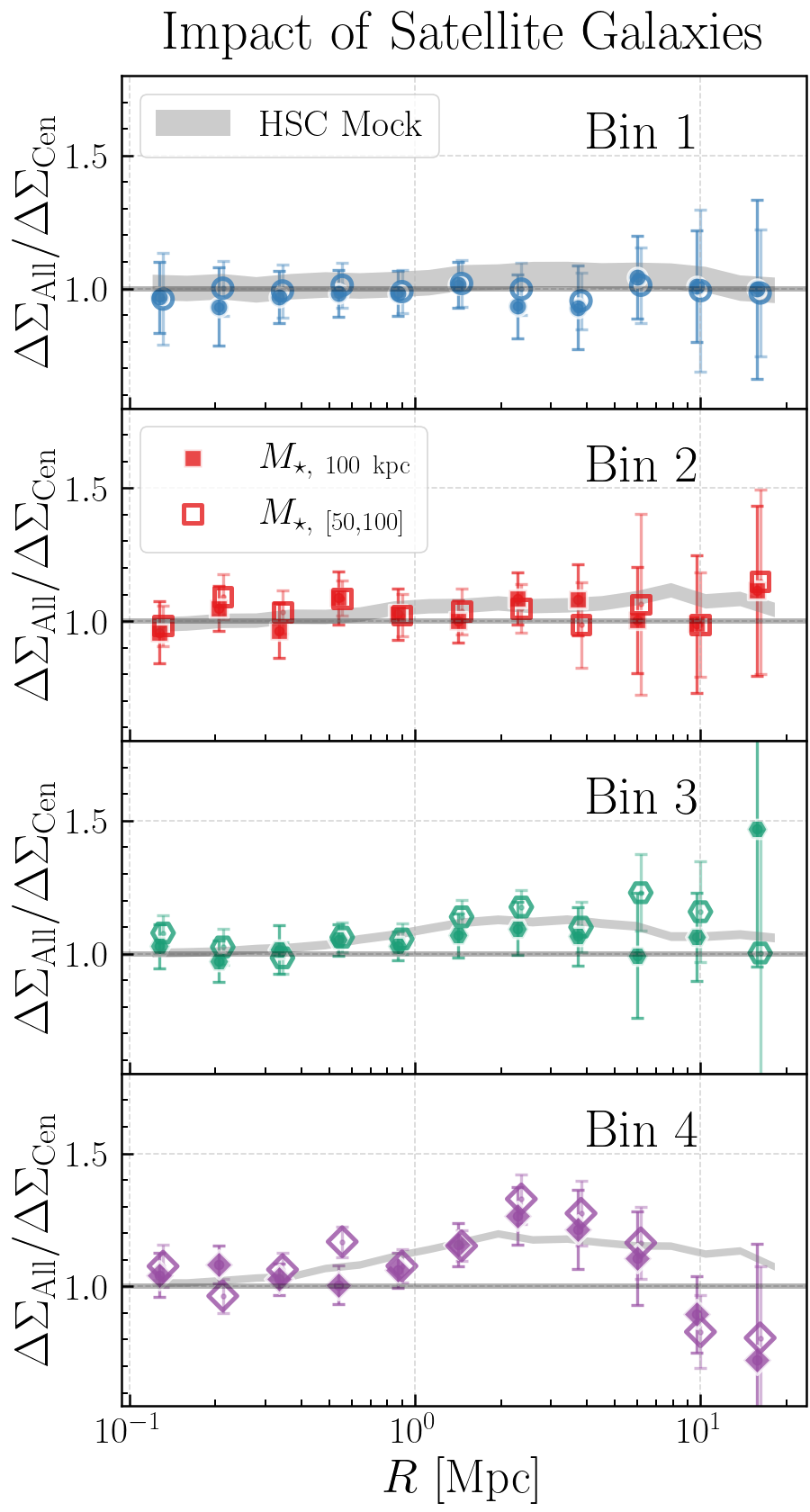}
    \caption{
        Ratio of \dsigma{} profiles for all galaxies (central \& satellite; $\Delta\Sigma_{\rm
        All}$) compared to central galaxies only ($\Delta\Sigma_{\rm Cen}$).
        Grey shaded regions show results from our fiducial HSC mock catalogue.
        Data points show results from HSC data where satellites are removed using a simple cylinder
        based technique. 
        Solid symbols correspond to galaxies selected by \maper{100}{}. 
        Open symbols correspond to galaxies selected by \menve{50}{100} (data points are slightly
        offset along the X--axis).
        Satellites have almost no impact in the first two bins (corresponding to \maper{100}$>
        10^{11.8} M_{\odot}$ and \menve{50}{100}$> 10^{11.1} M_{\odot}$). 
        The HSC mock suggests that satellites have a maximum impact of $\sim 15$\% ($\sim 20$\%) in
        Bin 3 (4) at $R\sim 2$--4 Mpc.
        The \texttt{Jupyter} notebook for reproducing this figure can be found here:
        \href{https://github.com/dr-guangtou/jianbing/blob/master/notebooks/figure/fig4.ipynb}{\faGithub}.
    }
    \label{fig:satellite}
\end{figure}

\subsection{Number Density Bins}
    \label{sec:binning}

    To perform the \topn{} tests using the previously mentioned proxies, we design four number
    density bins based on the richness of HSC \redm{} clusters ($\lambda_{\rm redMaPPer}$).
    These four bins correspond to the $\lambda_{\rm redMaPPer}$ ranges of $[35, 100], [20, 35),
    [10, 20), [6, 10)$ and have 50, 197, 662, \& 1165 objects in each bin.
    We refer to these bins as Bin 1 (richest clusters or highest average \mvir{}) through to Bin 4
    (least rich clusters or lowest average \mvir{}).
    The total number (2074) of objects is slightly smaller than the number of \logmaper{100}$\geq
    11.6$ galaxies (2247), which defines a \mstar{}-complete sample. 
    The area of the \texttt{S16A} data means that we do not have enough massive clusters to sample
    the high $\lambda_{\rm redMaPPer}$ range. 
    Hence, Bin 1 covers a fairly wide richness range.

    Figure \ref{fig:density_bins} illustrates how these bins in $\lambda_{\rm redMaPPer}$ correspond
    to bins in number density (top-left panel) and \mvir{} using the HMF from \mdpl2{} (top-right
    panel). 
    Assuming an ideal tracer with zero scatter, Bin 1, 2, \& 3 are well above the conventional
    standard for ``galaxy cluster'' (\logmvir{}$\geq 14.0$) while the mean \mvir{} of Bin 4 is on
    the boundary between a cluster and a ``massive group''. 
    In reality, the \mvir{} distributions will shift to lower values due to the scatter of the 
    \mvir{} -- observable relation.

    The $N_{\rm Mem} > 10$ threshold for the \camira{} clusters means it does not have enough
    objects for Bin 4, therefore we only consider the first three bins.
    Similarly, for the SDSS \redm{} catalogue, we only include Bins 1 \& 2, and we note that the
    richness range for Bin 4 is challenging even for deep HSC images.
    We must therefore take the results for \redm{} in Bin 4 with some caution.

    We were unable to measure \mstar{} for $\sim\,9$\% of galaxies due to excessive blending which
    prevented the extraction of a 1-D profile. 
    This reduces the effective area and volume of this sample.
    We do not correct for this when selecting the \topn{} galaxies. 
    The effect of this can only reduce the \dsigma{} amplitude, but we verify it does not affect our
    results.
    We summarize the key properties of these four bins in Table \ref{tab:summary}.

\section{Results}
    \label{sec:result}

    In this section we present our main results. 
    We begin in \S\ \ref{sec:satellite} with a qualitative evaluation of how satellite contamination
    impacts measurements of \dsigma{} for \mstar{}-selected samples.
    We then summarize the key findings from the \topn{} tests in \S\ \ref{sec:topn_results}. 
    First, we show how \scatterMhaloObsSym{} scales with number density for samples selected
    according to different choices of aperture mass (\S\ \ref{sec:m_aper}) and outskirt mass (\S\
    \ref{sec:m100_outskirt}).
    We also compare the \dsigma{} profiles of samples selected by \maper{100} relative to samples
    selected by \menve{50}{100} (\S\ \ref{sec:m100_outskirt}) and \mcmodel{} (\S\
    \ref{sec:m100_cmodel}).
    We summarize results related to \masap{} in \S\ \ref{sec:asap_result}.
    In \S\ \ref{sec:richness_results}, we examine the \topn{} results of richness-based cluster
    finders.
    We then show the behavior of \scatterMhaloObsSym{} for a series of different \mvir{} proxies
    (\S\ \ref{sec:trend}).
    Finally, in \S\ \ref{sec:mstar_vs_richness}, we compare the {\em shape} of the \dsigma{}
    profiles of \mstar{}- and richness-selected massive halos, and we assess the level of
    consistency of these profiles with theoretical expectations based on simulated halos that have
    been selected based on true halo mass. 
    The results shown in \S\ \ref{sec:result} focus on the most interesting cases, but the \topn{}
    results for all proxies are made publicly available here:
    \href{https://github.com/dr-guangtou/jianbing/tree/master/data/results}{\faGithub}
    
\begin{figure*}
    \centering
    \includegraphics[width=0.7\textwidth]{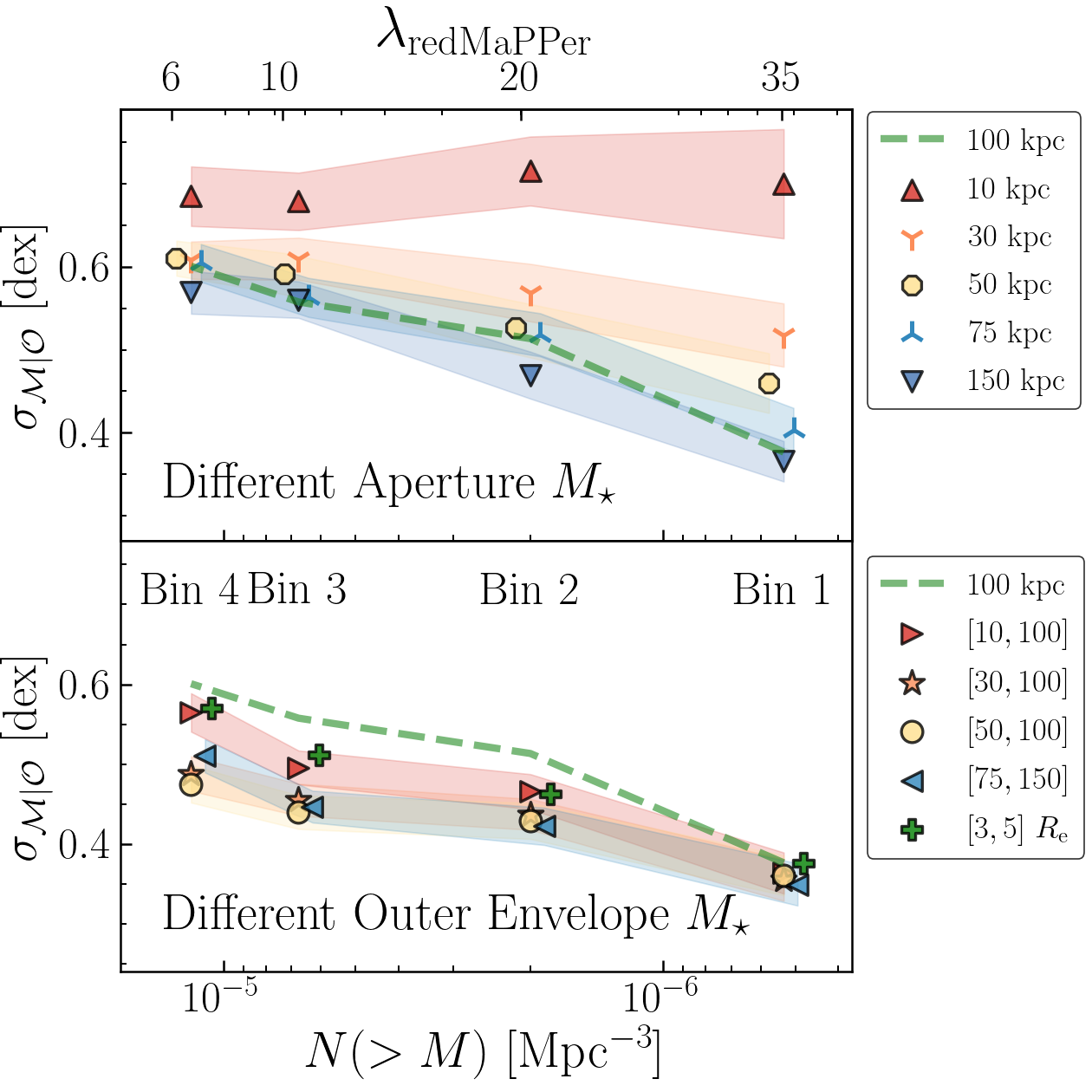}
    \caption{
        Scatter in \mvir{} at fixed observable versus number density bins. Aperture masses (the mass
        withing the indicated radius) are shown in the top panel and and outskirt \mstar{} is shown
        in the bottom panel. 
        The Y-axis shows the best-fit values of \scatterMhaloObsSym{} (data points) with
        uncertainties (shaded regions). 
        The X-axis labels on the top indicate the HSC \texttt{S16A} \redm{} richness thresholds
        corresponding to the four number density bins. The \maper{100} trend (green dashed-line) is
        used as a reference.
        We also slightly shift the symbols along the X-axis for visibility. 
        This figure shows that the outer mass is an excellent \mvir{} proxy.
        In contrast, inner stellar mass is a very poor tracer of present day \mvir{}.
        The \texttt{Jupyter} notebook for reproducing this figure can be found here:
        \href{https://github.com/dr-guangtou/jianbing/blob/master/notebooks/figure/fig5.ipynb}{\faGithub}.
    }
    \label{fig:scatter_trend}
\end{figure*}

\subsection{Impact of Satellite Contamination on \texorpdfstring{\dsigma{}}{DSigma}}
    \label{sec:satellite}

    Although satellite galaxies only make up a small fraction of \logmaper{100}$>$11.5 galaxies
    (e.g., \citealt{Reid2014, Saito2016, vanUitert2016, Huang2020}), they could affect our
    evaluation of the performance of \mstar{}-based \mvir{} proxies. 
    The \dsigma{} profiles of satellite galaxies show a unique ``bump"-like features at around $R
    \sim 1$ Mpc (e.g., \citealt{LiShan2014, LiShan2016, Sifon2015, Sifon2018}) which corresponds to
    the offset profile of the main parent halo. 
    Here, we evaluate the impact of satellites on \mstar{}-based proxies and \sigmh{} estimates by
    comparing the \dsigma{} profiles of a pure central sample to that of a central$+$satellites
    sample in the same \topn{} bin. 
    We use two methods: One based on a realistic mock catalogue with central/satellite assignments and
    one using satellites identification in real HSC data.
    
    We use a mock catalogue (see Appendix \ref{app:hsc_model}, DiMartino et al in prep) that was
    specifically designed to have realistic values for both \scatterMhaloObsSym{} and for the
    satellite fractions of massive galaxies. 
    This mock was constrained using the SMF and the two-point correlation functions (2PCF) of HSC
    massive galaxies.
    Using this mock, we select the pure central and the central$+$satellites samples based on 
    the model \mstar{} and the four number density bins defined in \S\ \ref{sec:binning}.
    In the four \topn{} bins, the HSC mock yields satellite fractions of [5.0\%, 6.9\%,
    8.9\%, \& 10.0\%], which are consistent with the expectation of low satellite fractions 
    among massive galaxies.
    
    Next, we attempt to classify massive satellite galaxies in our HSC sample directly by recursively
    identifying less massive galaxies around more massive ones.
    We start with the galaxy with the largest \mstar{} value and label all the other galaxies within
    a cylindrical region with a 1.0 physical Mpc radius in projection and a 30 comoving Mpc length
    in the line-of-sight (LOS) direction as satellites. 
    We then turn to the next most massive galaxy and repeat this exercise. 
    Using this simple strategy, the observed satellite fractions are [0.0\%, 3.6\%, 4.7\%, \& 8.8\%]
    in the four \topn{} bins. 
    These observed fraction values are slightly lower than when we use the mock catalogue but these
    differences are not large enough to affect any of our results. 
    Small variations of the radius and length of the cylinder do not change any results.

    Figure \ref{fig:satellite} shows the ratio of \dsigma{} profiles of the central$+$satellite 
    (\dsigma{}$_{\rm All}$) and the pure central (\dsigma{}$_{\rm Cen}$) samples using both 
    the mock catalogue and the HSC data.
    For HSC data, we test both the \maper{100} and the \menve{50}{100} sample. 
    In Bin 1 \& 2, the satellite fractions are low enough that there is no discernible impact on
    \dsigma{}.
    In Bin 3 \& 4, massive satellites lead to a small enhancement in the \dsigma{} profile at $R >
    500$ kpc.
    In Bin 3 (4), the mock catalogue predicts a maximum $\sim 10$\% ($\sim 20$\%) enhancement at $R
    \sim 2$-3 Mpc.
    Both \mstar{}-based proxies demonstrate behaviors that are statistically consistent with the
    mock catalogue despite our naive central/satellite classification scheme.
    The details of the fiducial mock catalogue do not affect this conclusion. 
    Even the simple $\alpha=1$ model with varying scatter values used for estimating
    \scatterMhaloObsSym{} can also lead to the same results\footnote{See the additional figure in
    this \texttt{Jupyter} notebook: 
    \href{https://github.com/dr-guangtou/jianbing/blob/master/notebooks/figure/fig4.ipynb}{\faGithub}.}

    In the rest of this paper, we include massive satellite galaxies both in our HSC samples and in
    the models that we draw from the mock catalogues. 
    Due to the low satellite fraction, the inclusion of satellite galaxies does not affect any 
    of our conclusions\footnote{The removal of massive satellite candidates usually only leads to
    0.01-0.02 dex variation in \scatterMhaloObsSym{} values.}.
    The impact of satellites is further discussed in Appendix \ref{app:sat_cen}.
    
\subsection{Amplitude of \texorpdfstring{\dsigma{}}{DSigma} and Inferred \texorpdfstring{\sigmvir{}}{SigMvir} Values}
    \label{sec:topn_results}

    We now present the \topn{} results with a focus on the overall amplitude of \dsigma{} for the
    \mvir{} proxies introduced in \S\ \ref{sec:proxies}. 
    We first describe the three proxies which yield the most interesting results: 
    aperture \mstar{}, outskirt \mstar{}, and the \mcmodel{}. 
    Then we compare all the \mvir{} proxies and present the inferred \sigmvir{} values.

\subsubsection{Aperture Stellar Masses}
    \label{sec:m_aper}

    We start by exploring the performance of aperture \mstar{}. 
    The upper panel of Figure \ref{fig:scatter_trend} shows the \sigmvir{} for various apertures. 
    It is clear that the \sigmvir{} decreases with increasing aperture size. 
    This decrease is particularly obvious in Bins 1 \& 2 (the most massive halos).
    In Bin 1 (2), \sigmvir{} decreases from 0.70 (0.72) dex for \maper{10}, to 0.52 (0.57) dex for
    \maper{30}, to 0.38 (0.51) dex for \maper{100}. 
    Figure \ref{fig:scatter_trend} shows that the aperture size used to estimate \mstar{} has
    significant impact on how well different samples trace \mvir{}. 
    More importantly, Figure \ref{fig:scatter_trend} shows that \textbf{the inner regions of massive
    galaxies (10 to 30 kpc) is a very poor tracer of present day halo mass}. 
    This is consistent with the SHMR constraints in \citet{GoldenMarx2019} where the authors focused
    more on the slope of the SHMR using different aperture \mstar{} (see their Figure 2).
    Using a sample of massive BCGs at $0.0 < z < 0.3$ in \mvir{}$>10^{14.0} M_{\odot}$ clusters, the
    authors find the slope of the SHMR increases from $\alpha \sim 0.1$ for \maper{10} to $\alpha
    \sim 0.4$ for \maper{100}.
    Assuming a constant \sigms{}$\sim 0.2$ dex value for the SHMR, such variation of slopes 
    correspond to \sigmvir{}$\sim 0.6$ dex for \maper{10} and $\sim 0.4$ dex for \maper{100},
    broadly consistent with the \topn{} results in Bin 1. 

    Further enlarging the aperture size to 150 kpc does not result in much improved \sigmvir{} in
    any of the bins (see Table \ref{tab:summary}).
    It is unclear whether the lack of improvement with apertures larger than 100 kpc reflects the
    intrinsic limitation of large aperture \mstar{} as a \mvir{} proxy or the statistical
    uncertainty of the current imaging data in the low surface brightness regime. 
    In the remainder of this paper, we use \maper{100} as the benchmark against which we will
    compare other \mvir{} proxies.
    In the Appendix \ref{app:size}, we also explore definitions of aperture \mstar{} based on
    $R_{50}$ but do not find any that has better performance.
    
\subsubsection{Outer Envelope Mass}
    \label{sec:m100_outskirt}

    Figure \ref{fig:scatter_trend} suggests that removing the inner portion of the galaxy and using
    only the outskirts could yield an improved \mvir{} proxy.
    From hydro-simulations or semi-empirical modelling of massive galaxy formation, we know that the
    accreted stellar component (or the \exsitu{} stars) dominates the \mstar{} budget, especially in
    the outskirt (\eg{} \citealt{RodriguezGomez2016}). 
    We will discuss this further in \S\ \ref{sec:outskirt_discussion}, but if the \exsitu{} stars
    have a tighter relation with \mvir{}, we should expect outskirt \mstar{} to be a better \mvir{}
    proxy. We test this hypothesis here.  

    Since there is no preference a priori for an optimal definition of what constitutes the
    ``outskirts" of a massive galaxy, we empirically study a number of different values. 
    The bottom panel of Figure \ref{fig:scatter_trend} shows one of the main findings of this paper. 
    Namely, \textbf{the \mstar{} in the outskirts of massive galaxies is an excellent proxy of halo mass,
    and largely outperforms any form of aperture mass, especially relative to masses defined by the
    inner regions of the galaxy}.
    Figure \ref{fig:scatter_trend} shows that \menve{50}{100} is the best \mvir{} proxy among the
    outer masses that we tested, and that \menve{30}{100} also displays comparable performance. 
    This figure also helps inform our understanding of the trade-off between retaining enough light
    to achieve a high \snratio{} measurement, while at the same time removing the \insitu{} stars
    that lie preferentially within the inner region. 
    For example, in Bins 2, 3, and 4, we can see that the performance of \menve{10}{100} is slightly
    worse than either \menve{30}{100} \menve{50}{100}. 
    This result taken together with the results shown in the top panel of Figure
    \ref{fig:scatter_trend} indicates that the stellar mass located within the inner 10-20 kpc does
    not correlate well with \mvir{}, and so should be excluded.
    Additionally, we can see from Figure \ref{fig:scatter_trend} that using outskirt masses that
    extend beyond 100 kpc (\eg{} \menve{75}{150}) does not improve the performance of the halo mass
    proxy.
    
    We have also explored outskirt \mstar{} defined using $R_{50}$, but fail to find one whose
    performance is as good as \menve{50}{100}. 
    These alternate definitions are discussed in Appendix \ref{app:size}. 
    
\begin{figure*}
    \centering
    \includegraphics[width=\textwidth]{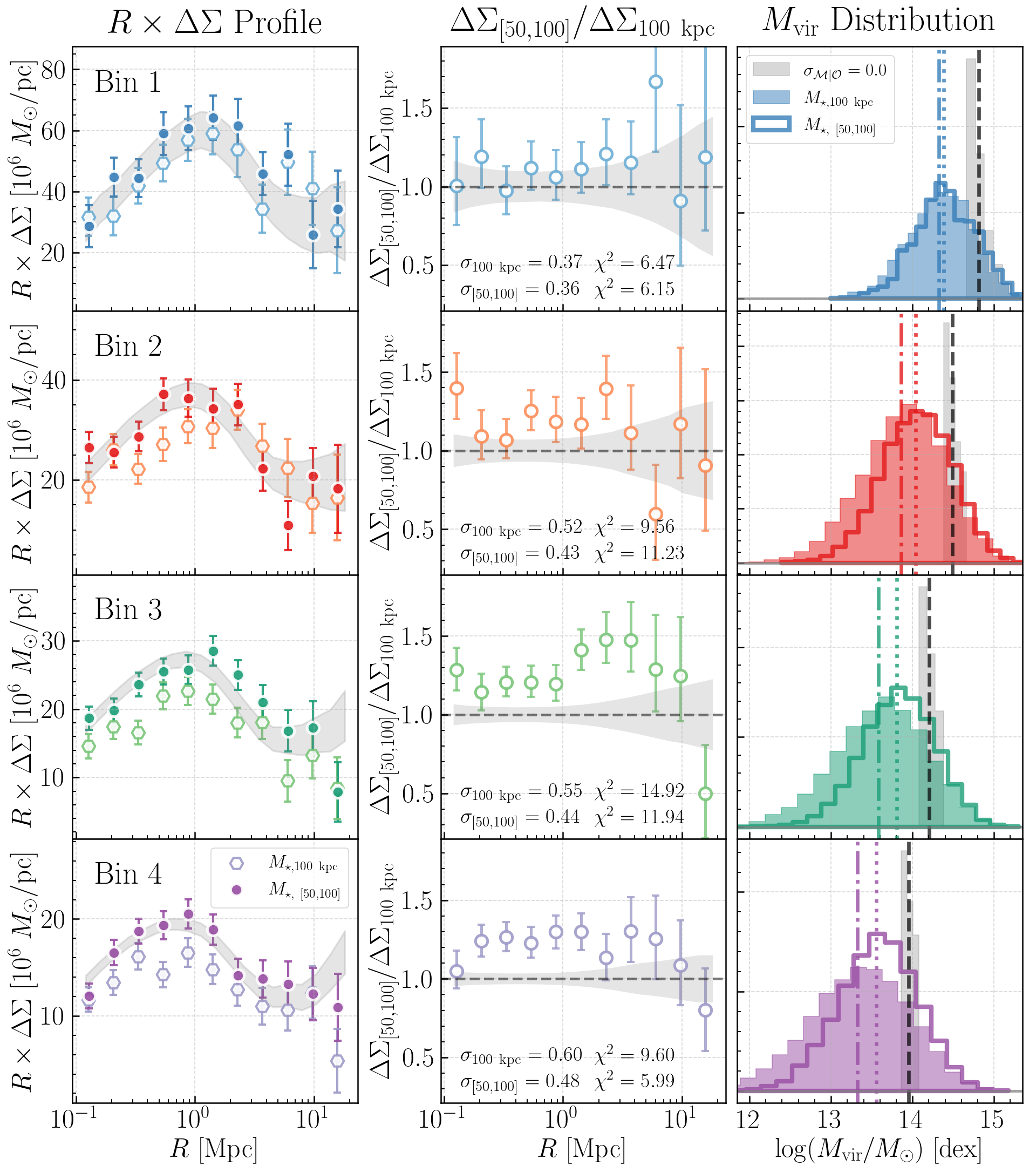}
    \caption{
        Comparison of the \topn{} results for \maper{100} and \menve{50}{100}. 
        Rows correspond to number density bins (see \S\ \ref{sec:binning}).  
        \textbf{Left} column: \rdsigma{} profiles of \menve{50}{100}-selected (circles) and
        \maper{100}-selected (hexagons) samples. 
        Grey shaded regions show the best-fit profiles and their associated uncertainties. 
        The overall amplitude of \dsigma{} profiles are similar in Bin 1.
        But \menve{50}{100} samples have consistently higher lensing amplitudes in the three other
        bins compared to the \maper{100} samples. 
        \textbf{Middle} column: ratio of \dsigma{} profiles. 
        Samples selected by \menve{50}{100} in Bin 2-4 show $\sim 20$--30\% higher \dsigma{}
        amplitudes at $R<2$ Mpc than the \maper{100} samples. 
        \textbf{Right} column: inferred \mvir{} distributions for the two samples using the model
        described in \S\ \ref{sec:estimate_scatter}. 
        Grey histograms indicate the \mvir{} distributions of an ideal tracer with 
        \scatterMhaloObsSym{}{}$=0$.
        Vertical lines indicate the average \mvir{}. 
        For Bins 2-4, \menve{50}{100} yield an average halo mass that is $\sim 0.2$ dex higher than
        for \maper{100}.
        The \texttt{Jupyter} notebook for reproducing this figure can be found here:
        \href{https://github.com/dr-guangtou/jianbing/blob/master/notebooks/figure/fig6.ipynb}{\faGithub}.
    }
    \label{fig:m100_mout}
\end{figure*}

    We now compare the \dsigma{} profiles of the \menve{50}{100} \topn{} samples to that of
    \maper{100} in Figure \ref{fig:m100_mout}.
    In general, the overall amplitudes confirm that \menve{50}{100} performs better than \maper{100}
    as a \mvir{} proxy.
    With the exception of Bin 1, the \dsigma{} profiles of the \menve{50}{100} samples show
    statistically higher amplitudes at $R < 2$ Mpc than the \maper{100} ones.
    The difference is more pronounced for lower mass bins.  
    The average \dsigma{}$_{100\ \rm kpc}/$\dsigma{}$_{[50,100]}$ ratios are
    $[0.90\pm0.17, 0.85\pm0.13, 0.82\pm0.09, 0.80\pm0.09]$ for Bin 1-4.
    As a result, \menve{50}{100} also has lower \sigmvir{} values than \maper{100}: 
    \sigmvir{}$=[0.36, 0.43, 0.44, 0.48]$ for \menve{50}{100} and 
    \sigmvir{}$=[0.38, 0.51, 0.56, 0.60]$ for \maper{100} in Bin 1-4.
    The right column of Figure \ref{fig:m100_mout} shows the \mvir{} distributions of these two
    samples.
    The \menve{50}{100} samples have higher mean \mvir{} (\logmvir{}$=[14.39, 14.04, 13.76, 13.52]$)
    values compared to \maper{100} samples (\logmvir{}$=[14.36, 13.86, 13.54, 13.32]$) too.
    These conclusions also qualitatively apply to \menve{50}{150}, and do not change if we switch
    \maper{100} with \maper{150}, or \mmax{}.

    We emphasis that both \maper{100} and \menve{50}{100} based selections yield \dsigma{} profiles
    that are statistically consistent with predictions based on our simple ``pure scatter'' model
    (\S\ \ref{sec:estimate_scatter}). 
    We can qualitatively confirm this conclusion using the left columns of Figure
    \ref{fig:m100_mout} and Figure \ref{fig:m100_mout} as there is no systematic deviations from the
    predicted \dsigma{} profiles (grey shaded regions).
    As for goodness-of-fit statistics, the \chisq{} values of \maper{100} samples are $[6.47, 9.56,
    14.92, 9.60]$\footnote{Each \dsigma{} profile has 11 data points and the scatter of \logmvir{}
    is the only ``free parameter'' in our model. 
    Therefore one can roughly estimate a reduced \chisq{} value using degree-of-freedom $\nu=10$.
    However, given the small sample size in our \topn{} bins, the simple resampling method used to
    estimate covariance matrix, we do not recommend to take the reduced \chisq{} values too
    literally. 
    Relative comparison is a more meaningful way to use these \chisq{} values.} and the values for
    \menve{50}{100} samples are $[6.15, 11.23, 11.94, 5.99]$.
    This conclusion is not only true for \menve{50}{100} and \maper{100}, but also valid for the
    majority of \mstar{}-based \mvir{} proxies with similar \sigmvir{} values. 
    Removing candidates of massive satellites using the method described in \S\ \ref{sec:satellite}
    can lead to marginal improvements of \chisq{} values, but does not change any conclusions.

    It is important to remember that a $\log$-normal \mvir{}-observable relation with \emph{just} a
    Gaussian scatter can already describe the \dsigma{} profiles of some promising \mstar{}-based
    \mvir{} proxies at the current \snratio{}.
    We will come back to this point when comparing with \topn{} results of richness-based 
    clusters (\S\ \ref{sec:mstar_vs_richness}). 
    And we discuss the origin and the implications of this result in \S\
    \ref{sec:outskirt_discussion}.
    
\subsubsection{The Case of \mcmodel{}}
    \label{sec:m100_cmodel}

    Since \cmodel{} is still widely used for galaxy luminosities and masses, it is of great
    interest to evaluate its performance as \mvir{} proxy. 
    
\begin{figure*}
    \centering
    \includegraphics[width=\textwidth]{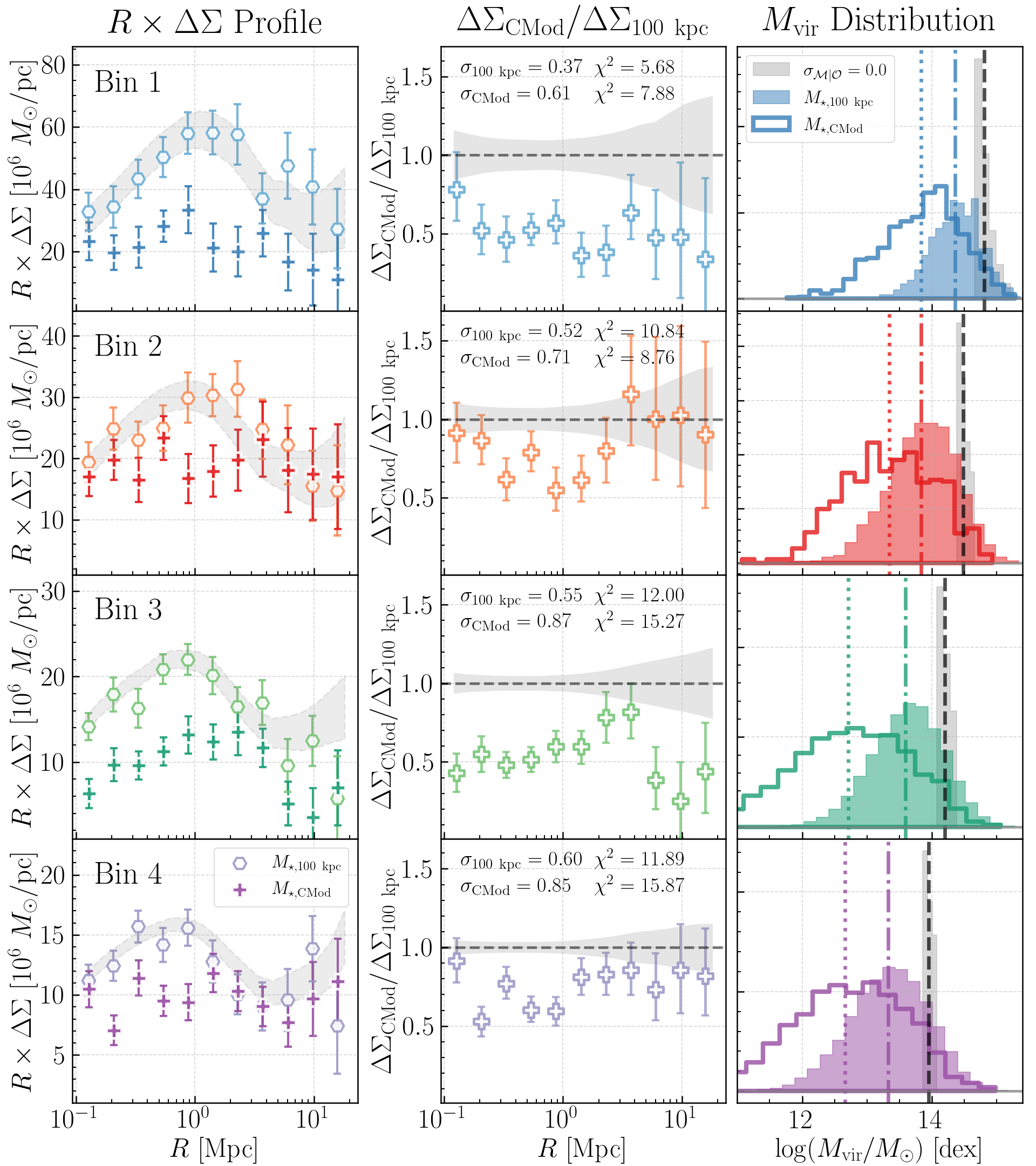}
    \caption{\topn{} results for \maper{100} and \mcmodel{}. 
        The format of this figure is the same as Figure \ref{fig:m100_mout}. 
        \textbf{Left} column: the \rdsigma{}
        profiles for the \maper{100}- (circles) and \mcmodel{}-selected (crosses) samples. 
        The lensing amplitude for \maper{100} is significantly higher than for \mcmodel{} in all
        four bins. 
        Even without fitting a model, it is obvious that \maper{100} is a much better tracer of
        \mvir{} than \mcmodel{}. 
        \textbf{Middle} column: the ratio of \dsigma{} profiles. 
        Samples selected by \mcmodel{} have lensing amplitudes $\sim$30-50\% lower than \maper{100}
        selected samples. 
        \textbf{Right} column: the inferred \mvir{} distributions for the two samples using the
        model described in \S\ \ref{sec:estimate_scatter}.
        Grey shaded regions indicate the \mvir{} distributions of an ideal tracer with
        \sigmh{}$=0$. 
        Vertical lines indicate mean \mvir{}. 
        The differences of the mean \mvir{} between \mcmodel{} and \maper{100} based selections
        range from 0.2-0.4 dex in Bin 1 \& 2 to 0.6-0.8 dex in Bin 3 \& 4.
        The \texttt{Jupyter} notebook for reproducing this figure can be found here:
        \href{https://github.com/dr-guangtou/jianbing/blob/master/notebooks/figure/fig7.ipynb}{\faGithub}.
        }
    \label{fig:m100_cmod}
\end{figure*}

    Figure \ref{fig:m100_cmod} compares the \topn{} results of the benchmark aperture stellar mass
    \maper{100} to that of \mcmodel{}.  
    The \dsigma{} profiles of \mcmodel{} selected samples have significantly lower lensing
    amplitudes (on average by 20-50\%) in all four bins over the entire radial range (left and
    middle panels of Figure \ref{fig:m100_cmod}). 
    The best-fit \sigmvir{} values for the \mcmodel{} samples are $[0.61, 0.71, 0.87, 0.85]$ and are
    much worse than those of the \maper{100} samples.
    Such large \sigmvir{} values also mean significantly lower \mvir{} values.
    As shown in the right panels of Figure \ref{fig:m100_cmod}, the mean \mvir{} of the \mcmodel{}
    \topn{} samples are $[0.52, 0.17, 0.79, 0.53]$ dex lower than that of the \maper{} ones.
    All these results make it  obvious that \cmodel{} is not a good \mvir{} proxy. 

    The \chisq{} values of the \mcmodel{}-based \dsigma{} profiles are $[7.88, 8.76, 15.27, 15.87]$.
    These values are statistically similar to those of \maper{100} samples in Bin 1-3, indicating 
    they are still broadly consistent with the ``pure scatter'' model.
    Note that the \sigmvir{} values for Bin 2-4 are so large that it includes halos with
    \logmvir{}$\leq 12.0$ which are unlikely to host real massive galaxies with 
    \logmcmodel{}$\geq 11.0$. 
    Given this, we suggest not taking the \sigmvir{} values for \mcmodel{} too literally.
    Instead, the main point here is that \mstar{} based on \cmodel{} photometry is not a promising
    \mvir{} proxy and the \sigmvir{} values associated with \mcmodel{} are much larger than other
    proxies.

\subsubsection{The Case of \masap{}}
    \label{sec:asap_result}

    As mentioned in \S\ \ref{sec:masap}, in \citet{Huang2020}, we proposed the \asap{} model that
    can use a linear combination of \maper{10} and \maper{100} (or \mmax{}) to improve the 
    prediction of \mvir{} for massive galaxies. 
    Here we briefly summarize the results.

\begin{figure*}
    \centering
    \includegraphics[width=0.85\textwidth]{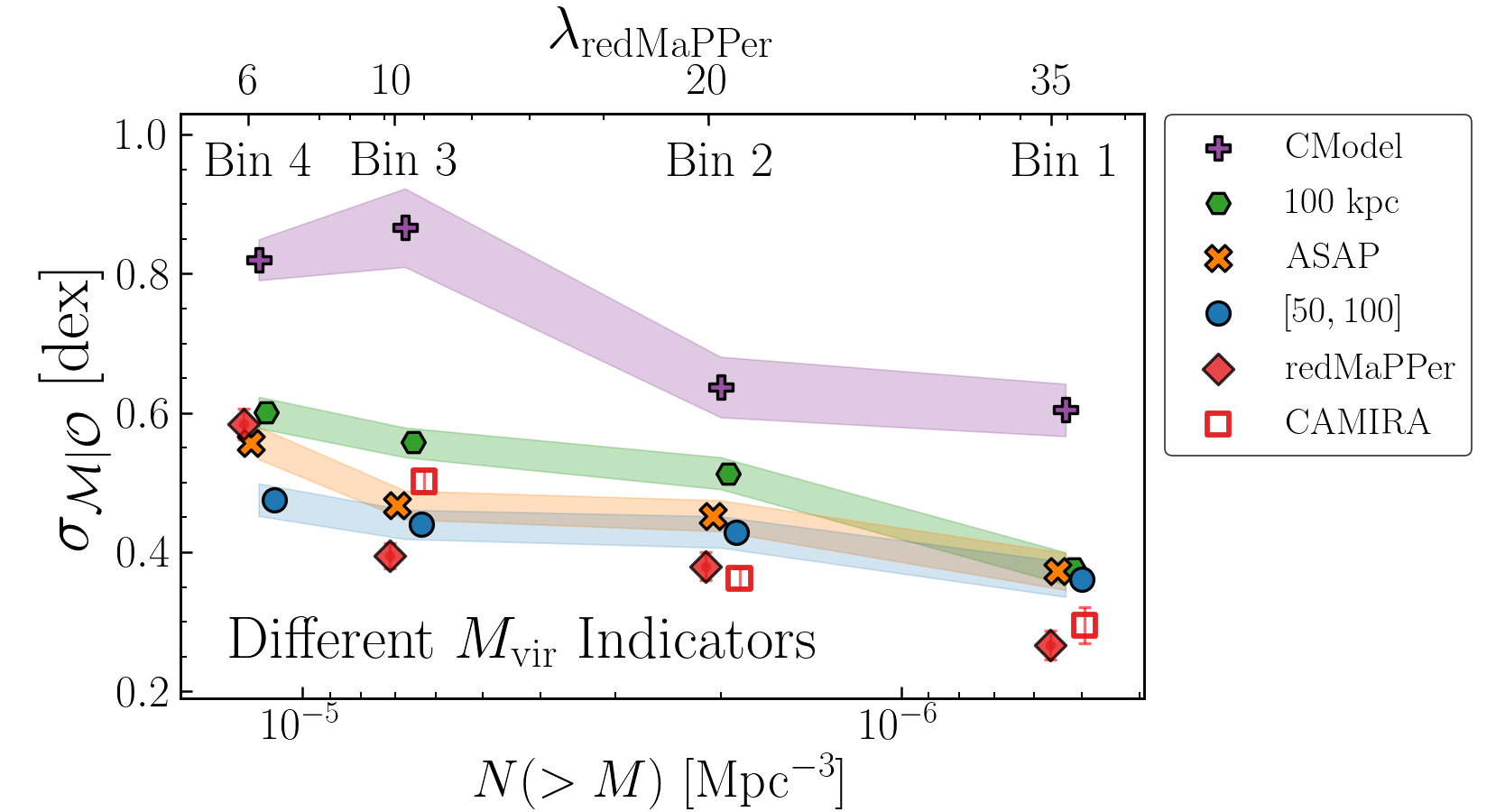}
    \caption{Number density--\sigmvir{} relation for
        six key \mvir{} proxies (similar format as Figure \ref{fig:scatter_trend}).
        For \camira{}, we only show \sigmvir{} in the first three bins since the 
        cluster sample size is not large enough to include Bin 4. 
        \menve{50}{100} shows performance comparable to richness-based proxies with lower \sigmvir{}
        values at the low-\mvir{} end. 
        While the formal values of \sigmvir{} are better for richness-based estimators for Bin 1,
        the quality of the fits for richness-based estimators is not as good as for \menve{50}{100}.
        The \texttt{Jupyter} notebook for reproducing this figure can be found here:
        \href{https://github.com/dr-guangtou/jianbing/blob/master/notebooks/figure/fig8.ipynb}{\faGithub}.
    }
    \label{fig:scatter_trend_2}
\end{figure*}

    In Figure \ref{fig:scatter_trend_2}, we compare the \sigmvir{} trends with number density 
    for a few important \mstar{}-based \mvir{} proxies including the \masap{}. 
    As expected, \masap{} shows improvement in \sigmvir{} values when compared to large aperture 
    \mstar{} such as \maper{100}.
    Meanwhile, in Bin 2-4, outskirt \mstar{} like \menve{50}{100} still displays small 
    advantage in \sigmvir{} over \masap{}, especially in Bin 4.
    We notice that the \sigmvir{}$=[0.38\pm0.03, 0.44\pm0.02, 0.48\pm0.02, 0.56\pm0.02]$ of \masap{}
    are very similar to the results of \menve{10}{100}: 
    \sigmvir{}$=[0.36\pm0.03, 0.47\pm0.02, 0.50\pm0.02, 0.56\pm0.02]$.
    This shows that the \asap{} model, which uses a linear combination of \maper{10}
    and \maper{100}, has minimal improvements over the one which just uses the difference between
    those masses. 
    This, along with the fact that the preferred definition of outskirt is between 50 and 100 kpc,
    strongly suggests that it is very difficult to gain additional information about \mvir{} using
    the inner regions ($R < 30$ kpc) of massive galaxies.
    
    We should note that the \dsigma{} profiles of \masap{} samples are also well described
    by the simple ``pure scatter'' model.
    
\subsubsection{Comparison with Richness-based Proxies}
    \label{sec:richness_results}

    We now compare the \mstar{}-based proxies with richness-based ones. 
    Figure \ref{fig:scatter_trend_2} compares the number density--\sigmvir{} trends for a 
    representative set of \mvir{} proxies in this work, including \mcmodel{} (default survey
    photometry), \maper{100} (large aperture \mstar{}), \menve{50}{100} (the best outer envelope
    mass), \masap{} (a combination of the inner and large aperture mass), and $\lambda_{\rm redM,\
    HSC}$ and $\lambda_{\rm CAMIRA,\ HSC}$ (richness of red-sequence galaxies).
    We summarize the \sigmvir{} values, along with the precise cuts that define the bins for each
    \mvir{} proxy, in Table \ref{tab:summary}. 
    We remind readers that the \mstar{}-based samples and the two richness-based
    cluster catalogues are independently selected from the HSC \texttt{S16A} dataset.
    While there is a considerable overlap, not all the \mstar{}-selected massive galaxies in the
    \topn{} samples belong to identified clusters, and not all cluster centrals are included in the
    parent sample of massive galaxies.
    We will briefly discuss this in \S\ \ref{sec:perfect_finder}.

    For \mstar{}-based proxies, we \emph{do not exclude} massive satellite galaxies from both
    observations and mock catalogues.
    For richness-selected clusters, we use a central-only mock catalogue to calculate their \dsigma{}
    profiles and estimate their \sigmvir{} values. 
    This assumes the cluster finders identify the correct central galaxies, which is not always the
    case, but as we showed in \S\ref{sec:satellite}, satellite contamination is not likely to affect
    any of the key results shown here.
    
    Judged solely by the \sigmvir{} values, the richness of red-sequence galaxies is an excellent
    \mvir{} proxy for massive halos.
    Both richness-based cluster finders show lower \sigmvir{} values in Bin 1 \& 2 than any of the
    \mstar{}-based \mvir{} proxies:
    \sigmvir{}$=[0.27, 0.38]$ for HSC \redm{} clusters (red diamonds in figure
    \ref{fig:scatter_trend_2}), $[0.30, 0.36]$ dex for the \camira{} \texttt{S16A} catalogues (red
    open squares).
    These two bins have \redm{} $\lambda > 20$ and \camira{} $N_{\rm mem} > 21$,
    corresponding to \mvir{}$\geq 2\times 10^{14} M_{\odot}$.
    In this \mvir{} range for typical galaxy clusters, such low values of \sigmvir{} computed with
    our simple model qualitatively agree with previous calibrations (e.g., \citealt{Murata2018,
    Murata2019}).

    The \sigmvir{} increases slightly for richness-based proxies towards the lower--\mvir{} end.
    The \sigmvir{} values for HSC \redm{} clusters are $[0.44, 0.62]$ dex in Bin 3 \& 4 and 0.52 dex
    in Bin 3 for \camira{} clusters.
    This trend with richness (or number density) is also qualitatively consistent with the results
    from \citet{Murata2018, Murata2019}.
    Taken at face value, the performance of the two richness-based proxies becomes comparable with
    the outskirt \mstar{} in Bin 3 \& 4.
    Note that the $\lambda_{\rm HSC}$ range in Bin 4 ($6 \leq \lambda_{\rm HSC} < 10$) is 
    very low and is challenging for any richness-based cluster finder.
    More importantly, we underscore the fact that the shapes of the \dsigma{} profiles of
    richness-selected clusters show systematic deviations from the pure scatter model. 
    This means that the inferred \sigmvir{} values for richness based selections could be
    underestimated. 
    We will return to this question in detail in \S\ \ref{sec:mstar_vs_richness}.

\subsubsection{Summary of \sigmvir{} Trends for Different \mvir{} Proxies}
    \label{sec:trend}

    We briefly summarize the number density - \sigmvir{} trends for different \mvir{} proxies.
    Note that not all \dsigma{} profiles are equally well described by the ``pure scatter'' model,
    but we only focus on the best-fit \sigmvir{} values here.

    \begin{itemize}
    
        \item The outer mass of massive galaxies is a promising \mvir{} proxy. 
            The \menve{50}{100} mass out-performs large aperture \mstar{} such as \maper{100}.
        
        \item Outer stellar mass is a competitive proxy with richness and may outperform richness in
            the low-\mvir{} regime (e.g. $\lambda \leq 20$, or $N_{\rm Mem} \leq 20$).
        
        \item Galaxy inner mass ($r< 10$-30 kpc) is a poor tracer of present day halo mass.
            For this reason, only the outskirt \mstar{} that excludes the inner 30 kpc demonstrates 
            clear improvement over large aperture \mstar{} (see Figure \ref{fig:scatter_trend}).
        
        \item Empirical model such as \asap{} that attempt to take advantages of more than one 
            aperture \mstar{} do show smaller \sigmvir{} values compared to single large aperture
            \mstar{}.
            However, the decision to use \maper{10} (which we now know only adds noise) in
            \citet{Huang2020} limits the level of improvement.

        \item Stellar masses derived from default survey photometry pipelines are likely to yield
            poor \mvir{} proxies as \mcmodel{} has the worst overall performance.
            This not only applies to \cmodel{}, but could be also true for the small aperture
            photometry, \texttt{SourceExtractor} \texttt{MAG\_AUTO}, or even single-\ser{} 2-D
            models.
            
    \end{itemize}

\begin{figure*}
    \centering
    \includegraphics[width=0.95\textwidth]{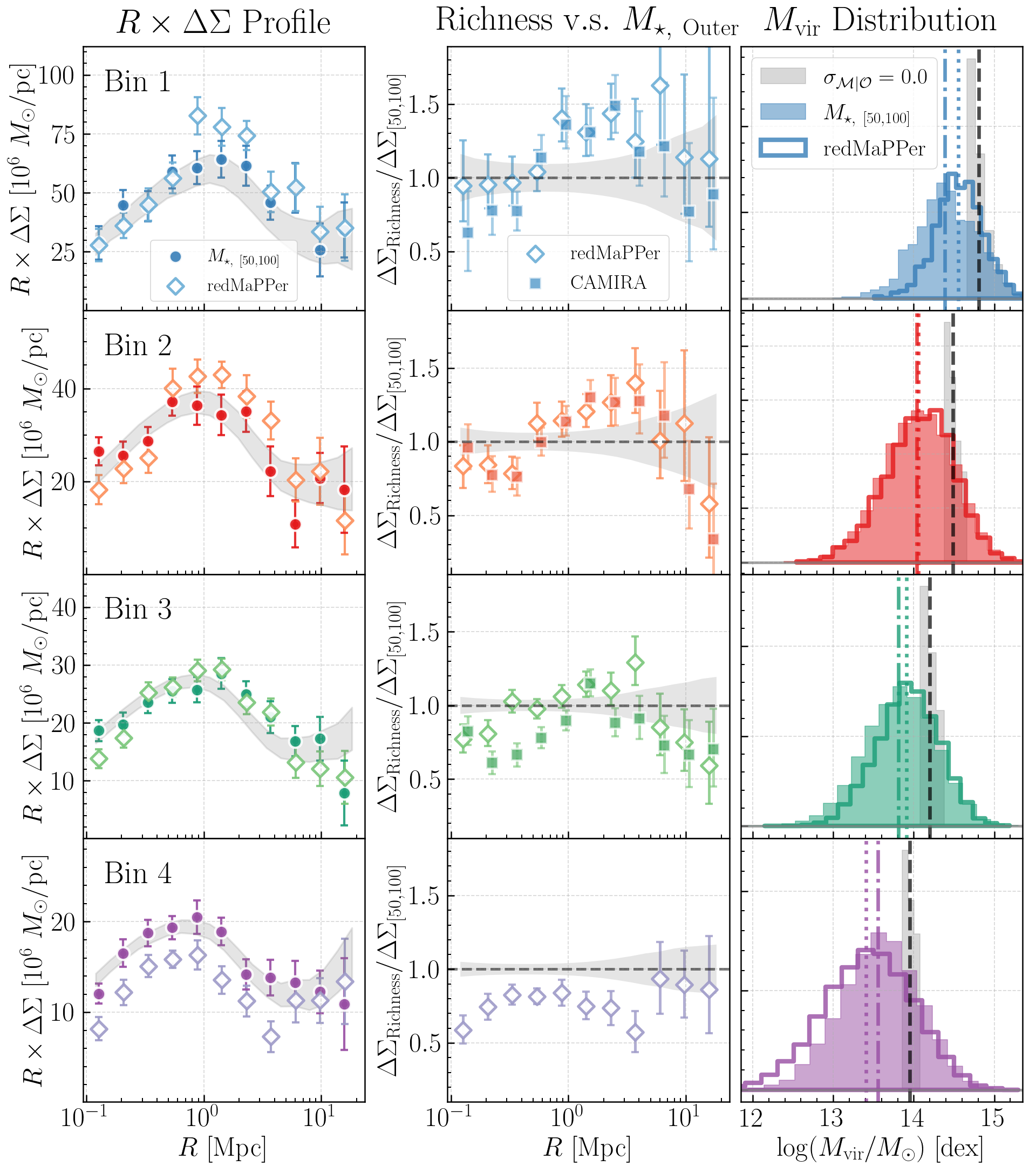}
    \caption{
        \topn{} comparisons between the richness-based optical clusters (\redm{} and \camira{})
        and the massive galaxies selected using outer envelope stellar mass (\menve{50}{100}).
        The layout is very similar to Figure \ref{fig:m100_cmod} and Figure \ref{fig:m100_mout}.
        \textbf{Left} column compares the \rdsigma{} profiles of the HSC \redm{} clusters
        (open diamond) and the \menve{50}{100} (solid circle) selected samples.
        The grey shaded region shows the best-fit profile of the \menve{50}{100} samples.
        In Bin 1-3, while the overall lensing amplitudes are similar, there are interesting
        scale-dependent differences that become more clear using the ratio of the \dsigma{}
        profiles (\textbf{middle} column):
        The lensing amplitudes of \redm{} clusters are systematically higher than the
        \menve{50}{100} samples around $\sim 1$-3 Mpc by $\sim$20--40\%.
        Meanwhile, the amplitudes of \redm{} \dsigma{} profiles are slightly lower than or
        similar to the \menve{50}{100} ones in the central ($R < 0.5$ Mpc) and outer
        ($R>6$-8 Mpc) regions.
        We also show the ratio of lensing profiles using the HSC \camira{} cluster samples
        (square filled with grey colour) to highlight the similar behaviour of these two
        richness-based cluster finders.
        In Bin 4, the \redm{} sample displays a $\sim 20-50$\% lower lensing amplitudes than
        the corresponding \menve{50}{100} sample.
        In the \textbf{right} column, we visualize the trend of the average \mvir{} in each bin:
        while the \redm{} samples show $\sim 0.2$ dex higher average \mvir{} value in Bin 1,
        the differences become smaller in Bin 2 \& 3.
        In Bin 4, the \menve{50}{100} selected sample shows a $\sim 0.2$ dex higher average
        \mvir{} values than the \redm{} one instead.
        The \texttt{Jupyter} notebook for reproducing this figure can be found here:
        \href{https://github.com/dr-guangtou/jianbing/blob/master/notebooks/figure/fig9.ipynb}{\faGithub}.
        }
    \label{fig:mout_richness}
\end{figure*}

\begin{figure*}
    \centering
    \includegraphics[width=\textwidth]{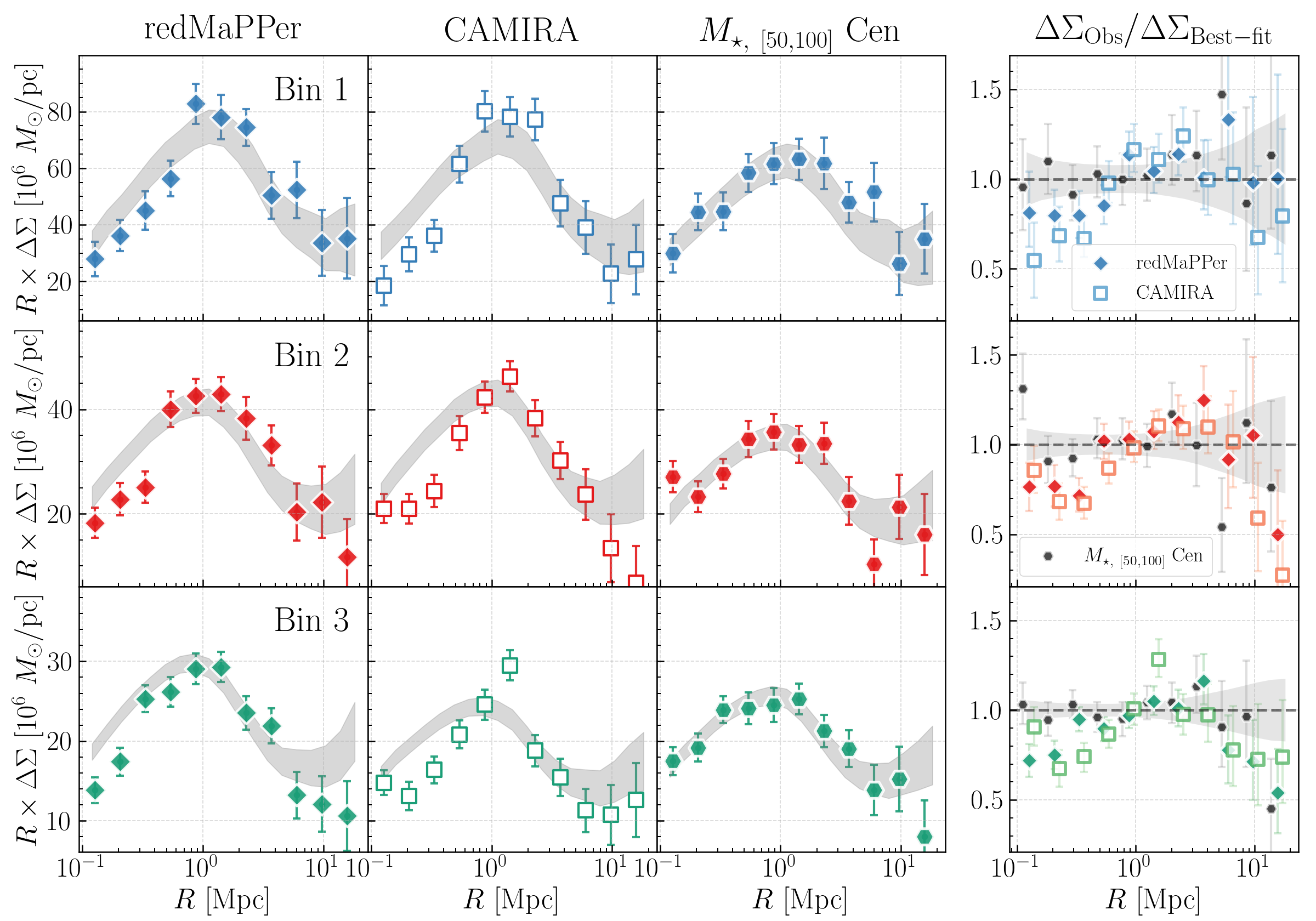}
    \caption{
        The comparisons between the shape of \rdsigma{} profiles and the best fit ``pure-scatter''
        model for richness based cluster finders. 
        \textbf{Left two} columns: the observed \rdsigma{} profiles (symbols) and their
        best--fit models (grey shaded regions) for \redm{} (left; filled diamonds) and \camira{}
        (middle; open squares) clusters in Bins 1-3. 
        We ignore Bin 4 because the cuts applied to the \camira{} catalogue preclude using this bin. 
        \textbf{The third} column from the left: same as the left two columns, but for 
        \menve{50}{100} as a reference for \mstar{}-based \mvir{} proxy (filled hexagons).
        \textbf{Right column}: ratio of the observed \dsigma{} profiles to their best-fit model.
        We also show the ratio for \menve{50}{100} (grey circles) as reference.
        The simple ``pure-scatter'' model is not a good fit and scale-dependent ``residuals'' are
        clearly visible for richness-based \mvir{} proxies when compared to \menve{50}{100}.
        While the exact values of the ratios are different, the two richness-based cluster finders
        display qualitatively similar behavior: the observed \dsigma{} profiles are lower than the
        best-fit models at $R<1$ Mpc by $\sim 30$\% but show higher amplitudes at $1-3$ Mpc. 
        This shape of \rdsigma{} may be a generic ``feature'' of richness-based cluster selections
        and could be due to mis-centering or projection effects.
        The \texttt{Jupyter} notebook for reproducing this figure can be found here:
        \href{https://github.com/dr-guangtou/jianbing/blob/master/notebooks/figure/fig10.ipynb}{\faGithub}.
    }
    \label{fig:richness_residual}
\end{figure*}

\subsection{Information Contained in the the Shape of \texorpdfstring{$\Delta\Sigma$}{DSigma}}
    \label{sec:mstar_vs_richness}

    In the previous section we focused on the overall amplitude of \dsigma{} and the inferred
    \sigmvir{} values. 
    Now we consider the shape of the \dsigma{} profiles. 
    We focus in particular on two questions:

    \begin{enumerate}

        \item Are there differences in the shape of the \dsigma{} profiles for samples selected by
            \mstar{}-based \mvir{} proxies and the clusters selected by richness?

        \item Which type of \mvir{} proxy can yield \dsigma{} profiles whose shapes are consistent
            with a ``clean selection" (we also use the term ``pure scatter") of massive halos? 
            ``Clean selection'' here means a selection based on a simple $\log$-linear \mvir{}-proxy 
            relation with Gaussian scatter.

    \end{enumerate}

    Figure \ref{fig:mout_richness} compares the \dsigma{} profiles of HSC \redm{}, \camira{}
    clusters (first three bins only), and \menve{50}{100}-selected massive halos.
    In Bin 1 to 3, both the \redm{} and \camira{} \dsigma{} profiles show similar systematic
    differences in shape compared to the \menve{50}{100} profiles.
    The most prominent difference is that, between $1 < R < 3$ Mpc, the richness-based \dsigma{}
    profiles demonstrate significantly enhanced ($\sim 30$-40\%) \dsigma{} amplitudes.
    On the other hand, at $R < 1$ Mpc, the \dsigma{} profiles of \redm{} and \camira{} samples
    are $\sim 20$-40\% lower than the outer envelope \mstar{} ones.
    At larger scale ($R > 5$ Mpc), we find that the richness-based and \menve{50}{100} \dsigma{}
    profiles become statistically similar but we are also limited by the low \snratio{} of the
    current profiles.
    Other \mstar{}-based proxies with similar \sigmvir{} values (\eg{} \maper{100}, \masap{},
    \menve{50}{150}) can also yield qualitatively same conclusions.
    In Appendix \ref{app:sdss_redm}, we show that the SDSS \redm{} clusters in the HSC
    \texttt{S16A} footprint display similar systematics in their \dsigma{} profiles using HSC
    lensing data. 
    In appendix \ref{app:des_redm}, we show that the DES \redm{} clusters in the same 
    redshift and richness bins show consistent shape in their lensing profile with the HSC one.
    The DES sample is not only based on a different imaging dataset, its lensing profile 
    is from an independent shear catalogue with different strategies for shape measurements and 
    lensing calibration.
    Both of these comparisons demonstrate the robustness of our result.
    The shape of \dsigma{} profiles robustly show that the \sigmvir{} values alone cannot fully
    explain the difference between \mstar{}- and richness-based \mvir{} proxies.
    Instead of the overall higher lensing amplitude expected from the lower \sigmvir{} values
    for \redm{} and \camira{} clusters, we see a ``bump''-like feature at $R \sim 1$-2 Mpc.

    In Bin 4, the \dsigma{} profile of \redm{} clusters shows lower amplitude than that of 
    \menve{50}{100} sample, consistent with its higher \sigmvir{} value in this bin. 
    The \dsigma{} profile in Bin 4 also does not show a clear 1 Mpc ``bump'' as in the other three
    bins.
    
    These results make question (ii) more interesting: are these richness-selected clusters
    consistent with a selection from a $\log$-linear \mvir{}--richness relation with Gaussian
    scatter? 
    To address this question, Figure \ref{fig:richness_residual} compares the observed \dsigma{}
    profiles of \redm{}, \camira{} clusters, and \menve{50}{100}-selected sample to each of their
    best--fit profiles using our ``pure scatter'' model (\S\ \ref{sec:estimate_scatter}). 
    In the larger, left-hand part of Figure \ref{fig:richness_residual}, each column shows results
    for a different \topn{} selection method, with results from different bins shown in different
    rows. 
    Within each panel, comparing the points with error bars to the shaded gray band allows us to 
    assess how the {\em shape} of the \dsigma{} profiles compare to theoretical expectations based
    on an unbiased selection of clusters based on true \mvir{} (i.e., the ``pure scatter'' model).
    In the single vertical column on the right-hand side of Figure \ref{fig:richness_residual}, we
    show the ratio of the observed \dsigma{} to the corresponding profile based on the ``pure
    scatter'' model; on the right-hand side, different symbols correspond to results based on
    different cluster-selection methods.\footnote{We remind the reader that for the case of
    \menve{50}{100}, we remove the possible massive satellites using the procedure described in \S\
    \ref{sec:satellite}.}    

    From Figure \ref{fig:richness_residual} it is visually apparent that clusters selected according
    to \menve{50}{100} exhibit a \dsigma{} profile that closely mimics the profile of cluster
    samples that have been selected according to \mvir{} in an unbiased fashion. 
    Relative to \menve{50}{100}-selected clusters, we can see that neither \redm{} nor \camira{}
    clusters have \dsigma{} profiles that are as well-described by the ``pure scatter'' model.
    For both \redm{} and \camira{}, the most prominent residual  is the steep drop in the profile in
    the $R < 1$ Mpc region (by up to $\sim 50$\%) relative to the profile of the corresponding
    best-fit ``pure scatter'' model. 
    In addition to this steep drop, the \dsigma{} profile of \redm{} and \camira{} clusters presents
    a distinct ``bump-like" feature around 1-2 Mpc in Bin 1 \& 2, with profiles that have a visibly
    larger lensing amplitude relative to the ``pure scatter'' model at this spatial scale.  
    Using \chisq{} to quantify the quality of the fits of the ``pure scatter" model, we find that
    the HSC \redm{} clusters have values of $[15.36, 55.92, 46.23]$ for Bins 1, 2, and 3,
    respectively; the \camira{} \texttt{S16A} samples have values of $[34.51, 35.65, 62.11]$; the
    quality of these fits is thus significantly poorer relative to \menve{50}{100}-selected
    clusters. 
    This exercise indicates that the \menve{50}{100}-based selection function closely resembles an
    unbiased selection of clusters based on \mvir{} with simple, Gaussian scatter, whereas the
    cluster selection function defined by \redm{} or \camira{} cannot be as well-described by such a
    simple model.
    
\section{Discussion}
    \label{sec:discussion}
    
    We now discuss the implications of the results presented in the previous section.
    We first discuss the connection between our findings based on the \topn{} tests  and the \asap{}
    model in \S \ref{sec:asap_discussion}.
    In \S\ \ref{sec:outskirt_discussion}, we focus on the potential of the outer envelope mass as a
    tool for identifying dark matter halos.
    \S\ \ref{sec:perfect_finder} discusses the possibility of using the techniques developed in this
    paper to search for even better \mvir{} proxies. 
    
\begin{figure*}
    \centering
    \includegraphics[width=\textwidth]{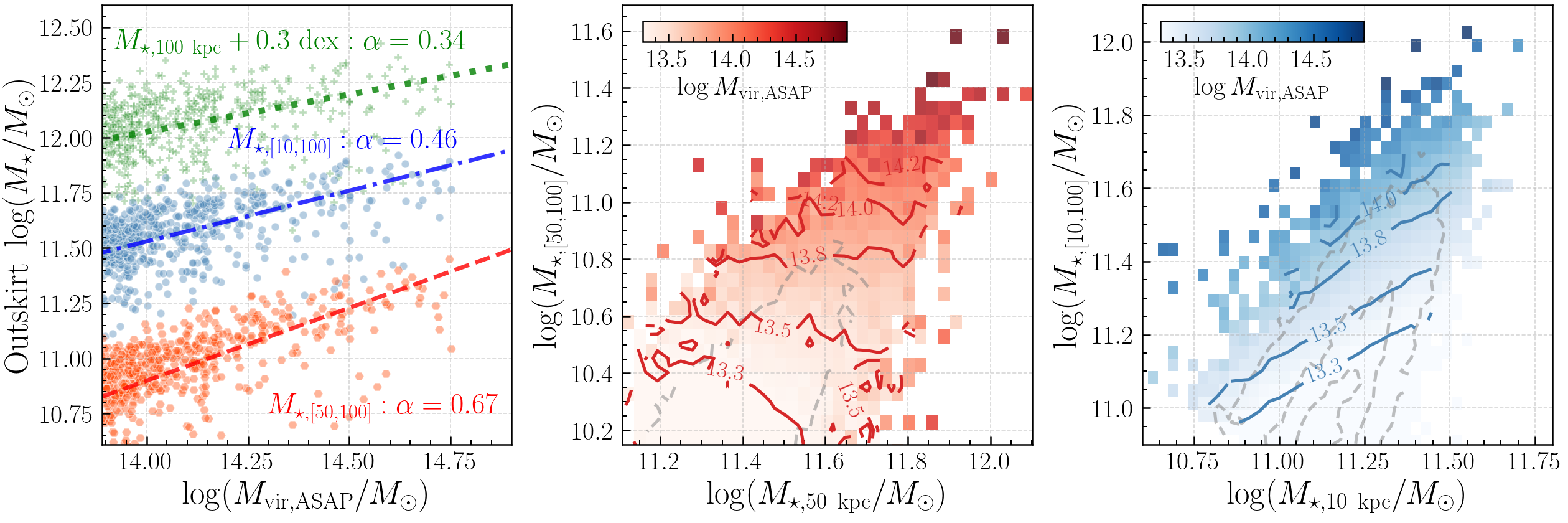}
    \caption{
        We explore the connections between the \asap{}-predicted \mvir{} and the outskirt 
        \mstar{} used in the \topn{} test. 
        \textbf{Left}: relations between the \asap{}-predicted \mvir{} and different stellar mass
        measurements, including \maper{100} (green; shifted up by 0.3 dex for better visibility),
        \menve{10}{100} (blue), and \menve{50}{100} (red).
        To help visualize these scaling relations, we highlight the best-fit $\log$--linear 
        relations at \masap{}$> 10^{13.6} M_{\odot}$.
        \menve{50}{100} displays steeper slopes than both \maper{100} and \menve{10}{100}.
        \textbf{Middle}: the 2-D plane of \maper{50} v.s.  \maper{50}{100} colour coded by the
        \asap{}-predicted \mvir{}.
        Several iso-\mvir{} contours are used to highlight the relation between \mvir{} and
        \menve{50}{100}.
        \textbf{Right}: similar to the middle panel, here we show the \maper{10} and
        \menve{10}{100} 2-D plane colour-coded by the \asap{}-predicted \mvir{}.
        Same set of iso-\mvir{} contours are used to compare with the middle panel.
        The \texttt{Jupyter} notebook for reproducing this figure can be found here:
        \href{https://github.com/dr-guangtou/jianbing/blob/master/notebooks/figure/fig12.ipynb}{\faGithub}.
    }
    \label{fig:outskirt_discussion}
\end{figure*}

\subsection{Relationship with Previous Work and the \asap{} model}
    \label{sec:asap_discussion}
    
    Using the \topn{} tests, we confirm that the \asap{} empirical model (\citealt{Huang2020}) has a
    lower \sigmvir{} value than large aperture stellar masses such as \maper{100} (see Figure
    \ref{fig:scatter_trend_2}). 
    However, we also find that \asap{} is outperformed by outer mass measured such as
    \menve{30}{100} and \menve{50}{100}.

    The \asap{} model uses \maper{10} and \maper{100} as rough proxies of the \insitu{} and
    ``total'' \mstar{} to take advantage of the additional \mvir{}-dependence of the stellar mass
    profiles of massive galaxies.
    However, the \topn{} test clearly shows that \maper{10} alone is a very poor \mvir{} proxy with
    \sigmvir{}$\sim 0.8$ in all four bins (see Figure \ref{fig:scatter_trend}).
    Therefore, the better performance of \masap{} must be driven by a tighter 
    \mvir{}--\menve{10}{100} relation than  \mvir{}-\maper{100}.
    
    Figure \ref{fig:outskirt_discussion} shows the \masap{}-observable relations for \maper{100},
    \menve{10}{100}, and \menve{50}{100} with their best--fit $\log$--linear relations at
    \masap$\geq 10^{13.6} M_{\odot}$\footnote{We use the \texttt{Python} implementation of the Least
    Trimmed Squares (LTS) algorithm (\href{https://pypi.org/project/ltsfit/}{\texttt{ltsfit}}) by
    \citet{Cappellari13b} for the fitting.}.
    Although \masap{} is not the true \mvir{}, it can be used to illustrate the underlying average
    SHMRs.
    The slope of the \masap{}-\menve{10}{100} relation is $\alpha = 0.47$, steeper than that of the
    \masap{}-\maper{100} relation $\alpha = 0.34$.
    Both relations have similar scatter values \sigms$\sim 0.15$ dex at the high-\masap{} end. 
    Hence, the \asap{} model predictions also suggest that outer mass \menve{10}{100} is a better
    \mvir{} proxy than \maper{100}.

    Figure \ref{fig:outskirt_discussion} also compares the \masap{}-\menve{10}{100} and the
    \menve{50}{100} relations. 
    Although \asap{} does not explicitly use \menve{50}{100}, it shows that \menve{50}{100} is a
    better \mvir{} proxy than \menve{10}{100} because the \masap{}-\menve{50}{100} relation has a
    steeper slope ($\alpha=0.67$) and slightly smaller \sigms$\sim 0.11$ dex at 
    \masap$\geq 10^{13.6} M_{\odot}$. 
    To further highlight this point, we also show the \maper{50}--\menve{50}{100} and
    \maper{10}--\menve{10}{100} 2-D planes.
    At \masap{}$>10^{13.5} M_{\odot}$, the iso--\masap{} curves runs almost parallel to
    \menve{50}{100}.
    When compared to the \maper{10}--\menve{10}{100} distributions in the right panel, it becomes 
    clear that at fixed outskirt \mstar{} value, \menve{50}{100} will yield smaller scatter of 
    \masap{}.
    This comparison reaffirms the finding that the $R > 30$--50 kpc outer mass is an excellent
    \mvir{} proxy and that adding \mstar{} from inner regions actually degrades the scatter.
    This also suggests that \asap{}-like models could be improved using an optimized choice for the
    outer mass tracer.

    The idea behind the original \asap{} model was to use two different radii to tracer regions 
    dominated by \insitu{} and \exsitu{} stars. Our \topn{} tests suggest the the \insitu{} may
    actually contain very little information about present day halo mass. 
    A next key step will be understanding the different SHMRs of \insitu{} and \exsitu{} stars
    better in the near future.

\subsection{Scatter in Stellar Mass at Fixed Halo mass}
    \label{sec:sigma_mstar}
    
    In this work, we focus on the \sigmvir{} values and their trends with number density. 
    Meanwhile, works about SHMR often focus on the \sigms{} values.
    For \maper{100}, assuming a SHMR with $\alpha \sim 0.35$ slope (\eg{} \citealt{GoldenMarx2019,
    Huang2020}), a \sigms{}$\sim 0.2$ dex scatter at high-\mvir{} end corresponds to a 
    \sigmvir{}$\sim 0.4$ dex scatter according to Equation \ref{eq:ratio_is_what_matters}.
    This is consistent with our results at high-\mvir{} end and with some recent modelling
    constraints (\eg{} \citealt{Kravtsov2018, Behroozi2018}). 
    Meanwhile, the \sigmvir{} values in lower-\mvir{} bins require \sigms{}$>0.3$ dex under the 
    same slope. 
    For the \camira{} clusters, if we adopt the \mvir{}-richness relation calibrated by 
    \citet{Murata2019} with a $\alpha = 0.6$ slope, the inferred \sigms{} values are around 
    0.25 to 0.35 dex. 
    For \menve{50}{100}, assuming the $\alpha \sim 0.67$ slope from the \asap{} model shown
    in \S\ \ref{sec:asap_discussion}, we derive a larger \sigms{} value ($\sim 0.4$ dex) for this
    outskirt \mstar{} than \maper{100} and richness measurements. 
    As the \sigms{} value inferred here is \emph{not} the intrinsic scatter value, this result 
    may reflect the nosier measurements of outer light profile.
    Since we do not directly constrain the SHMRs, the inferred \sigms{} values here depend on the
    assumed slope values and other systematics.
    They should be only used for relative comparisons within the \topn{} results.
    We will look into the constraint of SHMRs for different \mvir{} proxies in future works.

\subsection{Physical Insight: Why might the Outer Stellar Halo Trace Halo Mass Better than the Inner
    Mass or the Total Mass?}
    \label{sec:outskirt_discussion}

    Our \topn{} tests show that the outer mass of $z < 0.5$ massive galaxies are promising halo mass
    proxies with superior performance compared to ``total'' \mstar{} measurements using large 
    apertures.
    This result may be explained via the ``two-phase'' formation scenario of massive galaxies (e.g.,
    \citealt{Oser2010, vanDokkum2010, Moster2020}). 
    According to this picture, a massive galaxy consists of stars formed within the halo of its main
    progenitor at high redshift (\insitu{} component) and accreted stars from repeated mergers 
    (\exsitu{} component).
    In \citet{Bradshaw2020}, the authors explored the SHMRs of both components using the
    \texttt{UniverseMachine} semi-empirical model (e.g., \citealt{Behroozi2018}). 
    They showed that the \exsitu{} component displays a much tighter correlation with the current
    \mvir{} relative to either the \insitu{} component (see their Figure~9), or relative to the
    total \mstar{}.
    In the \texttt{UniverseMachine} model, the average \insitu{} \mstar{} at $z \sim 0.4$ is almost
    constant over a wide range of \mvir{} ($\sim 10^{10.9} M_{\odot}$), whereas the SHMR of
    \exsitu{} component shows a steep slope.
    In a recent analysis of the TNG-300 simulation, it was found that at fixed \mvir{}, both cluster
    richness and BCG \mstar{} exhibit residual correlations with halo assembly history
    \citep{Anbajagane2020}; our results provide motivation to consider whether such correlations
    persist for true \exsitu{} \mstar{}, and for \mstar{} estimations of the BCG that exclude the
    inner regions.
    Although different models and simulations display different scaling relations between \mvir{}
    and \exsitu{} \mstar{}, there is nonetheless compelling theoretical support for the idea that
    the scatter of SHMRs at high-\mvir{} end is closely tied to the assembly of the \exsitu{}
    component \citep[see, e.g.,][]{Gu2016}.

    Whereas massive galaxies at high redshift grow primarily by \insitu{} mass buildup, these
    galaxies are predominantly quenched at lower redshift, implying that massive galaxies grow
    primarily via merging at late times.  
    Under this picture, the \exsitu{} \mstar{} should scale with the number of accreted satellites,
    and can be considered as a measure of ``historical richness'' of the halo. 
    In simulations, the \exsitu{} component dominates the outskirts of massive galaxies, and its
    fraction increases with both stellar and halo mass (e.g., \citealt{Lackner2012,
    RodriguezGomez2016, Pulsoni2021, Pillepich2017b}). 
    For this reason, outer mass measures such as \menve{50}{100} are likely to scale better with the
    true \exsitu{} \mstar{} relative to mass estimates that include stars from the inner 30 kpc, and
    so \mstar{} measurements defined by the outskirts of a galaxy may be proxies of the ``historical
    richness'' of a parent halo.

    For low-$z$ galaxy clusters, multiple studies have explored the connection between halo
    properties and the flux, shape, and radial profile of the ICL --
    essentially the extended outer envelope around the central galaxy (or the BCG) of the cluster
    (e.g., \citealt{Montes2018, Montes2019, Zhang2019b, Furnell2021, Kluge2021, SampaioSantos2021}.
    While several of these studies demonstrate the tight correlation between the \mvir{} and the
    stellar mass/luminosity of the BCG$+$ICL component\footnote{This is sometimes referred as the
    ``diffuse stellar light'' component following the definition in the Illustris-TNG simulation
    (e.g., \citealt{Zhang2019b, SampaioSantos2021}).} (e.g., \citet{Zhang2019b, Kluge2021,
    SampaioSantos2021}), whether the ICL alone is a promising \mvir{} proxy is still under debate
    (e.g., \citealt{Furnell2021}).
    As discussed in \citet{Huang2018b} and \citet{Kluge2020}, the definition of ICL is often 
    ambiguous and somewhat arbitrary, but the light between 50 to 100 kpc around a BCG is often
    considered part of the ICL. 
    Our results support the idea that the ICL correlates with halo mass, but we also generalize this
    finding to all massive galaxies.

    \subsection{On the Possibility of Building Even Better Halo Mass Proxies}
    \label{sec:better_halo_proxy}
    
    Our work shows that a better understanding of the formation process of massive galaxies,
    together with high quality imaging data, offers the exciting prospect of developing better
    proxies of halo mass. 
    This work is only a first step in this direction and follow-up work may yield even better
    proxies than \menve{50}{100}. 
    Here we discuss possible improvements to the outskirt \mstar{}-based \mvir{} proxy.
    
    First, the outer stellar mass is estimated using the portion of the light profile that has low
    \snratio{}, which can potentially be affected by issues related to background subtraction, by
    contamination from other objects, and by galactic cirrus (\eg{} \citealt{Roman2020}). 
    Deeper images and improved data-reduction strategies should help improve the accuracy of
    measurements of outer mass.  
    Second, we have ignored the radial variation of \mlratio{}, and also the fact that galaxies are
    three-dimensional in nature. 
    
    More importantly, our results provide motivation for the quest for better proxies of \exsitu{}
    \mstar{}.
    For example, it could prove fruitful to explore improved definitions of outer stellar mass based
    on physical radial boundaries. 
    In this work, we have explored boundaries based on $R_{50}$ (Appendix \ref{app:size}) and found
    no substantial improvement beyond results based on \menve{50}{100}, but scaling boundaries
    according to the total \mstar{} might still be an interesting idea.
    This research direction is also motivated by recent simulations which are able to reproduce HSC
    light profiles fairly well (e.g., \citealt{Ardila2021}). 
    Ideally, we would like to be able to physically decompose massive galaxies into their \insitu{}
    and \exsitu{} components using real data. More careful approaches towards this decomposition
    that take these additional effects into account is worthy of exploration in future work and are
    well-motivated by our results.
    
\subsection{Implications for Optical Cluster Finding}
    \label{sec:perfect_finder}
    
    Our results have profound implications for cluster finding with optical surveys including the
    Vera Rubin Observatory's Legacy Survey of Space and Time
    (LSST)\footnote{\url{https://www.lsst.org}}, the \textit{Euclid}
    satellite\footnote{\url{https://sci.esa.int/web/euclid}}, and the Nancy Grace Roman space
    telescope (\textit{Roman})\footnote{\url{https://roman.gsfc.nasa.gov}}.  
    These results potentially open up a new way of approaching the problem of identifying massive
    halos from imaging surveys.

    Traditionally, optical/NIR cluster finders are mostly based on the relation between \mvir{} and
    some estimate of cluster richness.
    The \redm{} and \camira{} cluster finders rely on the number of quenched member galaxies on the
    red-sequence, while others use the over-density of galaxies within a narrow photo-$z$ range
    (e.g., \citealt{Wen2021, Zou2021}). 
    The prevalence of richness-based cluster identification reflects the widely-held expectation
    that the \mvir{}-richness relation has lower scatter than the SHMR.
    However, in this work we have shown that this comparative assessment has overlooked two critical
    aspects of the \mstar{} estimation of massive galaxies.
    1) Default photometry from data reduction pipelines often provide poor fits to the light 
    profiles of massive galaxies, and can be significantly impacted by issues related to background
    subtraction and deblending.
    Such photometry is a source of both bias and additional scatter in the \mstar{} estimates (see
    Figure \ref{fig:scatter_trend_2} and \S\ \ref{sec:m100_cmodel}). 
    2) Inner stellar mass is an {\em intrinsically} poor proxy of \mvir{} (see Figure
    \ref{fig:scatter_trend} and \S\ \ref{sec:m_aper}).
    Yet, commonly adopted photometry measures for massive galaxies focus on the bright, inner
    ``core'' region where the signal-to-noise is high. 
    Our \topn{} tests demonstrate that both of these issues have a strong influence on the level of
    scatter in the SHMR. 
    We have furthermore shown that through careful consideration of how the \mstar{}-based proxy is
    measured and defined, cluster samples selected by \mstar{} exhibit scatter in \mvir{} that is
    both tighter and simpler relative to richness-selected clusters. 
    Finally, it is important to note that customized estimates of the outer light profiles of
    massive galaxies are necessary to build \mvir{} proxies with comparable \sigmvir{} values with
    richness (at least with current versions of data reduction pipeline).
    
    Outer galaxy mass may offer distinct advantages over richness-based cluster finders with respect
    to two key systematics: mis-centering and projection effects.  
    Mis-centering bias occurs when the cluster finder identifies the wrong central galaxy, or when
    the central galaxy is not at the true center of the dark matter halo. 
    Projection effects have a variety of origins, such as the anisotropic distribution of satellite
    galaxies within the halo, and the presence of large-scale structure along the line-of-sight to
    the cluster. 
    While the calibration of the \mvir{}--richness relation now routinely includes mis-centering
    effects when modelling cluster \dsigma{} profiles (e.g., \citealt{Murata2018, Murata2019,
    McClintock2019}), projection bias is still a major issue (e.g., \citealt{Costanzi2019,
    Sunayama2020, DES2020, To2021b}). 
    In this paper, we show that \mstar{}-based proxies using both larger apertures or outer mass
    display stacked \dsigma{} profiles that are consistent with having negligible mis-centering
    effect and projection bias (see Figure \ref{fig:mout_richness} and \S\
    \ref{sec:mstar_vs_richness}) -- this is very exciting as it suggests that outer mass measures
    such as \menve{50}{100} directly trace central galaxies and could yield a more simple selection
    function than richness-based methods.  
    
    We have not identified the exact causes of the systematic differences in the \dsigma{} profiles
    between \mstar{}- and richness-selected samples (\S\ \ref{sec:mstar_vs_richness}).
    Recently, \citet{Sunayama2020} explored the impact of projection bias on cluster \dsigma{}
    profiles using mock catalogues based on N-body simulations.
    They show that projection effects can boost the stacked \dsigma{} profiles of clusters at $R >
    2$ Mpc by up to 20\%.
    Also, this ``bump" in the the outer \dsigma{} profile seems to increase with the intrinsic
    richness.
    While there are qualitative similarities between their Figure 4 and the middle panels of Figure
    \ref{fig:mout_richness}, the \citet{Sunayama2020} model cannot not fully explain the differences
    we see, and so a sophisticated modelling effort will be required in order to fully understand the
    origin of this feature in the lensing profiles of richness-selected clusters.
    
    Another key limitation of richness-based cluster finders stems from the difficulty of
    generating realistic mock catalogues of galaxies. 
    Such mocks are essential for the calibration of the \mvir{}-richness relation and for
    understanding the selection biases.
    However, considerable sophistication is required to produce these mock catalogues, since
    red-sequence richness estimation fundamentally requires that the mock galaxies have
    multi-wavelength properties such as a tight red-sequence and colour bimodality for {\em all}
    galaxies down to $\sim0.1L_{\star},$ and moreover that these features are realistically
    connected to the cosmic density field across a wide range of halo mass, redshift and
    larger-scale environment; thus the mocks used for cluster analysis by cosmological surveys have
    historically struggled to meet these challenges at the required levels of quantitative detail
    (e.g., \citealt{Trayford2015, Trayford2017, Nelson2018, DeRose2019}, although see
    \citealt{Hearin2020, DeRose2021} for recent progress). 
    It is also difficult to take systematics in colour measurements into account when building
    mocks.

    On the other hand, it is easier to calibrate \mstar{} estimates using the same definition of
    stellar mass (e.g., \citealt{Ardila2021}) and to account for uncertainties in \mstar{}. 
    It is also easier to reproduce the observed properties of galaxy samples composed primarily of
    centrals (e.g., \citealt{Moster2020}).
    To utilise a \mstar{}-based ``cluster finder'', we would nonetheless need to account for
    contamination from massive satellite galaxies (see \ref{sec:satellite}), but generating the
    required mock catalogues for this purpose is far simpler in comparison to the multi-wavelength
    needs of richness-based methods.  
    
    Lastly, the combination of \mstar{} together with richness-based cluster finders could help
    unveil more insight into the assembly history of massive dark matter halos. 
    Measures of ``historical'' and ``current'' richness might have different selection effects with
    regards to secondary properties (e.g., concentration, merging history) at similar \mvir{}.
    As mentioned previously, the \mstar{} and richness-selected samples do not fully overlap.
    Among the top 50 galaxies selected by \menve{50}{100}, 7 of them are not identified as the
    central of any \redm{} and \camira{} cluster.
    The non-overlap fraction increases with decreasing \mstar{} limits: among the top 200 (1000)
    \menve{50}{100}-selected galaxies, 49 (467) are not considered cluster centrals by \redm{} or
    \camira{}.
    If we limit the samples to the same \topn{} bin, the overlap fraction is even lower: within the
    top 50 (500) \menve{50}{100} galaxies, only 9 (158) of them are also included in the top 50
    (500) richest clusters selected by \redm{}. This suggests that the richness--\mstar{} relation
    for central galaxies has considerable scatter that would be worth exploring in further detail.
    The numerous advantages discussed above suggest that \mstar{}-based cluster finders could not
    only help us to understand the systematics of richness-based cluster finders, but that they may
    also yield competitive constraints on the growth of structure and on galaxy-halo connection
    models.

\section{Summary and Conclusions}
    \label{sec:summary}

    Taking advantage of the deep images and unprecedented lensing capabilities of the HSC survey
    (\S\ \ref{sec:hsc}), we show that the outer envelope of low-redshift massive galaxies is a
    promising \mvir{} proxy with scatter comparable to richness. 
    We further show that this proxy is less affected by systematics such as projection bias and
    mis-centering effects. 
    The outskirts of massive galaxies are dominated by \exsitu{} stars -- the stellar content
    accreted from previous satellites galaxies within the halo -- and thus the outer envelope
    \mstar{} provides an estimate of the ``historical richness'' of massive halos.
    This opens up new possibilities for tracing massive halos, studying their galaxy-halo connection,
    and investigating the assembly histories of massive galaxies.
    
    We have conducted our study by comparing the stacked \dsigma{} profiles (\S\ \ref{sec:dsigma}
    and Appendix \ref{app:dsigma_detail}) of massive halos selected by different \mvir{} proxies
    (\S\ \ref{sec:topn_intro} and \S\ \ref{sec:comp_scatters}) in four volume number density bins
    (\S\ \ref{sec:binning}).
    Assuming a simple $\log$-linear \mvir{}-observable relation model with Gaussian scatter, we
    estimate the scatter in \mvir{} in each bin by matching the observed \dsigma{} profiles to
    models generated from N-body simulations (\S\ \ref{sec:estimate_scatter}).
    Using this \topn{} methodology, we evaluate different \mstar{}-based and richness-based \mvir{}
    proxies for massive galaxies and halos at $0.2 < z < 0.5$ (\S\ \ref{sec:proxies}).
    These proxies include \mstar{} based on the default survey photometry (\cmodel{}), large
    aperture \mstar{} (\S\ \ref{sec:maper}) and outer envelope \mstar{} (\S\ \ref{sec:menvelope})
    based on deep 1-D surface mass density profiles (\S\ \ref{sec:1d_prof}).
    We also include richness estimates from the \redm{} (\S\ \ref{sec:cluster_redmapper}) and
    \camira{} cluster (\S\ \ref{sec:cluster_camira}) catalogues.
    The main results of this work are:
    
    \begin{itemize}
    
        \item Outer galaxy mass is an excellent tracer of halo mass (\S\ \ref{sec:m100_outskirt};
            Figure \ref{fig:scatter_trend}). 
            The performance of \menve{50}{100} and other similar outer envelope measures are 
            competitive with red-sequence cluster finders at the high-richness end (e.g. $\lambda >
            20$) and may outperform \redm{} or \camira{} at the low-richness regime 
            (see Figure \ref{fig:scatter_trend_2}).
            Since the outer envelope is likely to have been built from merging processes, we suggest
            that the outer envelope mass serves as an estimate of the ``historical richness" of a
            cluster, and so could serve as a proxy for \mvir{} that is complementary to the 
            ``current richness'' measurements used by contemporary cluster finders.
        
        \item While both richness-based \mvir{} proxies (\redm{} and \camira{}) have
            impressively low inferred \sigmvir{} values, they result in stacked \dsigma{} profiles
            that are not consistent with predictions based on a ``pure scatter'' model (see Figure
            \ref{fig:mout_richness}).
            Instead, the \dsigma{} profiles of richness-selected clusters have enhanced amplitudes
            around $R\sim 1$-2 Mpc and suppressed inner profiles at $R < 1$ Mpc when compared to
            ``pure scatter'' models (\S\ \ref{sec:mstar_vs_richness} and Figure
            \ref{fig:richness_residual}).
            These results indicate that the richness-based \mvir{} proxies have additional
            systematics (e.g., mis-centering, projection effect) that need to be accounted for. 
            In contrast, the \dsigma{} profiles of galaxies selected according to their outer mass
            are very well described by a ``pure scatter'' model, suggesting that cluster samples
            selected by stellar mass of the outer envelope suffer from little to no mis-centering
            effects or projection effects.
            
        \item Inner galaxy mass (e.g. \mstar{} within 10 to 30 kpc) is a very poor tracer of halo
            mass (Figure \ref{fig:scatter_trend}).
            The total \mstar{} enclosed within a large aperture such as \maper{100} shows much
            better performance, but is still not as effective as the outer envelope mass in terms of
            its ability to serve as a proxy for \mvir{} (Figure \ref{fig:m100_mout}).
            This indicates that there is a comparatively weak physical correlation between the stars
            in the inner region of massive galaxies and the total mass of their dark matter halos. 
        
        \item Stellar masses based on \cmodel{} or any other default photometry from a generic data
            reduction pipeline do not yield good halo mass proxies (see \S\ \ref{sec:m100_cmodel}
            and Figure \ref{fig:m100_cmod}).
            \cmodel{} does not accurately account for the flux in the extended stellar halo of
            massive galaxies, which is the specific region of the galaxy that has the tightest
            connection with the underlying \mvir{}.
            We therefore caution against the use of \cmodel{}-like photometry in studies of the
            galaxy-halo connection. 
            LSST will eventually be even deeper than HSC and currently shares a very similar data
            reduction pipeline. 
            In order to realise the scientific potential afforded by accurate measurements of the
            masses and profiles of massive galaxies, it will be critical to improve the image
            deblending and galaxy modelling algorithms in \texttt{lsstPipe}. 
            It would highly beneficial to the wider cosmology community if \texttt{lsstPipe} could
            accurately perform the required measurements without the need for the custom pipelines
            that were needed for the present work.

        \item We have shown that satellite galaxies do not have a strong impact on the stacked
            \dsigma{} profile of the highest \mstar{} samples (due to low satellite fraction).
            Satellites do impart a mild effect on \dsigma{} towards the lower-\mstar{} end (\S
            \ref{sec:satellite}) -- further work will be required in order to model this effect for
            cosmological applications.

    \end{itemize}

    Motivated by these results, we plan to further explore \mstar{}-based \mvir{} proxies for
    studies of the galaxy-halo connection and for cosmology.
    Recent HSC data releases (\texttt{S20A} or \texttt{PDR3} in 2021) have increased sky coverage to
    $> 600$ deg$^2$, four times larger than the current sample.
    Not only would these larger samples improve the statistical uncertainty of the \topn{} tests,
    they would also enable us to explore the high-\mvir{} regime in much finer detail than is
    permitted by the broad richness bins used here.
    Moreover, the new data releases come with improved background subtraction that will improve
    measurements of the outer profile of massive galaxies.
    We are also working on an improved outer envelope \mstar{} measurement using a more accurate
    \mlratio{} and a more sophisticated modelling approach.
    On the theoretical side, we will use state-of-the-art hydro-simulations and semi-empirical
    models to investigate the connection between the outer envelope of massive galaxies and the
    assembly history of their dark matter halo. 
    It would also be interesting to compare cluster samples selected by richness- and \mstar{}-based
    methods.
    In addition to the selection biases of different methods, this could yield further insight into
    the distribution of halo properties at the high-\mvir{} end. 
    Finally, for purposes of developing an \mstar{}-based cluster finder, it will also be fruitful
    to compare the properties of \mstar{}-selected clusters to samples identified by other
    multi-wavelength methods that are less sensitive to projection effect (e.g., samples identified
    in X-ray or microwave bands).

    As outlined in \citet{Bradshaw2020}, we also suggest that a ``hybrid'' cluster finder that
    combines the advantages of richness- and \mstar{}-based \mvir{} proxies may be possible by
    simply combining the \mstar{} of the central galaxy and a few (e.g., top 2 or 3) massive
    satellite galaxies.
    Such a ``Cen$+N$'' method could be an excellent \mvir{} proxy, with low \sigmvir{} values in a
    given number density bin, while also being minimally impacted by projection effects.
    
    Both the \mstar{}-based and the ``Cen$+N$'' methods require accurate identification of massive
    satellite galaxies.
    This is a challenging task when using photometric redshift from imaging surveys.  Spectroscopic
    surveys such as DESI (e.g., \citealt{DESI2016}) will greatly improve the situation.
    Using images from the DECam Legacy Survey (DECaLS, e.g., \citealt{Dey2019})\footnote{
        \url{https://www.legacysurvey.org/}
    }, we will measure large aperture and outer envelope \mstar{} of $z<0.5$ massive galaxies 
    out to $\sim 100$ kpc (e.g., Li et al. in prep.).
    When combined with their DESI spec-$z$ in the next few years, this much larger DECaLS ($\sim
    9000$ deg$^2$) survey will provide us an ideal sample to constrain galaxy-halo connection
    models.
    We will extend our \topn{} tests to include group/cluster catalogues for DECaLS (e.g.,
    \citealt{Yang2020, Zou2021}), and apply our ``Cen$+N$'' method to define a sample of massive
    halos that are suitable for unbiased cosmological analysis.
    
\section*{Data Availability Statements}

    The data underlying this article are available in Zenodo at 
    \url{https://doi.org/10.5281/zenodo.5259075}.
    The \texttt{Python} code, \texttt{Jupyter} notebooks, and the data files for reproducing the 
    results and figures of this work can be found on \texttt{Github} at
    \url{https://github.com/dr-guangtou/jianbing}.
    The Hyper Suprime-Cam Subaru Strategi Program data used in this work are included in the 
    Public Data Release 2 at \url{https://hsc-release.mtk.nao.ac.jp/doc/}.

\section*{Acknowledgements}

  The authors would like to thank Benedikt Diemer and Matthew Becker for useful discussions and 
  suggestions.

  This material is based upon work supported by the National Science Foundation under
  Grant No. 1714610.

  The authors acknowledge support from the Kavli Institute for Theoretical Physics.
  This research was also supported in part by National Science Foundation under Grant
  No. NSF PHY11-25915 and Grant No. NSF PHY17-48958

  We acknowledge use of the lux supercomputer at UC Santa Cruz, funded by NSF MRI grant AST
  1828315. AL is supported by the U.D Department of Energy, Office of Science, Office of High
  Energy Physics under Award Number DE-SC0019301. AL acknowledges support from the David and
  Lucille Packard foundation, and from the Alfred .P Sloan foundation.
  
  JD is supported by the Chamberlain Fellowship at Lawrence Berkeley National Laboratory.
  
  The Hyper Suprime-Cam (HSC) collaboration includes the astronomical communities of
  Japan and Taiwan, and Princeton University.  The HSC instrumentation and software were
  developed by National Astronomical Observatory of Japan (NAOJ), Kavli Institute
  for the Physics and Mathematics of the Universe (Kavli IPMU), University of Tokyo,
  High Energy Accelerator Research Organization (KEK), Academia Sinica Institute
  for Astronomy and Astrophysics in Taiwan (ASIAA), and Princeton University.
  Funding was contributed by the FIRST program from Japanese Cabinet Office,  Ministry
  of Education, Culture, Sports, Science and Technology (MEXT), Japan Society for
  the Promotion of Science (JSPS), Japan Science and Technology Agency (JST), Toray
  Science Foundation, NAOJ, Kavli IPMU, KEK, ASIAA, and Princeton University.

  Funding for SDSS-III has been provided by Alfred P. Sloan Foundation, the
  Participating Institutions, National Science Foundation, and U.S. Department of
  Energy. The SDSS-III website is http://www.sdss3.org.  SDSS-III is managed by the
  Astrophysical Research Consortium for the Participating Institutions of the SDSS-III
  Collaboration, including University of Arizona, the Brazilian Participation Group,
  Brookhaven National Laboratory, University of Cambridge, University of Florida, the
  French Participation Group, the German Participation Group, Instituto de Astrofisica
  de Canarias, the Michigan State/Notre Dame/JINA Participation Group, Johns Hopkins
  University, Lawrence Berkeley National Laboratory, Max Planck Institute for
  Astrophysics, New Mexico State University, New York University, Ohio State University,
  Pennsylvania State University, University of Portsmouth, Princeton University, the
  Spanish Participation Group, University of Tokyo, University of Utah, Vanderbilt
  University, University of Virginia, University of Washington, and Yale University.

  The Pan-STARRS1 surveys (PS1) have been made possible through contributions of
  Institute for Astronomy; University of Hawaii; the Pan-STARRS Project Office;
  the Max-Planck Society and its participating institutes: the Max Planck Institute
  for Astronomy, Heidelberg, and the Max Planck Institute for Extraterrestrial Physics,
  Garching; Johns Hopkins University; Durham University; University of Edinburgh;
  Queen's University Belfast; Harvard-Smithsonian Center for Astrophysics; Las
  Cumbres Observatory Global Telescope Network Incorporated; National Central
  University of Taiwan; Space Telescope Science Institute; National Aeronautics
  and Space Administration under Grant No. NNX08AR22G issued through the Planetary
  Science Division of the NASA Science Mission Directorate; National Science
  Foundation under Grant No. AST-1238877; University of Maryland, and Eotvos
  Lorand University.

  This research makes use of software developed for the Large Synoptic Survey
  Telescope. We thank the LSST project for making their code available as free
  software at http://dm.lsstcorp.org.

  The CosmoSim database used in this research is a service by the Leibniz-Institute for
  Astrophysics Potsdam (AIP).
  The MultiDark database was developed in cooperation with the Spanish MultiDark
  Consolider Project CSD2009-00064.

  This research made use of:
  \href{http://www.stsci.edu/institute/software_hardware/pyraf/stsci\_python}{\texttt{STSCI\_PYTHON}},
      a general astronomical data analysis infrastructure in Python.
      \texttt{STSCI\_PYTHON} is a product of the Space Telescope Science Institute,
      which is operated by Association of Universities for Research
      in Astronomy (AURA) for NASA;
  \href{http://www.scipy.org/}{\texttt{SciPy}},
      an open source scientific tool for Python (\citealt{SciPy});
  \href{http://www.numpy.org/}{\texttt{NumPy}},
      a fundamental package for scientific computing with Python (\citealt{NumPy});
  \href{http://matplotlib.org/}{\texttt{Matplotlib}},
      a 2-D plotting library for Python (\citealt{Matplotlib});
  \href{http://www.astropy.org/}{\texttt{Astropy}}, a community-developed
      core Python package for astronomy (\citealt{AstroPy});
  \href{http://scikit-learn.org/stable/index.html}{\texttt{scikit-learn}},
      a machine-learning library in Python (\citealt{scikit-learn});
  \href{https://ipython.org}{\texttt{IPython}},
      an interactive computing system for Python (\citealt{IPython});
  \href{https://github.com/kbarbary/sep}{\texttt{sep}}
      Source Extraction and Photometry in Python (\citealt{PythonSEP});
  \href{https://jiffyclub.github.io/palettable/}{\texttt{palettable}},
      colour palettes for Python;
  \href{http://dan.iel.fm/emcee/current/}{\texttt{emcee}},
      Seriously Kick-Ass MCMC in Python;
  \href{http://bdiemer.bitbucket.org/}{\texttt{Colossus}},
      COsmology, haLO and large-Scale StrUcture toolS (\citealt{Colossus}).

\bibliographystyle{mnras}
\bibliography{topn}

\appendix

\section{Derivation of the galaxy--galaxy lensing profiles}
    \label{app:dsigma_detail}

    Here we walk through the derivation of the final \dsigma{} profile used in the \topn{} test.
    As mentioned in \S\ \ref{sec:dsigma}, we adopt a slightly modified version of the methodology
    from \citet{Singh2017} to measure the excess surface mass density (ESD or \dsigma{}) profiles
    around massive galaxies or clusters.
    This method emphasises the importance of subtracting lensing signals around large number of
    random positions from the signals for real lenses to achieve unbiased measurement.
    The \dsigma{} signal at a physical radius $R$ is:

    \begin{equation}
        {\Delta\Sigma}_{\rm LR}(R) =
        f_{\rm bias}({\Delta\Sigma}_{\mathrm{L}}(R) - {\Delta\Sigma}_{\mathrm{R}}(R))
        \label{eq:ds1}
    \end{equation}

    Here, $L$ indicates measurements for the lens galaxies while $R$ is for random positions.
    For each \dsigma{} profile, we use a set of $1.5 \times 10^5$ random points whose redshift
    distribution is matched to the lenses.
    The number of random points is at least 100 time larger than the largest \topn{} sample.
    The \dsigma{} profile around lenses is:

    \begin{equation}
        {\Delta\Sigma}_{\rm L}(R) = \frac{1}{2 \mathcal{R}(R) [1+\mathcal{K}(R)]}
            \frac{\Sigma_{\rm Ls} w_{\rm Ls} \gamma_{t}^{(\rm Ls)}
            \Sigma_{\rm crit}^{(\rm Ls)}}{\Sigma_{\rm Ls} w_{\rm Ls}}
        \label{eq:ds2}
    \end{equation}

    \noindent where $\gamma_{t}$ is the tangential shear component, $\Sigma_{\rm crit}$ is the
    critical surface density, $w_{\rm Ls}$ is the weight used for each lens-source pair.
    Following the calibration strategy outlined in \citet{HSC-WLCALIB}, we also include
    the shear responsivity factor $\mathcal{R}(R)$ and the correction for the multiplicative
    shear bias $[1+\mathcal{K}(R)]$.
    Here $\Sigma{\rm Ls}$ represents the summation over all lens-source pairs.
    We perform the same measurements for random points, so replacing L with R in
    Equation \ref{eq:ds2} will form the estimator for randoms.

    The critical surface density is:

    \begin{equation}
        \Sigma_{\rm crit}=\frac{c^2}{4\pi G} \frac{D_A(z_s)}{D_A(z_l) D_A(z_l, z_s) (1+z_l)^2}
        \label{eq:sigcrit}
    \end{equation}

    \noindent where $D_A(z_L)$, $D_A(z_s)$, and $D_A{z_L, z_s}$ represent the angular diameter
    distances to the lens, source, and the distance between the lens-source pair.

    The weight applied to each lens-source pair is described by:

    \begin{equation}
        w_{\rm Ls} = \frac{\Sigma_{\rm crit}^{-2}}{\sigma^2_{e, {\rm Ls}} + \sigma^2_{\rm rms}}
            \equiv \frac{\Sigma_{\rm crit}^{-2}}{\sigma^2_{{\rm Ls}}}
        \label{eq:weight}
    \end{equation}

    \noindent where $\sigma_{\rm rms}$ represents the intrinsic shape dispersion while
    $\sigma_{e, \rm Ls}$ is the per-component shape measurement error.

    Meanwhile, the shear responsivity factor is defined by:

    \begin{equation}
        \mathcal{R}(R) = 1 - \frac{\Sigma_{\rm Ls} w_{\rm Ls} \sigma^2_{e, {\rm Ls}}}{\Sigma_{\rm Ls} w_{\rm Ls}}
        \label{eq:rfactor}
    \end{equation}

    \noindent And the multiplicative shear bias correction is defined as:

    \begin{equation}
        \mathcal{K}(R) = \frac{\Sigma_{\rm Ls} w_{\rm Ls} m_{\rm s}}{\Sigma_{\rm Ls} w_{\rm Ls}}
        \label{eq:kfactor}
    \end{equation}

    \noindent where $m_{\rm s}$ is multiplicative shear bias value for each source.
    The shape catalogue provides estimates of $\sigma_{\rm rms}$, $\sigma_{e, \rm Ls}$, and $m_{\rm
    s}$, while \citet{HSC-WLCALIB} provides in-depth discussion of these calibration related
    issues.

    In difference with \citet{Singh2017}, we do not use boost factor to account for the photo-$z$
    dilution effect.
    Following the strategy in \citet{Leauthaud2017}, we develop a correction
    factor, $f_{\rm bias}$, to account for it.
    We define $f_{\rm bias}$ as the ratio between the \dsigma{} profile calculated using
    the real redshift and the one using photo-$z$ from the COSMOS photo-$z$ calibration
    catalogue\footnote{
    \url{https://hsc-release.mtk.nao.ac.jp/doc/index.php/s17a-wide-cosmos/}}.

    In practice, it is estimated based on:

    \begin{equation}
        f_{\rm bias} = \frac{
            \sum_{\rm Ls} w_{\rm Ls} w_{\rm sys} \left(\Sigma_{{\rm crit, T}} / \Sigma_{{\rm crit, P}} \right)}
            {\sum_{\rm Ls} w_{\rm Ls} w_{\rm sys}}
        \label{eq:fbias}
    \end{equation}

    based on the photo-$z$ calibration sample in the COSMOS field for
    such purpose
    (e.g., \citealt{Mandelbaum2008, Nakajima2012, Leauthaud2017}).

    As for the $f_{\rm bias}$,

    \noindent where $\Sigma_{{\rm crit, T}}$ is the critical surface density estimated using the
    ``true'' redshift in the calibration catalogue (can be spec-$z$ or COSMOS 30-band photo-$z$),
    while $\Sigma_{{\rm crit, P}}$ is the one using \texttt{frankenz} photo-$z$.
    $w_{\rm sys}$ is the systematic photo-$z$ weight in the calibration catalogue that matches the
    colour-magnitude distribution of the COSMOS photo-$z$ calibration catalogue to the same
    distribution of the source catalogue (e.g., \citealt{Mandelbaum2008, Nakajima2012}).
    Note that the estimator shown here is different from the ones in \citet{Leauthaud2017} and
    \citet{Speagle2019}, and it accounts for the photo-$z$ dilution effect more accurately.
    The $f_{\rm bias}$ level for galaxies in our sample is general very low ($\sim 1$-2\%).

    To estimate the covariance matrix of a \dsigma{} profile, we use both jackknife and
    bootstrap resampling method.
    For the jackknife case, we assign lens and randoms into the same 45 jackknife regions with
    similar area around 2.5 \sqdeg{} and regular shapes.
    The covariance matrix from the jackknife resampling is:

    \begin{equation}
        \mathrm{Var}_{\rm Jk}(\widehat{\Delta\Sigma}) = \frac{N_{\rm Jk} - 1}{N_{\rm Jk}} \sum\limits_{i=1}^{N_{\rm Jk}} (\Delta\Sigma_{i} - \overline{\Delta\Sigma})^2
    \end{equation}

    \noindent here $N_{\rm Jk}=45$, $\Delta\Sigma_{i}$ represents the \dsigma{} profile from each
    Jackknife region, and $\overline{\Delta\Sigma}$ is the mean profile of all regions.

    For the \topn{} test, the small sample size in Bin 1 \& 2 sometimes make it difficult to assign
    jackknife regions. Therefore we also calculate the covariance matrix using bootstrap resampling
    with $N_{\rm Bt}=5000$ iterations:

    \begin{equation}
        \mathrm{Var}_{\rm Bt}(\widehat{\Delta\Sigma}) = \frac{1}{N_{\rm Bt} - 1} \sum\limits_{i=1}^{N_{\rm Bt}} (\Delta\Sigma_{i, \rm Bt} - \overline{\Delta\Sigma})^2
    \end{equation}

    \noindent The two methods provide consistent measurements of covariance matrix.

  \begin{figure*}
      \centering
      \includegraphics[width=\textwidth]{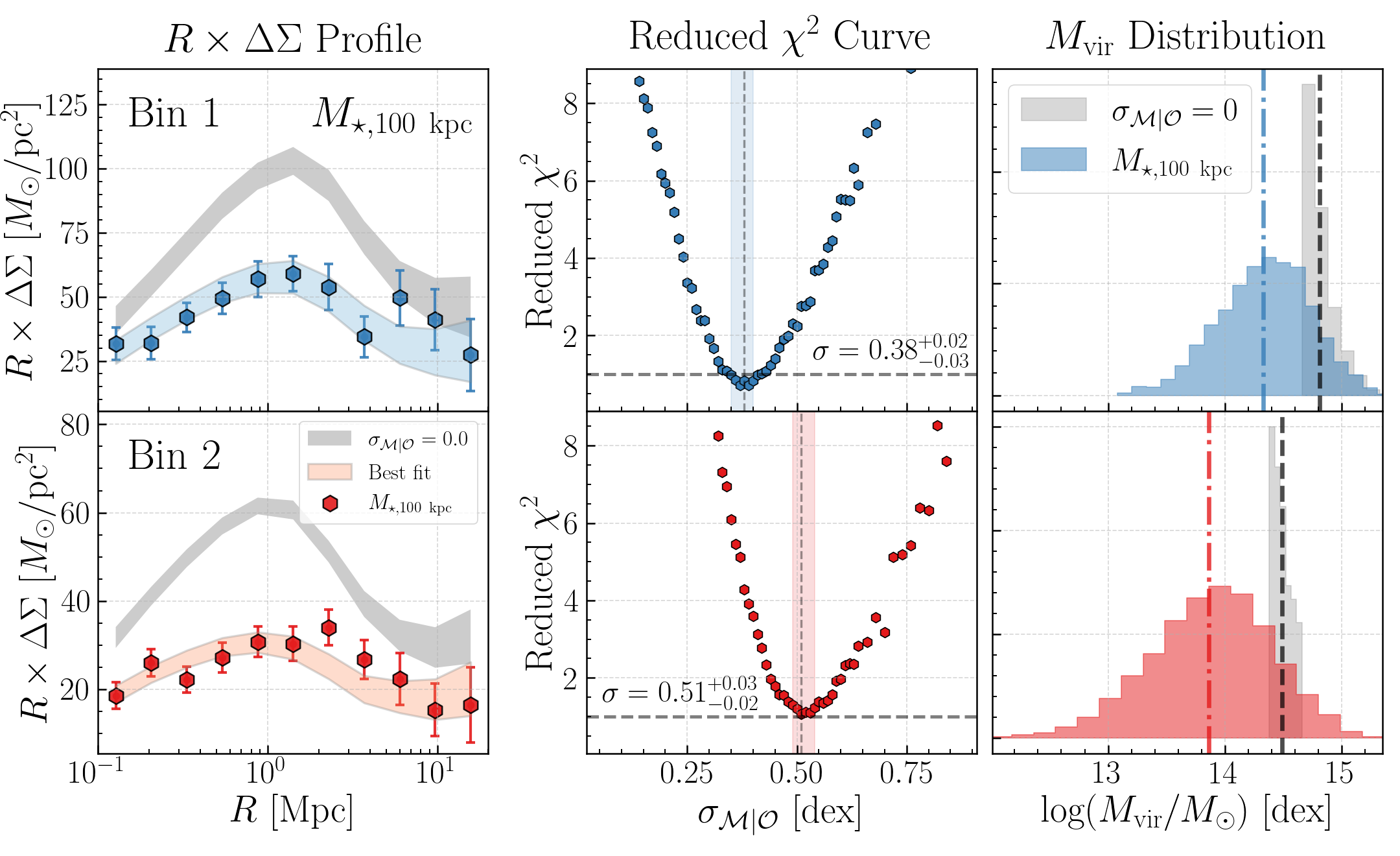}
      \caption{
          Illustration of the scatter-matching process in the four \topn{} bins using the
          \maper{150} as an example.
          \textbf{Left} column shows the observed \rdsigma{} profiles (solid hexagon), the best-fit
          profiles using \mdpl2{} simulation (shaded regions with similar colours), and the
          lensing profiles of the ``perfect'' \topn{} sample.
          The uncertainties of the model lensing profiles are inflated to match the expected
          statistical uncertainties of the volume occupied by the HSC data.
          \textbf{Middle} column shows the reduced $\chi^{2}$ curve of the fitting process.
          A horizontal dashed-line highlights where $\chi^{2}=1$.
          The vertical dashed-line and the shaded region show the best-fit scatter value and the
          associated uncertainty. We also display these values on the figure.
          \textbf{Right} column shows the best-fit \mvir{} distributions in each bin predicted
          by the \mdpl2{} simulation (coloured histograms; dot-dashed lines label the average
          \mvir{} values).
          We also compare them with the ``true'' \mvir{} distributions in each number density bin
          (grey histograms; grey dashed-lines label the average \mvir{} values).
          The \texttt{Jupyter} notebook for reproducing this figure can be found here:
          \href{https://github.com/dr-guangtou/jianbing/blob/master/notebooks/figure/figA1.ipynb}{\faGithub}.
      }
      \label{fig:fitting}
  \end{figure*}

\section{Matching the lensing profiles}
    \label{app:fitting}

    As described in \S\ \ref{sec:estimate_scatter}, we estimate the \sigmh{} value
    from an observed \dsigma{} profile from the \topn{} test through matching it to a densely
    sampled grid of model \dsigma{} profiles that cover a wide range of \sigmh{}
    values.
    For each pair of observed and predicted \dsigma{} profiles, we define a simple $\chi^2$
    statistic (Equation \ref{eq:chi2}) to describe the ``similarity'' between them.

    In Figure \ref{fig:fitting}, we use the \topn{} result for \maper{150} stellar mass as example
    to visualise the ``scatter matching'' procedure, which produce a well-behaved reduced $\chi^2$
    curves with a clear minimum.
    For \maper{150}, the reduced $\chi^2$ values in all four \topn{} bins are reasonably close
    to 1.0 ([0.65, 0.88, 1.31, 0.90]).
    In line with this impression, the left panels show that the best-fit ``scatter only'' \dsigma{}
    profile is fully consistent with the observed one.
    As discussed in \S\ \ref{sec:mstar_vs_richness}, this is not always the case (see
    Figure \ref{fig:richness_residual}).
    However, even when the best-fit model is not satisfying (e.g., reduced $\chi^2 >2$), we still
    estimate the ``best-fit'' \sigmh{} value.

    Since we only calculate the $\chi^2$ on a grid of \sigmh{} values and the statistical
    uncertainties of the predicted \dsigma{} profiles cannot be completely ignored, we did not
    just report the \sigmh{} value with the lowest $\chi^2$.
    Instead, we interpolate the normalised cumulative distribution of the likelihood $\equiv
    \exp{(-0.5 \times \chi^2)}$ to derive the \sigmh{} at 50th percentile as the ``best-fit''
    scatter value.
    We estimate the 1-$\sigma$ uncertainty range in the same way.

    We should note that the choice of covariance matrix (Jackknife v.s. bootstrap) does not
    affect any results of this work.
    We also attempted to include the uncertainties of the predicted \dsigma{} profile to the
    covariance matrix as additional diagonal term, and verify it has no impact on any
    conclusions.

    In the figure, we inflate the error bars of the
    model profiles to reflect the volume difference between the HSC data and the simulation
    used. For \mdpl2{} simulation, the volume is about $\sim 25 \times$ larger than the HSC
    volume. We therefore increase the error bar by a factor of 5. However, we did not
    include the model uncertainty during the fitting process. 

\section{Scaling relation model calibrated to match HSC observations}
    \label{app:hsc_model}

\begin{figure*}
  \centering
  \includegraphics[width=0.9\textwidth]{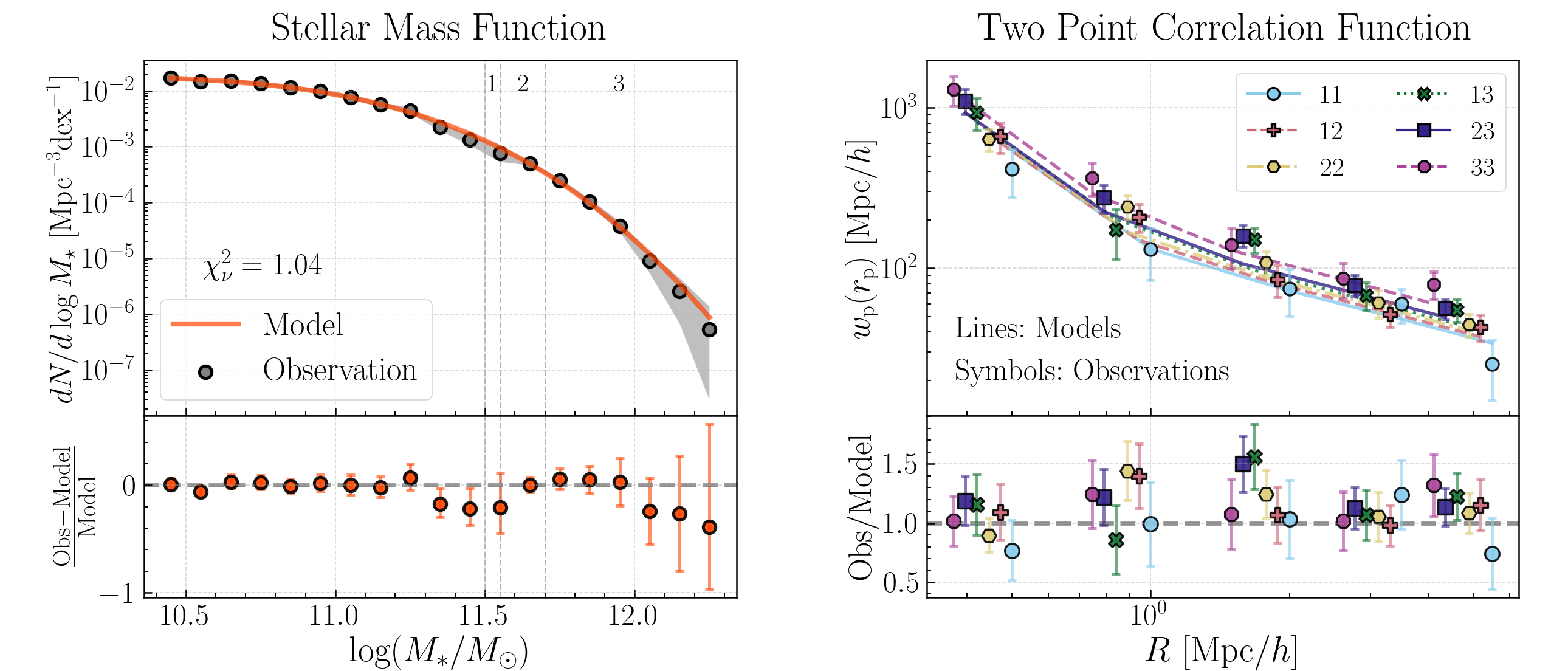}
  \caption{
      \textbf{Left} panels demonstrate how well the model (red line) can fit the observed SMF
      (grey symbols; shaded regions are uncertainties) that combines data from HSC at high-\mstar{}
      end and \texttt{PRIMUS} survey at low-\mstar{} end.
      The bottom sub-panel shows the relative residual of the best-fit SMF.
      The three vertical dashed-lines highlight the \mstar{} boundaries of the three \mstar{}-bins
      ([$11.50$, $11.55$, $11.70$]) used for measuring the two--point correlation functions of massive
      HSC galaxies.
      \textbf{Right} panels summarise the observed clustering of HSC massive galaxies (symbols)
      and their best-fit models (lines).
      [11, 22, 33] are the auto-correlation functions of the three \mstar{} bins, while
      [12, 13, 23] indicate the cross-correlation functions among the three bins.
      The bottom sub-panel shows the ratio between the observed and the model clustering signals.
        The \texttt{Jupyter} notebook for reproducing this figure can be found here:
        \href{https://github.com/dr-guangtou/jianbing/blob/master/notebooks/figure/figB1.ipynb}{\faGithub}.
    }
  \label{fig:best_mock}
\end{figure*}

    In \S\ \ref{sec:estimate_scatter}, we describe the method to predict the stacked \dsigma{}
    profile of a sample of number density selected halos with certain \scatterMhaloObsSym{} value
    based on a $\log$-normal scaling relation with fixed slope ($\alpha=1$).
    This simple model helps us predict the stacked \dsigma{} profile of a specific \topn{} bin
    (see Figure \ref{fig:mdpl2}).
    Meanwhile, to evaluate the impact of satellite galaxies on the \topn{} tests, we still need
    a mock catalogue from simulation that can fit basic HSC observations of massive galaxies and
    have realistic satellite fraction at high-\mstar{} end.

    Taking advantage of the work by (DeMartino \etal{} in prep.), we create such a mock catalogue 
    that can reproduce the SMF and clustering statistics of HSC massive galaxies using a sub-halo 
    abundance matching model (SHAM) based on peak halo mass ($M_{\rm Peak}$).
    We also use this model to constrain the SHMR and its scatter at high-\mhalo{} end.
    In particular, we model the SHMR using the functional form from \citet{Behroozi2013} but
    fixing the slope at low-\mhalo{} end ($\beta$).
    In total, the model has five free parameters:
    1. The four parameters that govern the mean SHMR at high-\mhalo{} end from 
    \citet{Behroozi2013};
    2. And the scatter of \mstar{} at fixed \mhalo{}.

    As shown in Figure \ref{fig:best_mock}, the best-fit model can reproduce the observed
    mass function and clustering statistics of massive galaxies reasonably well.
    To ensure the model can fit the SMF beyond just the high-\mstar{} end, we adopt a ``hybrid''
    SMF: we use the complete sample of HSC massive galaxies at $0.2 < z < 0.5$ to cover the
    \logms{}$>11.5$ range, and use the \texttt{PRIMUS} $0.3 < z < 0.4$ SMF
    (\citealt{Moustakas2013}) for the $10.5 <$\logms{}$<11.5$ range.
    Both the HSC and the \texttt{PRIMUS} \mstar{} are from the \texttt{iSEDfit} code under very
    similar assumptions of stellar population properties.
    The \mstar{} of the HSC sample is based on our customised 1-D profile that capture the
    luminosity beyond 100 kpc, while the \texttt{PRIMUS} sample is based on small aperture
    photometry.
    Using the \texttt{PRIMUS} galaxies that also have the HSC 1-D \mstar{} measurements from
    \citet{Huang2018b}, we derive a simple constant offset term that help us ``stitch'' the two
    SMFs together.
    We note that this just ensures a smooth SMF for the fitting, and does not affect any results
    in this work.
    As for the clustering signals of HSC massive galaxies, we compute the auto- and
    cross-correlation signals after separating the sample into three \mstar{} bins: $11.50
    <$\logms{}$\leq 11.55$, $11.55 <$\logms{}$\leq 11.70$, and \logms{}$> 11.70$.
    The best-fit SHMR is broadly consistent with previous works including the scatter of
    \mstar{} value ($\sim 0.2$ dex).
    More importantly, the satellite fraction at high-\mstar{} end is between 5 and 10\%, which is also
    similar to the results of previous works.
    We also verify that the satellite fraction value is robust to small changes in abundance
    matching methodology.

\section{\texorpdfstring{$\Delta\Sigma$}{DSigma} profiles of massive satellite galaxies}
	\label{app:sat_cen}

\begin{figure}
    \centering
    \includegraphics[width=0.47\textwidth]{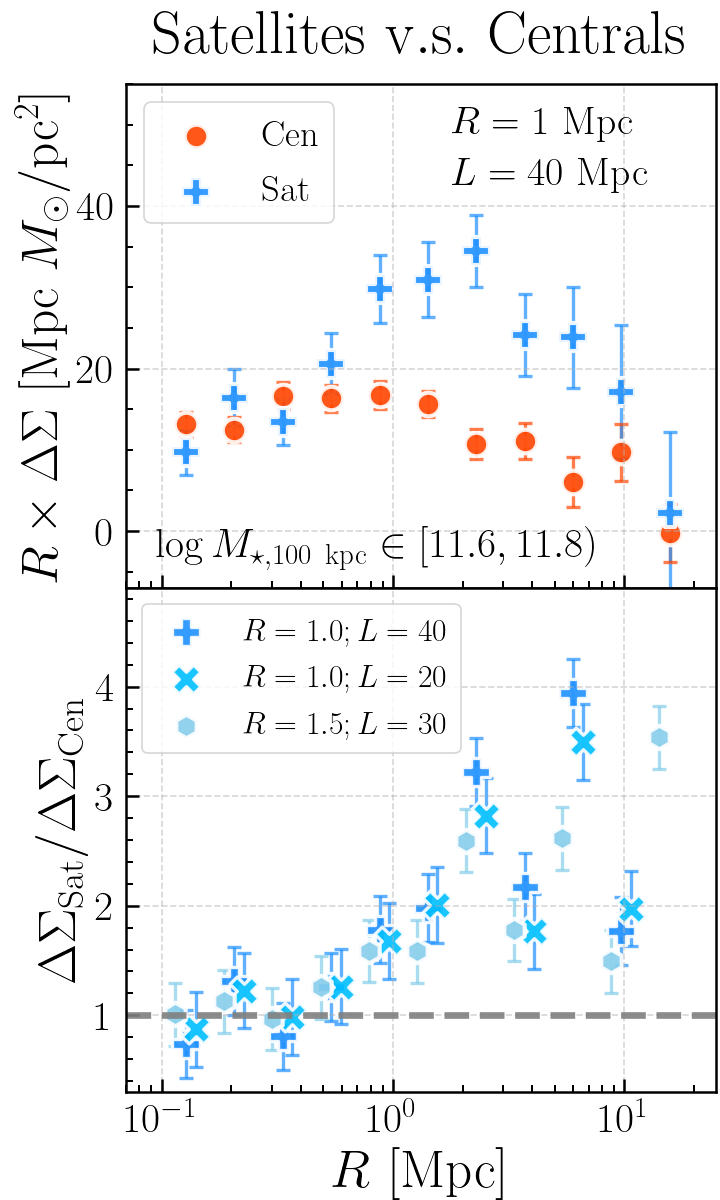}
    \caption{
        The comparison of \dsigma{} profiles of centrals and satellites at similar \mstar{}.
        \textbf{Top} panel: Blue crosses show the \dsigma{} profile of satellites within 
        $11.6 < \log (M_{\star, 100\ \rm kpc}/M_{\odot} < 11.8$ selected using a cylinder with $R=1$
        Mpc and $L=40$ Mpc. 
        The red circles show the \dsigma{} profile for a sample of central galaxies with matched 
        \mstar{} distributions.
        \textbf{Bottom} panel shows the ratios of the stacked \dsigma{} profiles for satellites 
        and centrals selected using different definitions of cylinders.
        The \texttt{Jupyter} notebook for reproducing this figure can be found here:
        \href{https://github.com/dr-guangtou/jianbing/blob/master/notebooks/figure/figD1.ipynb}{\faGithub}.
    }
    \label{fig:sat_cen}
\end{figure}

    In \S\ \ref{sec:satellite}, we introduced our method for identifying candidates of massive
    satellite galaxies from our sample and investigated their impacts on the stacked \dsigma{}
    profile.
    In Figure \ref{fig:sat_cen}, we compare the stacked \dsigma{} profiles of massive satellite
    galaxies within $11.6 < \log (M_{\star, 100\ \rm kpc}/M_{\odot} < 11.8$ to that of the central 
    galaxies in the same \mstar{} bin.
    As explained in \S\ \ref{sec:satellite}, for massive galaxies in our sample, we iteratively
    identify satellite galaxies with lower \maper{10} in a cylinder with radius $R=1$ Mpc and LOS
    length of $L=40$ Mpc.
    We ignore the redshift and \mstar{} uncertainties during this procedure, so strictly speaking 
    these are just candidates of satellite galaxies.
    Within the \maper{100} bin, we find 161 massive satellite galaxies and 1804 central galaxies.
    The \maper{100} distribution of satellite clearly skews toward lower values than the one
    for centrals.
    To make it a fair comparison, we match the centrals to satellites in the 2-D 
    \maper{50}--\menve{50}{100} plane: using a k--d tree, we search for the nearest 7 centrals 
    around each satellite and keep the unique centrals.
    This yield 765 central galaxies with similar \maper{100} and \maper{50}--\menve{50}{100} 
    distributions to the satellites. 
    They also share very similar redshift distributions.
    In the top panel of Figure \ref{fig:sat_cen}, we compare their \dsigma{} profiles.
    While the centrals and satellites share similar profiles within inner 500 kpc, the satellites 
    display clearly enhanced \dsigma{} signals at $R>1$ Mpc.
    We highlight this result in the bottom panel of Figure \ref{fig:sat_cen} using the ratio 
    of the satellite \dsigma{} profile to that of the centrals. 
    We also show the ratios for satellites selected using different cylinders. 
    This comparison shows that small variation of the radius (from 1.0 to 1.5 Mpc) and length
    (20 to 40 Mpc) of the cylinders will not affect the results.

    Figure \ref{fig:sat_cen} shows that, at the same \mstar{}, massive satellite galaxies show 
    very different \dsigma{} profiles with centrals due to the strong impact from their host 
    dark matter halos.
    Despite the small impact on the stacked \dsigma{} profiles due to the low satellite fraction 
    value, the \dsigma{} profile of massive satellite galaxies alone contains valuable information 
    about the galaxy--halo connection of massive galaxies (\eg{} \citealt{Sifon2015, Li2016,
    Dvornik2020}).
    We will aim to understand it more so that we can deal with satellites better when using 
    \mstar{}-based \mvir{} proxies.

\section{Galaxy size as \texorpdfstring{\mvir{}}{Mvir} indicator}
	\label{app:size}

    In this work, we have explored different aperture and outskirt stellar mass defined using 
    fixed physical radius (\eg{} 100 kpc, 50 to 100 kpc).
    They provide unambiguous definitions of apertures, which is important when comparing results
    from different imaging data or between simulation and observation.
    But, for galaxies with very different size, \mstar{} defined using fixed radius could have 
    very different physical meanings. 
    For example, while \menve{50}{100} is a good measurement of outer envelope \mstar{} for very 
    massive elliptical galaxies, it is not even practical to apply it to Milky Way-mass galaxies.
    
\begin{figure}
    \centering
    \includegraphics[width=\columnwidth]{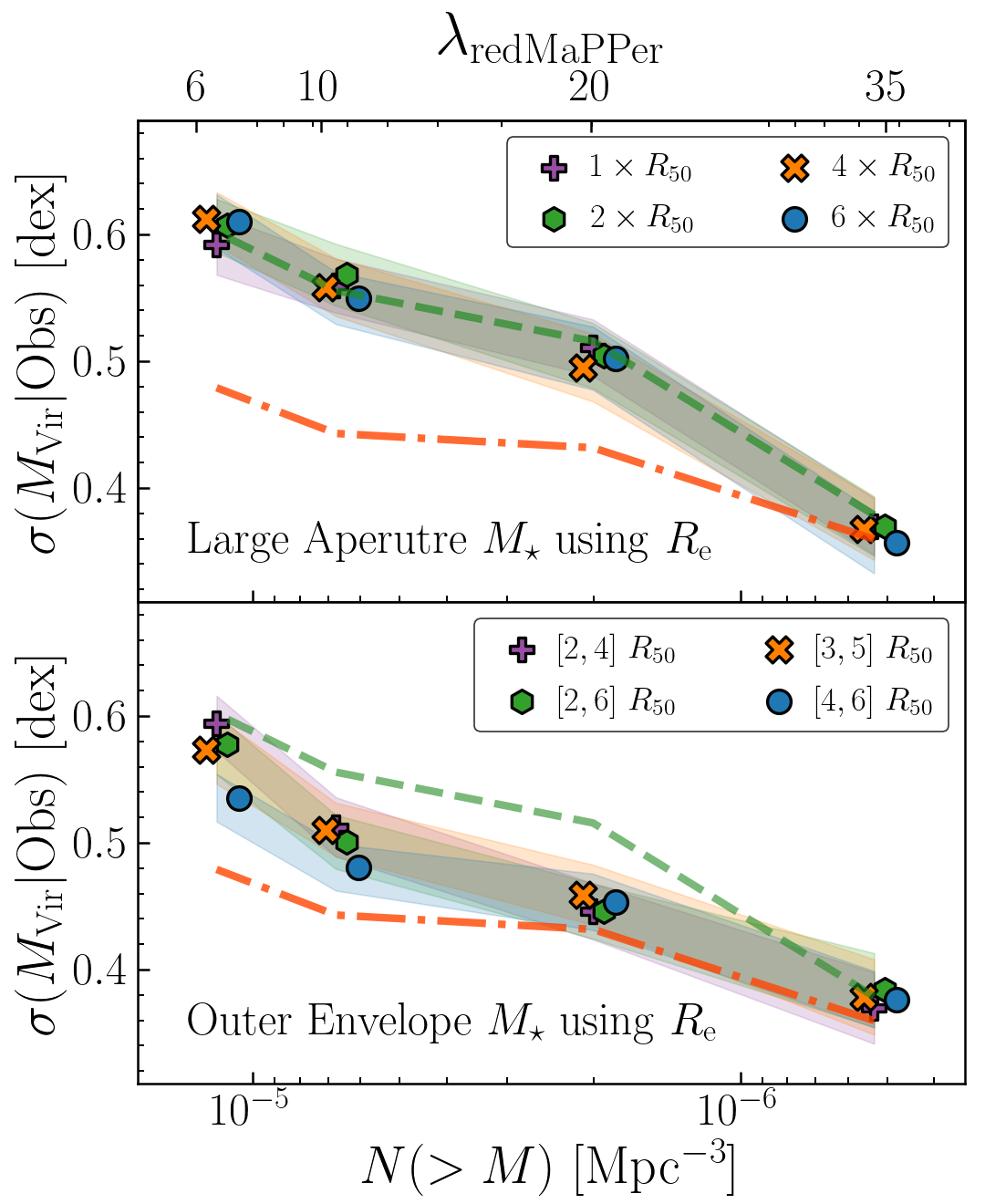}
    \caption{
        The relations between the cumulative number density of each \topn{} bin and \sigmvir{} 
        for \mstar{}-based \mvir{} proxies defined using $R_{50}$.
        The format is the same with Figure \ref{fig:scatter_trend_2} and Figure 
        \ref{fig:scatter_trend}.
        \textbf{Top} panel shows the \topn{} results for different aperture \mstar{} defined using 
        $R_{50}$ while the \textbf{bottom} panel is for different outskirt \mstar{} defined using 
        $R_{50}$.
        We use the \sigmvir{} trends for \maper{100} (green dashed line) and \menve{50}{100} 
        (red dot--dashed line) as the references.
        The \texttt{Jupyter} notebook for reproducing this figure can be found here:
        \href{https://github.com/dr-guangtou/jianbing/blob/master/notebooks/figure/figE1.ipynb}{\faGithub}.
    }
    \label{fig:scatter_trend_size}
\end{figure}

    The half-light (-mass) radius ($R_{50}$), or the effective radius ($R_{\rm e}$), is a commonly 
    adopted galaxy size measurement. 
    It naturally provides another way to define aperture and outskirt for galaxies. 
    In Figure \ref{fig:scatter_trend_size}, we summarise the \topn{} results for a few different 
    aperture (top panel) and outskirt (bottom panel) \mstar{}. 
    In this work, the $R_{50}$ is measured using the $i$-band integrated 1-D intensity profiles 
    (also known as the curve-of-growth) along the major axis (so it is not ``circulized''). 
    It is defined as the radius that contains 50\% of light within 100 kpc radius.
    We choose this definition because the surface brightness profile at $R>100$ kpc becomes 
    less reliable, but replacing 100 kpc with larger radius such as 150 kpc will not change our
    results.
    
    Aperture \mstar{} defined using $R_{50}$ show similar performance with \maper{100}. 
    This is expected for $M_{\star, R_{50}}$ as it represents 50\% of \maper{100} by definition. 
    Meanwhile, none of the other larger aperture masses using $R_{50}$ show any improvement.
    Outskirt masses using $R_{50}$ do have lower \sigmvir{} values than \maper{100}.
    While this confirms the result using fixed radius, none of the outskirt \mstar{} using $R_{50}$ 
    has performance as good as \menve{50}{100} especially in Bin 3 \& 4.
    
    We note that the stellar masses defined by $R_{50}$ directly ties to the measurement of galaxy 
    size, which is not an easy task.
    Replacing the $R_{50}$ from 1-D curve-of-growth with the $R_{\rm e}$ from single \ser{} fitting 
    could lead to different results.
    We will explore more \mvir{} proxies related to galaxy size in future works.
    
\section{The comparison of \texorpdfstring{\dsigma{}}{DSigma} profile between the HSC and SDSS \texorpdfstring{\redm{}}{redMaPPer} clusters}
	\label{app:sdss_redm}

\begin{figure}
    \centering
    \includegraphics[width=\columnwidth]{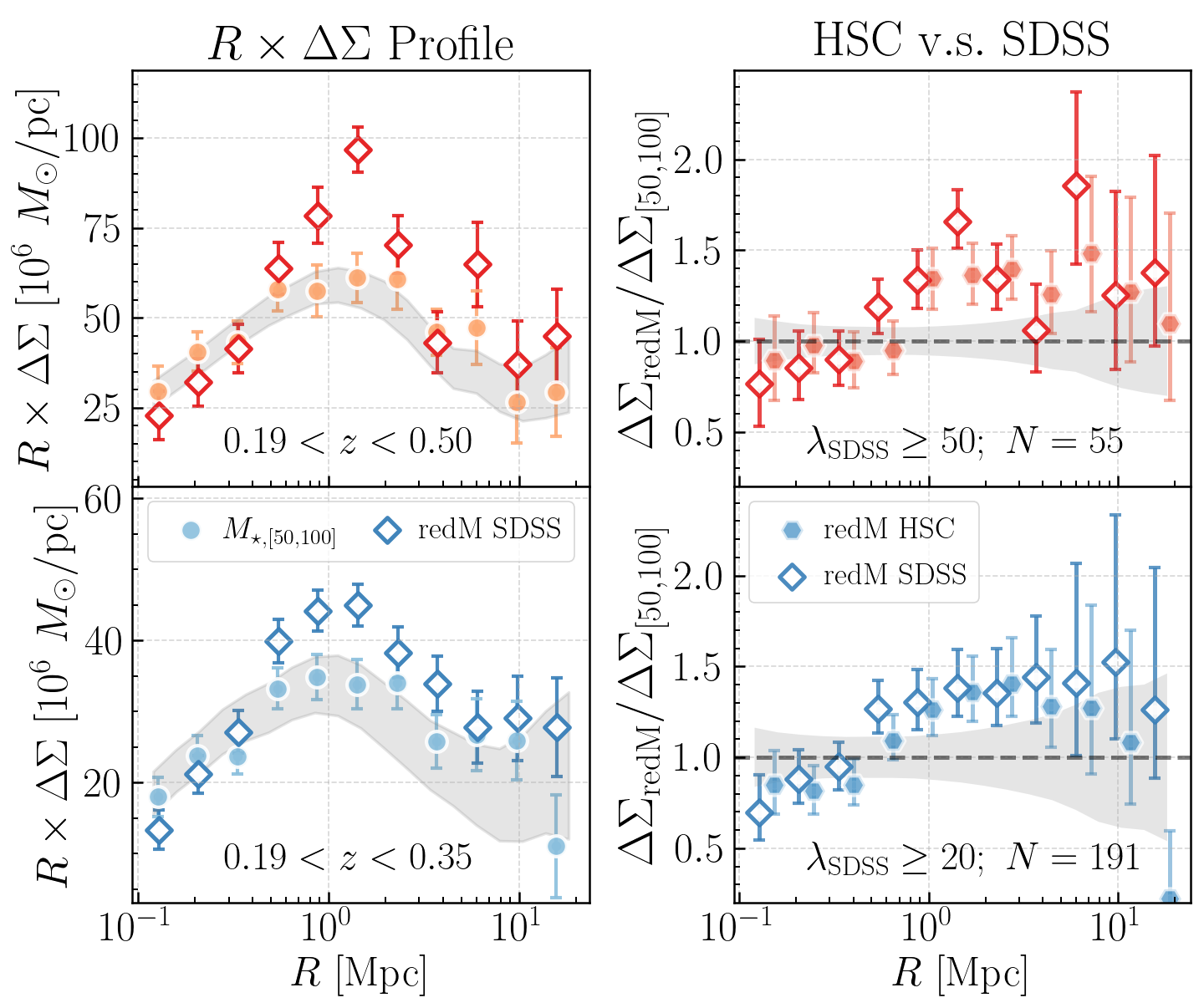}
    \caption{
        Similar to Figure \ref{fig:mout_richness}, here we compare the \dsigma{} profiles of
        SDSS \redm{} clusters (empty diamonds) to those of HSC \redm{} clusters (solid hexagons)
        and \menve{50}{100}-selected HSC massive galaxies (solid circles). 
        Given the completeness of SDSS \redm{} clusters, we show the comparison in two richness
        and redshift bins: \textbf{top} panel is for $\lambda_{\rm SDSS} \geq 50$ clusters at
        $0.2 < z < 0.5$ while the \textbf{bottom} panel is for $\lambda_{\rm SDSS} \geq 20$
        clusters at $0.2 < z < 0.35$.
        \textbf{Left} panels show the comparisons of lensing profiles. 
        The shaded regions display the best-fit model profile of the \menve{50}{100} samples.
        \textbf{Right} panels show the ratio of \dsigma{} profiles between richness- and 
        \menve{50}{100}-selected samples.
        The SDSS \redm{} clusters' \dsigma{} profiles show similar systematic differences 
        with the \menve{50}{100} profile just like their HSC counterparts.
        The \texttt{Jupyter} notebook for reproducing this figure can be found here:
        \href{https://github.com/dr-guangtou/jianbing/blob/master/notebooks/figure/figF1.ipynb}{\faGithub}.
    }
    \label{fig:sdss_redm}
\end{figure}

    In Figure \ref{fig:sdss_redm}, we compare the \dsigma{} profiles of \redm{} clusters from SDSS
    survey to those of HSC data, and also to the HSC massive galaxies selected using \menve{50}{100}.
    We are using the \href{http://risa.stanford.edu/redmapper/}{\texttt{v6.3} catalogue for SDSS DR8}.
    While the SDSS images are much shallower than HSC, they also suffer less from the over-deblending
    issue that affects the red--sequence cluster finders using deeper data.
    The $u$-band image could also help improve the red--sequence redshift of low redshift clusters.
    
    Given the redshift coverage and richness completeness of SDSS \redm{}, we do not have enough 
    objects to perform \topn{} tests except for Bin 1. 
    We therefore define two SDSS \redm{} samples for our comparison: 
    1) 55 clusters in $0.19 < z < 0.50$ and $\lambda_{\rm SDSS} \geq 50$ (top panel of Figure 
    \ref{fig:sdss_redm}; 
    2) 191 clusters in $0.19 < z < 0.35$ and $\lambda_{\rm SDSS} \geq 20$ (bottom panel).
    We then use the same redshift bins and number density to select HSC \redm{} clusters and 
    massive galaxies.
    
    In Figure \ref{fig:sdss_redm}, we show that the \dsigma{} profiles of SDSS \redm{} clusters are
    not only consistent with the HSC \redm{} ones, they also display very similar systematic 
    differences with the \menve{50}{100}--selected massive galaxies.
    This reinforces our conclusions in \S\ \ref{sec:mstar_vs_richness} and 
    \S\ \ref{sec:richness_results}.

\section{The comparison of \texorpdfstring{\dsigma{}}{DSigma} profile between the HSC and DES \texorpdfstring{\redm{}}{redMaPPer} cluster}
	\label{app:des_redm}
	
	The Dark Energy Survey (DES) has adopted the \redm{} algorithm for finding galaxy clusters
	(\eg{} \citealt{Rykoff2016}).
	It would be interesting to compare the lensing profiles of HSC and DES \redm{} clusters.
	Since the overlapping area between DES Y1 and HSC \texttt{S16A} is very small, here we 
	directly compare the stacked \dsigma{} profile of DES \redm{} clusters at $0.20 \leq z < 0.55$
	and $20 \leq \lambda_{\rm DES} < 100$ presented in \citet{Chang2018} to their HSC counterparts.
	We ignore the small offset between richness measurements using different data and select
	285 HSC \redm{} clusters in the same richness and redshift bin.
	This roughly corresponds to the combination of the Bin 1 \& 2 in our \topn{} tests.
	In \citet{Chang2018}, the authors adopted the same cosmology but use comoving coordinates 
	instead. 
	Therefore we calculate a new \dsigma{} profile for HSC clusters using comoving coordinates as
	well. 
	
\begin{figure}
    \centering
    \includegraphics[width=\columnwidth]{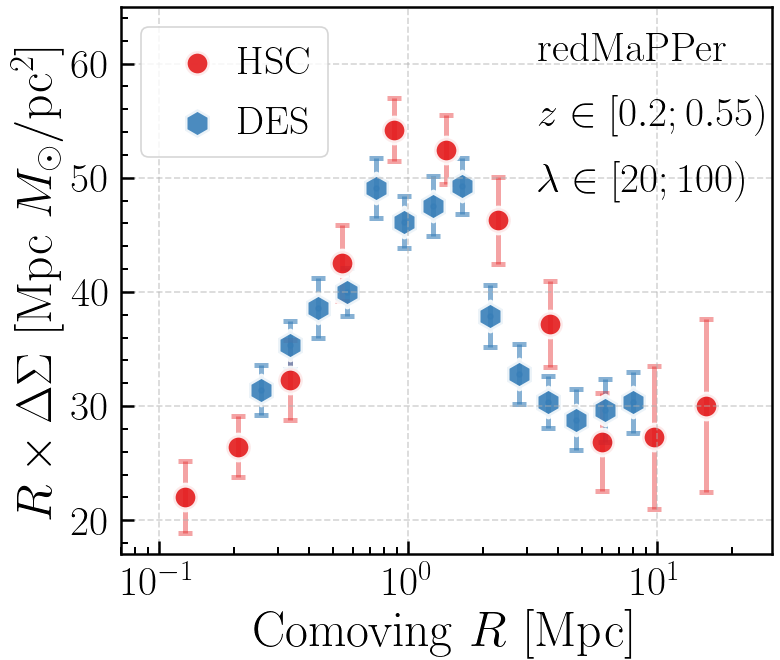}
    \caption{
        Comparison of the stacked \dsigma{} profiles of HSC and DES \redm{} clusters within 
        the same richness ($20 \leq \lambda < 100$) and redshift ($0.2 \leq z < 0.55$) bin.
        The DES \redm{} \dsigma{} profile is from \citet{Chang2018}.
        In different with the other \dsigma{} profiles in this work, we use \emph{comoving}
        coordinate here to be consistent with \citet{Chang2018}.
        The \texttt{Jupyter} notebook for reproducing this figure can be found here:
        \href{https://github.com/dr-guangtou/jianbing/blob/master/notebooks/figure/figG1.ipynb}{\faGithub}.
    }
    \label{fig:des_redm}
\end{figure}

	Figure \ref{fig:des_redm} shows that the \dsigma{} profiles for HSC and DES \redm{} clusters
	are broadly consistent with each other.
	Note that the DES \dsigma{} profile is based on an completely independent lensing catalogue 
	using different algorithms for shear measurement, lensing calibration, and photo-$z$ estimation.
	While the two profiles show subtle difference at $1 < R < 4$ Mpc, they show very similar overall
	shapes and amplitudes.
	This again shows that our results about the shape of lensing profiles of \redm{} clusters 
	should be robust against imaging dataset and lensing measurements.
	
\bsp
\label{lastpage}
\clearpage\end{CJK*}
\end{document}